\documentclass{article}
\usepackage{authblk}
\usepackage[utf8]{inputenc}
\usepackage{amssymb}
\usepackage{amsmath}
\usepackage[colorlinks=true,linkcolor=blue,citecolor=blue,urlcolor=blue,unicode=true]{hyperref}
\usepackage[capitalise]{cleveref}
\usepackage[paper=a4paper,text={15cm,20cm},centering]{geometry}

\usepackage[sorting=none,backend=bibtex]{biblatex}
\DeclareFieldFormat[article]{volume}{\textbf{#1}}
\renewbibmacro{in:}{}
\DeclareFieldFormat{pages}{#1}

\DefineBibliographyStrings{english}{%
  andothers = {\emph{et\addabbrvspace al\adddot}}
}
\usepackage{xpatch}
\usepackage{etoolbox}

\AtEveryBibitem{%
  \ifentrytype{article}{%
  }{}%
}

\usepackage[absolute,overlay]{textpos}
\textblockorigin{10mm}{10mm} 

\usepackage{mathtools}
\usepackage{amsfonts}
\usepackage{mathrsfs}
\usepackage{slashed}

\usepackage{graphicx}
\usepackage{color}
\usepackage{array}

\usepackage{placeins}
\usepackage{booktabs}
\usepackage{makecell}
\usepackage{subcaption}
\usepackage{floatrow}
\usepackage{siunitx}
\usepackage{rotating}

\usepackage{bm}
\usepackage{todonotes}

\title{\vspace{-100pt}\textbf{\LARGE{EMMI Rapid Reaction Task Force}}\\\vspace{20pt}\textbf{\huge{Few is different: deciphering many-body dynamics in mesoscopic quantum gases}}\\\vspace{12pt}}

\author[1]{Juergen Berges}
\author[2,28]{Sandra Brandstetter}
\author[3]{Jasmine Brewer}
\author[4]{Georg Bruun}
\author[1]{\underline{Tilman Enss} (organizer)}
\author[5]{Stefan Floerchinger}
\author[1,6,7]{Keisuke Fujii}
\author[2]{Maciej Ga{\l}ka}
\author[8,1]{\underline{Giuliano Giacalone}\thanks{Corresponding authors \href{mailto:giuliano.giacalone@cern.ch}{giuliano.giacalone@cern.ch} and \href{mailto:a.mazeliauskas@thphys.uni-heidelberg.de}{a.mazeliauskas@thphys.uni-heidelberg.de}} (organizer)}
\author[9]{Qingze Guan}
\author[2]{Carl Heintze}
\author[10,1]{Lars H. Heyen}
\author[11]{Ilya Selyuzhenkov}
\author[2]{\underline{Selim Jochim} (organizer)}
\author[12]{Jesper Levinsen}
\author[2]{Philipp Lunt}
\author[2,8]{\underline{Silvia Masciocchi} (organizer)}
\author[1]{\underline{Aleksas Mazeliauskas}\raisebox{-0.4ex}{\textsuperscript{*}}
 (organizer)}
\author[13,14]{Nir Navon}
\author[15]{Alice Ohlson}
\author[12]{Meera Parish}
\author[16]{Stephanie M. Reimann}
\author[17,18]{Francesco Scazza}
\author[19]{Thomas Sch{\"a}fer}
\author[20]{Derek Teaney}
\author[21]{Joseph Thywissen}
\author[22]{Raju Venugopalan}
\author[23]{Yangqian Yan}
\author[24,25]{Matteo Zaccanti}
\author[26,27]{Torsten V. Zache}

\affil[1]{\small{Institut f\"ur Theoretische Physik, Heidelberg University, Philosophenweg 16, 69120 Heidelberg, Germany}}
\affil[2]{Physikalisches Institut der Universit\"at Heidelberg, Im Neuenheimer Feld 226,\hspace{200pt}69120 Heidelberg, Germany}
\affil[3]{Rudolf Peierls Centre for Theoretical Physics, University of Oxford, Oxford OX1 3PU, United Kingdom}
\affil[4]{Department of Physics and Astronomy, Aarhus University, Ny Munkegade 120,\hspace{200pt}DK-8000 Aarhus C, Denmark}
\affil[5]{Theoretisch Physikalisches Institut, Friedrich-Schiller-Universität Jena,\hspace{200pt}Max-Wien-Platz 1, 07743 Jena, Germany}
\affil[6]{Department of Physics, Institute of Science Tokyo, Ookayama, Meguro, Tokyo 152-8551, Japan}
\affil[7]{Department of Physics, The University of Tokyo, Hongo, Bunkyo, Tokyo 113-0033, Japan}
\affil[8]{Theoretical Physics Department, CERN, 1211 Geneva 23, Switzerland}
\affil[9]{Department of Physics and Astronomy, Washington State University, Pullman, WA 99164, USA}
\affil[10]{Scientific Computing Center, Karlsruhe Institute of Technology, Hermann-von-Helmholtz-Platz 1, 76344 Eggenstein-Leopoldshafen, Germany}
\affil[11]{GSI Helmholtzzentrum für Schwerionenforschung, Planckstraße 1, 64291 Darmstadt, Germany}
\affil[12]{School of Physics and Astronomy, Monash University, Victoria 3800, Australia}
\affil[13]{Department of Physics, Yale University, New Haven, Connecticut 06520, USA}
\affil[14]{Yale Quantum Institute, Yale University, New Haven, Connecticut 06520, USA}
\affil[15]{Department of Physics, Lund University, Box 118, SE-221 00 Lund, Sweden}
\affil[16]{Mathematical Physics and NanoLund, Lund University, LTH, Box 118, 22100 Lund, Sweden }
\affil[17]{Department of Physics, University of Trieste, 34127 Trieste, Italy}
\affil[18]{CNR-INO -- Istituto Nazionale di Ottica, Consiglio Nazionale delle Ricerche, 34149 Trieste, Italy}
\affil[19]{Department of Physics, North Carolina State University, Raleigh, NC 27695, USA}
\affil[20]{Center for Nuclear Theory, Department of Physics and Astronomy,\hspace{200pt}Stony Brook University, New York 11794-3800, USA}
\affil[21]{Department of Physics, University of Toronto, Ontario M5S 1A7, Canada}
\affil[22]{Physics Department, Brookhaven National Laboratory, Upton, NY 11973, USA}
\affil[23]{Department of Physics, The Chinese University of Hong Kong, Shatin,\hspace{200pt}New Territories, Hong Kong, China}
\affil[24]{Istituto Nazionale di Ottica del Consiglio Nazionale delle Ricerche (CNR-INO),\hspace{200pt}50019 Sesto Fiorentino, Italy}
\affil[25]{European Laboratory for Non-Linear Spectroscopy (LENS), Università di Firenze,\hspace{200pt}50019 Sesto Fiorentino, Italy}
\affil[26]{Institute for Theoretical Physics, University of Innsbruck, Innsbruck, 6020, Austria}
\affil[27]{Institute for Quantum Optics and Quantum Information of the Austrian Academy of Sciences, Innsbruck, 6020, Austria}
\affil[28]{Current address: Department of Physics, Harvard University, Cambridge, Massachusetts 02138, USA}

\begin{document}

\date{}

\thispagestyle{empty}

\maketitle

\thispagestyle{empty}

\begin{abstract}
Emergent macroscopic descriptions of matter, such as hydrodynamics, are central to our description of complex physical systems across a wide spectrum of energy scales. The conventional understanding of these many-body phenomena has recently been shaken by a number of experimental findings. Collective behavior of matter has been observed in \emph{mesoscopic} systems, such as high-energy hadron-hadron collisions, or ultra-cold gases with only few strongly interacting fermions. In such systems, the separation of scales between macroscopic and microscopic dynamics (at the heart of any effective theory) is inapplicable. To address the conceptual challenges that arise from these observations and explore the universality of emergent descriptions of matter, the EMMI Rapid Reaction Task Force was assembled. This document summarizes the RRTF discussions on recent theoretical and experimental advances in this rapidly developing field. Leveraging technological breakthroughs in the control of quantum systems, we can now quantitatively explore what it means for a system to exhibit behavior beyond the sum of its individual parts. In particular, the report highlights how the (in)applicability of hydrodynamics and other effective theories can be probed across three principal frontiers: the size frontier, the equilibrium frontier, and the interaction frontier.
\end{abstract}
\newpage

\tableofcontents

\thispagestyle{empty}

\newpage

\section{Science and motivation of the RRTF}

\subsection{The physical world seen through the lens of emergent phenomena}

A recurring and yet profound theme in physics is that simple, elegant laws often enable us to transcend the extraordinary complexity of the physical phenomena that we observe in our universe. From the shape of atomic nuclei to the evolution of matter in galaxies, our description of natural phenomena is expressed in the language of emergent collective phenomena, that is, effective theories, such as thermodynamics or hydrodynamics, that summarize the behavior of complex many-body systems in terms of a few macroscopic variables. 
Building on universal notions such as symmetry and conservation laws, the simplifications brought by effective descriptions of matter are not mere approximations. They often capture phenomena that become manifest only when a system becomes complex enough. In 1972, P. W. Anderson famously stated that \textit{More is different} \cite{Anderson:1972pca}, which encapsulates the idea that adding more particles to a system does not merely scale up its behavior, but can fundamentally change it, creating new kinds of patterns that are invisible in the constituent parts.

Today, we are in a unique historical moment. Thanks to recent technological breakthroughs that allow the preparation and manipulation of quantum systems under controlled conditions, we are in a position to take Anderson’s maxim as the trigger for a new program of quantitative research. With tools ranging from quantum simulators and ultracold atoms to high-energy collider experiments, we can engineer systems where the notion of \textit{more} can be tuned systematically. We can incrementally add complexity, control the strength and nature of interactions, break scale separations, and ultimately observe in real time how \textit{different} emerges or fails to emerge. The experimental findings in mesoscopic systems are
challenging the textbook knowledge of hydrodynamics and the microscopic interpretation of transport phenomena.

Therefore, the ability to dissect the emergence of collective behavior opens a new avenue of research. 
The precise understanding of the boundaries of validity of effective theories, or what governs the transition between microscopic chaos and macroscopic order, become now the subject of data-driven inquiry. Once more, this does not simply amount to testing equations, but rather to thoroughly assessing what it means for something to be more than the sum of its parts. 

This fresh and ambitious perspective motivated the formation of the EMMI Rapid Reaction Task Force \textit{Deciphering many-body dynamics in mesoscopic quantum gases}, whose discussions took place at Heidelberg University in March 2024~\cite{EMMI_RRTF}. This document is the summary of the workshop, presenting central ideas from this emerging area of research, which we believe holds the promise of establishing a new frontier within the discipline of physics as a whole.

\subsection{Hydrodynamic behavior across the temperature  spectrum: from the ultra-hot QGP to ultra-cold atoms}\label{sec:hydro_2}

The focus of the RRTF was largely on the emergence of hydrodynamics as an effective collective description of matter. Hydrodynamics provides a powerful framework to encapsulate the complex dynamics of a system of essentially infinite constituents into simple equations that are solely based on scale separation and conservation laws \cite{Rezzolla:2013dea}. The universal character of the hydrodynamic description is particularly striking in the two extreme  regimes of quantum matter discussed below.

Around the start of the new century discoveries of two new states of matter were announced: $(i)$ The experimental realization of a Bose-Einstein condensate, enabled by advances in the cooling and manipulation of atoms in table-top setups \cite{Anderson:1995gf,Davis:1995pg}; $(ii)$ The experimental evidence of the formation of a quark-gluon plasma (QGP) \cite{STAR:2005gfr,PHENIX:2004vcz,PHOBOS:2004zne,BRAHMS:2004adc} following the start of the high-energy nuclear collision program at the BNL Relativistic Heavy Ion Collider (RHIC).

\begin{figure}[t]
    \centering
    \includegraphics[width=0.85\linewidth]{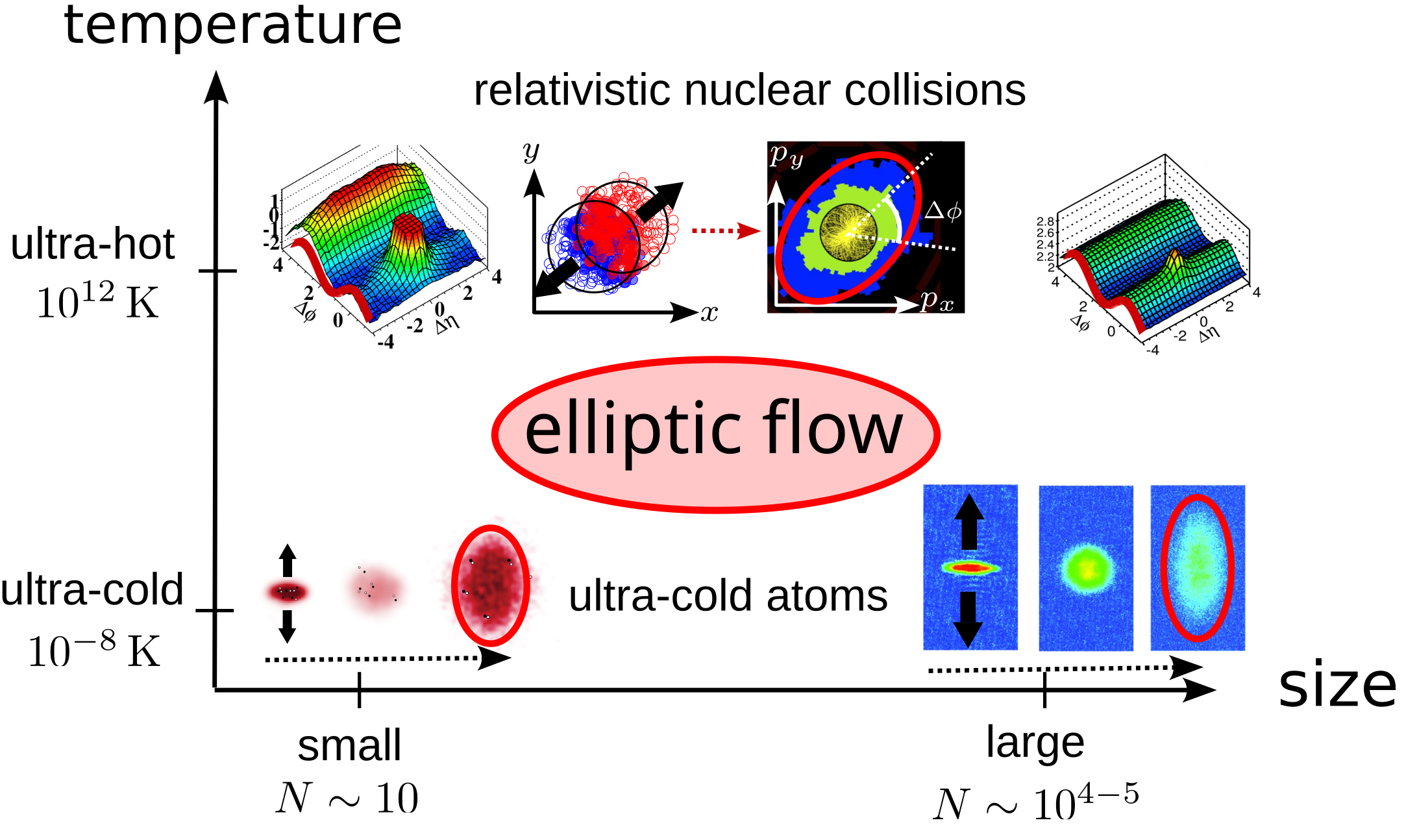}
    \caption{Signatures of emergent hydrodynamic behavior are observed in microscopic systems ranging from ultra-cold to ultra-hot temperatures. While first established in \textit{large} systems with tens of thousands of constituent particles, recent experiments imply the emergence of collective phenomena in the \textit{mesoscopic} regime. In all cases, emergent collective behavior has been signaled by the observation of interaction-driven elliptic flow. 
    Figure elements adapted from Refs.~\cite{CMS:2012xss,CMS:2010ifv,O_Hara_2002,Brandstetter:2023jsy,Velkovska2021}.
    }
    \label{fig:hydro_vs_T_N}
\end{figure}

Remarkably,  both systems have become platforms for testing universality of the hydrodynamic description in extreme conditions, as illustrated in \cref{fig:hydro_vs_T_N}~\cite{Schafer:2009dj,Adams:2012th}. Both ultracold atomic gases and the quark-gluon plasma are systems with a relatively small number of constituents, typically $N\approx10^4-10^6$, which is much smaller compared to Avogadro’s number, $N\approx10^{23}$. On the other hand these systems probe opposite extremes of the temperature scale: trillions of kelvin for the QGP, while ultracold gases are cooled to the nanokelvin regime. 

\paragraph{Heavy-ion collisions.} In collisions of heavy nuclei at ultrarelativistic energies an exotic state of matter known as the QGP is created in which quarks and gluons (collectively known as partons) are no longer confined into hadrons~\cite{Busza:2018rrf,Elfner:2022iae}.  During the initial stages of a heavy-ion collision, the system is characterized by an inhomogeneous initial energy density resulting from the random positions of the colliding protons and neutrons (nucleons) within the incoming nuclei.  This initial energy distribution rapidly evolves into a QGP phase, which then continues to cool and expand, ultimately transitioning to a gas of hadrons that fly to the detectors of the experiment.  

In these collisions, the overlap region of the two nuclei is typically not rotationally symmetric in the plane transverse to the beam axis. To first order, the overlap region can be described by an almond shape or ellipse, given by the geometrical (optical) overlap of the nuclei (see the sketch in the upper part of \cref{fig:hydro_vs_T_N}). Because of this, pressure gradients build up in the QGP in an \textit{anisotropic} fashion, causing emitted particles to be boosted preferentially along some direction dictated by the asymmetry of the initial density profile. Therefore, initial asymmetry in position space converts into anisotropy in momentum space~\cite{Ollitrault_1992}.  For an initially elliptical shape, this corresponds to the \textit{elliptic flow} phenomenon, which was the main discovery of the heavy-ion program at the 
Relativistic Heavy Ion Collider (RHIC)~\cite{PHENIX:2003qra,
STAR:2004jwm}. This discovery was later confirmed at the 
Large Hadron Collider (LHC) \cite{ALICE:2010suc}. The 
quantitative analysis of the transverse expansion of 
the fireball created in a heavy ion collision is based
on the flow coefficients $v_n(p_T)$, defined by 
\begin{align}
\label{v_2}
 \left. p_0\frac{dN}{d^3p}\right|_{p_z=0} &=& 
 \left. p_0\frac{dN}{\pi dp_T^2dp_z}\right|_{p_z=0}
 \Big( 1 + 2v_1(p_T)\cos(\phi-\Psi_1) 
 + 2v_2(p_T)\cos(2(\phi-\Psi_2)) +\ldots \Big) .
\end{align}
Here, $p_z$ is the momentum in the beam direction, $p_T$ 
is the transverse momentum, and $\phi$ is the angle in 
the transverse plane relative to the direction of the
impact parameter. The coefficient $v_2$ is referred to
as elliptic flow, and the higher moments are termed
triangular, quadrupolar flows, etc.~\cite{Ollitrault:2023wjk}. The flow angles 
$\Psi_i$ take into account the fact that the flow angles 
need not be aligned with the impact parameter plane. Alternatively, the harmonic flows can be studied by Fourier decomposition of the two-particle correlation function. As depicted in the upper right corner of \cref{fig:hydro_vs_T_N}, the modulations in the relative azimuthal angle $\Delta \phi$ that are long range in the longitudinal angle (pseudorapidity) $\Delta \eta$ are interpreted as the hydrodynamic response to the initial spatial deformations.

Substantial elliptic flow, reaching about $v_2
(p_T\!=\!2\,{\rm GeV}) \simeq 20\%$ in semi-central 
collisions, was observed in the early data from RHIC, 
and interpreted as the consequence of nearly ideal 
hydrodynamic expansion driven by the pressure gradients
in the almond-shaped overlap region of the two nuclei \cite{Romatschke:2007mq}. 
 Later, it was realized that fluctuations in the positions of the initial nucleon-nucleon collisions lead to non-elliptical, higher-order shapes of the initial energy density distribution, which are the source of the substantial higher harmonics, 
including odd Fourier moments such as $v_3$
\cite{Alver:2010gr}, that are observed experimentally.

   The quantitative analysis of the flow data is 
based on relativistic fluid dynamics, combined with 
models of the initial state, and kinetic theories 
that describe the evolution of the hadronic phase 
after the quark gluon fluid freezes out~\cite{Romatschke:2017ejr,Kolb:2003dz}. These studies
show that some amount of dissipation is needed to 
describe the data. In particular, a non-zero shear 
viscosity is required to understand the $p_T$ dependence
of elliptic flow, the relative magnitude of higher 
flow harmonics, and the evolution of flow in going
from semi-central to peripheral collisions. Recent Bayesian studies find that the shear viscosity to entropy
density ratio near the deconfinement transition is 
$\eta/s\simeq 0.15$, with uncertainties of order 50\%
\cite{Bernhard:2019bmu,JETSCAPE:2020shq,Nijs:2020ors,Heffernan:2023gye,Virta:2024avu}.
Remarkably, this value is just above the value found in infinitely strongly interacting quantum field theories~\cite{Kovtun:2005}.

\paragraph{Ultracold atomic gases.}   Shortly after the discovery at RHIC,   
large elliptic flow was also observed in a completely 
different system: ultracold atomic Fermi gases in 
which the interaction between the atoms has been 
tuned to the limit of infinite scattering length.
In the original setup atoms were optically trapped 
in a deformed potential. The experiment then studied
the time evolution of the aspect ratio $A_R=(\langle
x_T^2\rangle/\langle z^2 \rangle)^{1/2}$ after release from
the trap \cite{O_Hara_2002}. Elliptic flow is reflected
in a shape inversion from $A_R<1$ at early time to 
$A_R>1$ at late time (see lower right corner of \cref{fig:hydro_vs_T_N}). These experiments provide a 
more direct observation of pressure driven collective 
flow than the heavy ion experiments, because we can
study the entire time evolution including the initial
state, and do not have to rely solely on the observation of 
final state particles. 

 The Fermi gas experiments have been analyzed using
viscous fluid dynamics. As in the heavy ion case, the 
data require some viscosity to be present. At unitarity, 
where the system is expected to be scale invariant, the 
data favor a non-zero shear viscosity but vanishing bulk
viscosity \cite{Elliott_2014}. An important consistency 
check for the flow analysis is that one can show that 
the high temperature limit of the viscosity, $\eta = 
(0.265 \pm 0.020)(mT)^{3/2}$ \cite{Bluhm:2017rnf}, agrees 
very well with the kinetic theory prediction $\eta=15/
(32\sqrt{\pi})(mT)^{3/2}$ \cite{Bruun_2005}. This result
increases our confidence in the determination of the shear
viscosity near the critical temperature for superfluidity,
which is found to be $\eta/s\simeq 0.5\pm 0.1$
\cite{Bluhm:2017rnf}. Considerable uncertainty remains 
regarding the behavior of the viscosity below $T_c$. 
Elliptic flow experiments indicate that the viscosity 
drops below $T_c$ \cite{Hou:2021xra}, but this result 
is not consistent with the expectation from kinetic theory
\cite{Rupak:2007vp}, nor does it agree with experimental
studies of sound attenuation below $T_c$ \cite{Patel:2019udb}.
 The most recent generation of Fermi gas experiments 
is based on ``box traps'' that enable experiments on 
a uniform Fermi gas. This avoids difficulties related
to the theoretical treatment of the dilute corona 
of a trapped Fermi gas. A number of groups have 
performed linear response and collective mode 
experiments on uniform Fermi gases, see 
\cite{Patel:2019udb,Baird_2019,LiPan2022,Li_2024,YanZwierlein2024}.
These experiments enable the separate extraction of 
shear viscosity and thermal conductivity, and provide
access to the full set of transport coefficients in 
the superfluid phase. Linear response experiments can
map the full response as a function of wave number, 
and shed light on the breakdown of hydrodynamics 
at short wavelengths.

\subsection{The small system revolution}

In 2010 the observation of anisotropic flow in the form of long-range azimuthal correlations of final-state hadrons had been viewed for a decade as a signature of the formation of a strongly-interacting fluid-like medium in heavy-ion collisions. It was a striking surprise that the first LHC results obtained in $p$+$p$ collisions~\cite{CMS:2010ifv} and later in $p$+$^{208}$Pb collisions~\cite{ALICE:2012eyl} also showed evidence of the same behavior. Namely, pairs of particles were correlated in azimuthal angle ($\phi$) despite being separated in the longitudinal direction by a large pseudorapidity ($\eta$) gap, see the \emph{ridge} structures in the two-particle correlation functions displayed in the top row of \cref{fig:hydro_vs_T_N}. Originally, small collision systems were meant to provide a baseline of cold-matter effects that one could compare nucleus-nucleus collisions to. It was indeed entirely unexpected that collective flow phenomena may become manifest in systems that are in principle too \textit{small} to produce a QGP \cite{Nagle:2018nvi,Schenke:2021mxx}, or to build up a flow field defined by a global plane of symmetry.

These observations have naturally triggered a vast research program. Detailed measurements of $v_n$ coefficients have since then been made across a wide range of system sizes, including light-heavy ion collisions at RHIC (such as $d$+Au, $^{3}$He+Au \cite{PHENIX:2018lia,STAR:2022pfn}), with more expected to come from the recent light-ion collisions at LHC ($p$+$^{16}$O, $^{16}$O+$^{16}$O, and $^{20}$Ne+$^{20}$Ne collisions \cite{Brewer:2021kiv,Loizides:2025ule}). On the side of theory, phenomenological models that implement a hydrodynamic description of $p$+$p$ or $p$+$A$ collisions have been able to explain the observed trends in the data, confirming the \textit{qualitative} evidence that the observed flow anisotropies emerge as a response to the geometry of the initial conditions of these systems \cite{Schenke:2020mbo}. In addition, remarkable progress has been made in elucidating the applicability of hydrodynamics under conditions that extend beyond those dictated by conventional textbook criteria~\cite{Berges:2020fwq,Soloviev:2021lhs}.

Fifteen years later, the interpretation of flow-like signals in small collision systems remains a very active area of research and development \cite{Nagle:2018nvi,Noronha:2024dtq,Grosse-Oetringhaus:2024bwr}. These findings give evidence of emergent collective behavior, largely consistent with hydrodynamic expectations, in experimental conditions where such an effective description should not work. However, understanding the detailed dynamical processes that lead to emergent collective behavior in a collider experiment is hard, as one has only a limited control of the system that is produced because variables such as the impact parameter of the collisions or the degree of anisotropy of the initial conditions cannot be directly controlled. In this respect, the unique capabilities of experimental platforms with cold atomic gases help fill an important gap. As we shall see, they enable us to address the emergence of collective phenomena under conditions that are fully tunable. For instance, in the experiments performed at Heidelberg University \cite{Brandstetter:2023jsy}, discussed in the next sections, evidence of elliptic flow has been reported in systems made of as few as 10 fermions, as depicted in the lower left corner of \cref{fig:hydro_vs_T_N}, with tunable initial geometry and interaction strength.

\subsection{Breaking textbook criteria: scale analysis}

As this work focuses on high-temperature QCD matter (QGP) and ultra-cold quantum gases, it is worth drawing a parallel between these two areas and providing more quantitative arguments through a scale analysis to clarify what we mean by a \emph{small system puzzle} in both fields.

\paragraph{\textbf{Ultra-hot quark-gluon plasma.}}

The matter formed in the overlap of two nuclei at high energy appears to reach a state close to local thermal equilibrium which enables a hydrodynamic description. This is reached within the first $\mathcal{O}(1)$ fm/$c$ of the collision, a scale much shorter than the typical system size, which is $\mathcal{O}(10)$ fm. This separation of scales ensures that the system reaches local equilibrium before it falls apart because of transverse expansion.

Another way to see this is by considering that the QGP behaves like a collisional fluid. In the non-relativistic limit, its dynamics is governed by the Navier-Stokes equation (neglecting the bulk viscosity for simplicity),
\begin{equation}
    \rho \frac{d\mathbf{v}}{dt} = -\nabla P + \eta \nabla^2 \mathbf{v}, \tag{2.34}
\end{equation}
where $\rho$ is the mass density, $d/dt = \left( \frac{\partial}{\partial t} + \mathbf{v} \cdot \nabla \right)$, $P$ is the pressure, and $\eta$ is the shear viscosity parameter, which is zero for an ideal fluid. The viscous correction acts against the effect of the pressure gradient. A hydrodynamic description is meaningful as long as these corrections are small compared to the acceleration term that comes from the pressure gradients. This is equivalent to 
\begin{equation}
    {\rm Kn} \equiv \frac{\eta}{\rho \, v} \frac{1}{R} \ll 1,
\end{equation}
where $R$ is the system size and $v$ is the typical thermal velocity. This quantity is also known as the Knudsen number, defined as the ratio between the mean free path in the gas ($\lambda = \eta / (\rho\,v)$) and the macroscopic size of the system ($R$).

In relativistic systems the corresponding control parameter of viscous corrections is~\cite{Teaney:2009qa}
\begin{equation}
    {\rm Kn} = \frac{\eta}{sT} \frac{1}{R}
\end{equation}
where $\eta/s$ is the shear viscosity over entropy ratio, which is constant in conformal systems. The characteristic temperature of QGP for a fixed size system can be related to the final number of  charged hadrons $N_\text{ch}\propto T^3$ measured perpendicular to the beam axis over a finite window of the longitudinal angle~\cite{Arslandok:2023utm}. Therefore parametrically the Knudsen number scales with the number of particles as
\begin{equation}
    {\rm Kn} \sim N_\text{ch}^{-1/3}.\label{eq:Knrel}
\end{equation}
More detailed analysis by Teaney in \cite{Arslandok:2023utm} for a relativistic system shows that viscous corrections are controlled by the opacity $\chi\sim \text{Kn}^{-1}$ of the system. For favorable experimental conditions such as in large nuclei collisions at top LHC energy, and for $\eta/s\approx 0.16$, one has $\chi\approx 3$. While this is reasonably larger than unity, one generally does not have the strong separation of scales underlying everyday classical fluids. For this reason, we often refer to the hydrodynamic description of the QGP as \textit{unreasonably effective}. 

How about small systems?  The dramatic reduction in system size implies that one cannot talk anymore about a separation of scales.  The size of the system is of the same order as the time that is needed to locally equilibrate. Recent kinetic theory calculations \cite{Kurkela:2019kip,Ambrus:2022koq,Ambrus:2022qya,Arslandok:2023utm} suggest indeed that there is no scenario in which a system as small as that formed in $pp$ collisions can reach local equilibrium before breaking apart under the effect of the transverse expansion.

\paragraph{\textbf{Ultra-cold atom gases.}}

We perform a similar analysis in the context of ultra-cold quantum gases. We consider systems of bosons (or paired fermions) at $T=0$ that belong to a condensate state. The standard formalism describing the emergent fluid-like dynamics of the field representing the macroscopic condensate wavefunction corresponds to the Gross-Pitaevskii equation (GPE \cite{Dalfovo:1999zz}), that is,
\begin{equation}
i\hbar \frac{\partial \psi}{\partial t} = \left( -\frac{\hbar^2}{2m} \nabla^2 + V_{\text{ext}}(\mathbf{r}) + g|\psi|^2 \right) \psi.
\end{equation}
Here $g = 4\pi \hbar^2 a_s / m$ is the interaction strength, $a_s$ is the $s$-wave scattering length parameter, $m$ is the boson mass, and $V_{\text{ext}}$ is the external trapping potential. The particle density is $n=|\psi|^2$.

Analogously to the Navier-Stokes hydrodynamics discussed above, the ideal hydrodynamics is recovered in the limit where the kinetic term, which contains an additional gradient of the density, is negligible compared to the interaction term, i.e., the Thomas-Fermi condition \cite{Dalfovo:1999zz}. The resulting macroscopic description is typically referred to as \textit{superfluid hydrodynamics}, which differs from the usual collisional hydrodynamics derived from kinetic theory. If the quantum pressure term is not neglected, solving the GPE amounts to finding the solution to a second-order fluid dynamical problem. In either case, the validity of the GPE is only granted if a diluteness condition applies to the system under consideration,
\begin{equation}
n a_s^3 \ll 1,
\end{equation}
implying a scale separation between the typical distance associated with particle interactions ($a_s$) and the particle spacing, $n^{-1/3}$. This condition ensures that only two-body local interactions contribute to the Hamiltonian. 
Note that diluteness should still be accompanied by a well-defined notion of a \textit{condensate} made of a macroscopic number of bosons, $N\gg1$. 

Therefore, even at zero temperature, a large number of constituents as well as a strong separation of scales in place are required to warrant a macroscopic description based on the GPE.
Corrections to the Thomas-Fermi limit scale with the ratio of the boundary region (healing length $\xi\sim\mu^{-1/2}$) to the size of the cloud, $\xi/R \sim \mu^{-1} \sim N^{-2/(D+2)}$. Therefore we conclude that the superfluid limit should be reached faster in lower dimensions. For the applicability of ordinary hydrodynamics, we can consider how the mean free path and the system size scale with the particle number in a harmonic trap. The typical size is given by the Thomas-Fermi radius $R\sim \mu^{1/2}\sim N^{1/2D}$, where the chemical potential for a trapped degenerate Fermi gas scales as $\mu\sim N^{1/D}$~\cite{Ketterle:2008}.  At resonant scattering, the mean free path $\lambda\sim n^{-1/D}$ is given by the average particle spacing \cite{Schafer:2009dj}, where $n\sim N/R^D \sim N^{1/2}$ denotes the peak number density in the system.
In this unitary regime, the Knudsen number scaling thus simplifies to
\begin{equation}
    \text{Kn} \sim N^{-1/D},\label{eq:knnonrel}
\end{equation}
which agrees with the scaling in relativistic systems in \cref{eq:Knrel} for $D=3$. We conclude that for the same number of atoms in a trap, the ordinary fluid dynamic behaviour would be reached faster for $D=2$.

How about small systems? They are rather new in the field of cold atoms but are nevertheless expected to attract growing attention, due to recent technological breakthroughs enabling single-atom-resolved imaging \cite{Parish}, as well as deterministic control over their numbers (down to one atom) and interactions. For the experiments conducted at Heidelberg University discussed later in this manuscript, all the features of the quantum problem (number of quanta, interaction strength, system geometry) can be deterministically controlled down to a single atom. A first campaign of hydrodynamic-behavior searches has established that elliptic flow, $v_2$, is prominent in a strongly-interacting Fermi gas at zero temperature with only 10 atoms \cite{Brandstetter:2023jsy}. The emergence of a GPE-type dynamics has yet to be thoroughly assessed. These experiments are devised with the specific purpose of breaking a separation of scales: the scattering length is tuned to be of the same order as the mean free path and system size, dictated by the trapping potential. In addition, the small number of quanta should rule out any mean-field-based picture. The observation of emergent hydrodynamic behavior in these systems opens, thus, a new \textit{small system puzzle} at ultra-cold temperatures.

\paragraph{\textbf{A new beginning.}} In summary, on the side of high-energy collisions and the study of ultra-hot QCD matter, we have data from a comprehensive program of experiments related to collectivity phenomena across different energies and system sizes, from large to small. At the same time, on the side of ultra-cold gases, we now have the tools to engineer quantum systems with fully tunable setups where the standard conditions of many-body theories are violated.

All in all, this reinforces our initial claim that the physics community as a whole is in a unique position to study the validity of notions and effective descriptions of matter that have been central to the understanding of macroscopic physical phenomena for decades, if not centuries. This calls for a better understanding of the limitations of textbook criteria based on scale separation, as well as new ideas to develop modern definitions of terms such as \textit{hydrodynamics} or \textit{collective behavior}. This is the motivation driving the EMMI RRTF discussions and the writing of this report, which continues previous EMMI supported efforts of connecting the two fields \cite{EMMI_previous}, albeit focusing on a completely new direction of investigation.

\subsection{Structure of the report}

This document is not intended to be a comprehensive review, but rather as a survey of perspectives, which hopefully provides a sufficiently diverse snapshot of current outstanding questions while emphasizing relevant future research directions. The report is divided into two parts.

\bigskip
\textbf{Cold atoms.}~In the first part, \cref{sec:statusCA}, we discuss the status and prospects of collectivity studies using different ultra-cold atom systems. Particular emphasis is on the applicability of many-body physics in mesoscopic systems. We start with \cref{sec:fewatoms}, where we present remarkable results indicating that for some static properties of Fermi gases, two-body systems already present features experimentally observed in the many-body limit. Along this line, we present further experimental and theoretical perspectives.
\begin{itemize}
    \item Experiment --  In \cref{sec:exp_overview} we provide an overview of experimental platforms that probe the collective dynamics at different extremes. \Cref{sec:flow} presents the observation of elliptic flow in the $5+5$ atom system. \Cref{sec:diffusivity} explores the applicability of hydrodynamics at high frequencies and short distances. \Cref{sec:uniformFG} discusses the emergence of hydrodynamics with increasing interaction strength. \Cref{sec:mixedFG,sec:composite} present two new cold atom platforms that are currently being developed, using different isotopes or atom species mixtures to study emergent collective phenomena.
    \item Theory -- The theory perspectives are given in \cref{sec:theory_overview}. In \cref{sec:microscopic} we discuss the state-of-the-art of exact few-body computations. \cref{sec:CAfluid} presents the fluid dynamic interpretation of the elliptic flow observations discussed in \cref{sec:flow}. \cref{sec:ECG} presents variational methods for solving few-body systems, while \cref{sec:dissipation} is the theory counterpart of \cref{sec:diffusivity}. \cref{sec:nonhydro,sec:attractor} discusses the far-from-equilibrium behavior of cold atomic gases, as the interplay of exponentially decaying non-hydrodynamic modes and a hydrodynamic attractor component. Finally, \cref{sec:entanglement} discusses thermalization in a few-body system from the perspective of entanglement entropy.
\end{itemize}

\textbf{Heavy ions.}~The second part of the report, \cref{sec:high-energy}, briefly addresses the same types of issues from the perspective of high-energy nuclear and hadron collisions. Contrary to the field of cold atoms, small system collectivity in high-energy collisions has been an important topic of research and debate for a number of years, and we refer to existing comprehensive reviews for broader perspectives on the topic~\cite{Nagle:2018nvi,Noronha:2024dtq,Schenke:2021mxx,Grosse-Oetringhaus:2024bwr}. 
\begin{itemize}
    \item Experiment -- We summarize the current puzzle on small system collectivity in high-energy collisions in \cref{sec:smallpuzzle}, while \cref{sec:multi,sec:HIprospects} give experimental prospects on upcoming measurements. 
    \item Theory -- We address questions related to the thermalization of the QGP within the framework of the hydrodynamic attractors in \cref{sec:HIattractor}, as well as recent insights into novel probes of the quantum entanglement of hadrons in ultra-peripheral collisions in \cref{sec:UPCent}.
\end{itemize}

Finally, in \cref{sec:conclusions} we recap the rationale of the RRTF and its driving questions, and outline future research directions.

\section{Status and prospects with ultra-cold atomic gases}\label{sec:statusCA}

\subsection{Apparent applicability of many-body physics with few atoms. Importance of two-body physics.}
\label{sec:fewatoms}

It is well established that few-body problems can provide important insights into nonrelativistic quantum many-body systems. A prominent example is the fractional quantum Hall effect, which was explained by Laughlin using a wave function based on a near-exact solution of the electronic three-body problem~\cite{Laughlin1983}. Similarly, the development of the theory of superconductivity relied on the understanding provided by the Cooper pair problem, which involved two interacting electrons above an inert Fermi sea. 

Most recently, investigations with ultracold atomic gases have suggested that the properties of strongly interacting two-component ($\uparrow,\downarrow$) Fermi systems can be extracted from just a small number ($\leq$ 10) of fermionic atoms~\cite{Wenz2013,Holten2022}.
Indeed, experimental studies of the one-dimensional (1D) Fermi gas in the limit of large spin imbalance have observed a fast convergence of the interaction energy to the predicted many-body limit~\cite{Astrakharchik2013} for remarkably few fermions~\cite{Wenz2013}. This has been further bolstered by follow-up theoretical works on 1D Fermi gases showing a similarly fast convergence with particle number for arbitrary polarization and interaction strength~\cite{Grining2015,Laird2017}, and there are indications that the behavior of two-dimensional Fermi gases can also be captured by a few particles~\cite{Holten2022,Laird2024}.

\begin{figure}[t]
\centering
\includegraphics[width=0.6\textwidth]{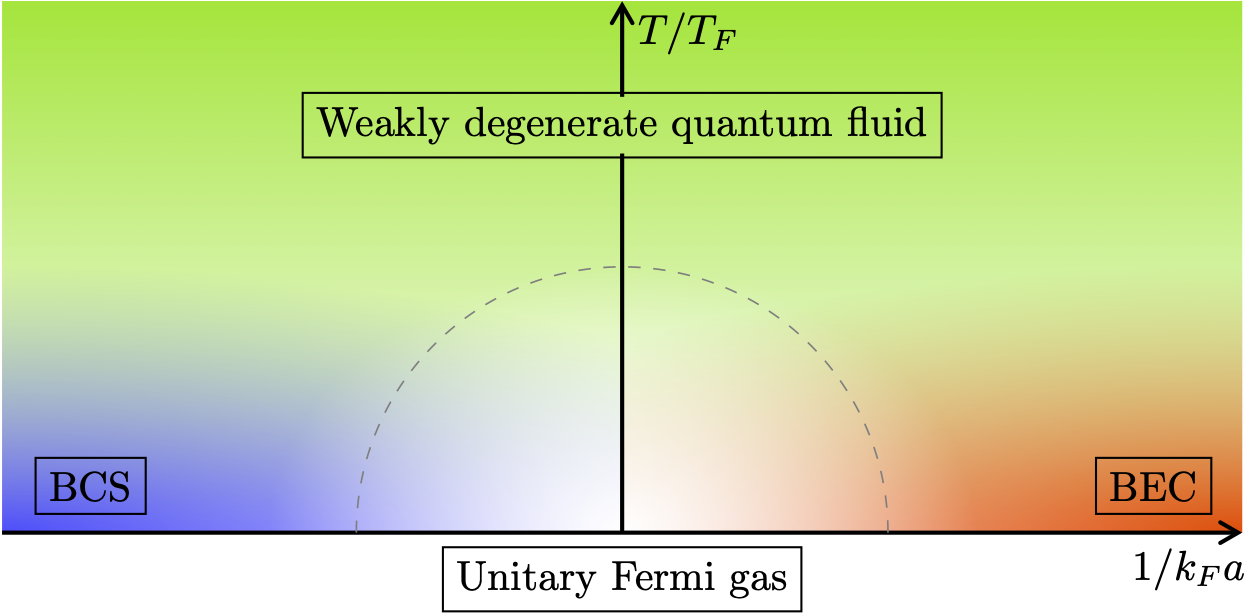}
\caption{\label{fig:UFGsketch} 
Regimes of the two-component 3D Fermi gas, with density parameterized by the Fermi momentum $k_F$ and Fermi temperature $T_F$. In the BCS regime $1/k_Fa \to -\infty$, there are weakly bound pairs of $\uparrow$ and $\downarrow$ fermions at low temperature $T\ll T_F$, while the opposite regime $1/k_Fa \to \infty$ corresponds to tightly bound $\uparrow\downarrow$ bosons that can form a Bose-Einstein condensate (BEC). In the high-temperature regime, $T\gtrsim T_F$, the system is a weakly degenerate quantum fluid whose properties are governed by few-body clusters. The strongest correlations occur at unitarity $1/k_Fa=0$, making the low-temperature unitary Fermi gas  (region enclosed by dashed line) challenging to model.}
\end{figure}

Here we review the case of the three-dimensional (3D) two-component Fermi gas, as depicted in Fig.~\ref{fig:UFGsketch}, where the short-range interactions are parameterized by a single length scale --- the two-body $s$-wave scattering length $a$. In the unitarity limit $a \to \pm \infty$, the interaction length scale drops out of the problem, yielding a paradigmatic example of a universal strongly correlated system~\cite{Bloch2008mbp,Giorgini2008RMP,Zwerger2012tbb,Chevy2012}. The lack of a small parameter at low temperatures $T$ also makes the unitary Fermi gas challenging to describe theoretically. On the other hand, the few-body physics at unitarity is more tractable and has been extensively studied (see, e.g., Ref.~\cite{Blume2012} and references therein). Furthermore, our understanding of the regimes surrounding the unitary Fermi gas in the strongly correlated quantum degenerate limit is based heavily on few-body physics (Fig.~\ref{fig:UFGsketch}): From pairing at the Fermi surface in the BCS limit and Bose-Einstein condensation of tightly bound pairs in the BEC limit, to the high-temperature case where we have a weakly degenerate quantum fluid whose deviations from Boltzmann statistics are determined by few-body virial coefficients~\cite{Liu2013virial}. Therefore, it is natural to ask whether the few-body limit can yield quantitative insight into the ground state of the unitary Fermi gas. We address this question by following the treatment in Ref.~\cite{Levinsen2017}, which illustrates how many-body physics seems to emerge from a few atoms.

We focus on the most symmetric scenario of an $N$-fermion system with equal numbers of each spin, i.e., $N_\uparrow=N_\downarrow = N/2$, and equal masses $m_\uparrow = m_\downarrow \equiv m$. To connect the results of few-body calculations with many-body physics, one needs to consider particles confined to a finite volume, such that the few-body system has an effective density. This can, for instance, be achieved by applying an isotropic harmonic potential $V(\mathbf{r}) = \frac{1}{2}m\omega^2 r^2$, where $\omega$ is the trapping frequency. An appealing approach, carried out in Ref.~\cite{Levinsen2017}, is to then apply the local density approximation (LDA) ``in reverse'', in the sense that we infer the uniform-space thermodynamic properties from those of the trapped system. Within the LDA, one assumes that the system adiabatically adjusts to 
a local chemical potential $\mu(\mathbf{r}) = \mu - V(\mathbf{r})$~\cite{PethickSmith2008book}, where the global chemical potential $\mu = \frac{\partial \mathcal{E}}{\partial N}$ is derived from the total energy $\mathcal{E}$ of the trapped gas. 

Of particular interest is the so-called Bertsch parameter $\xi$~\cite{Heiselberg2012}, a dimensionless quantity which, due to scale invariance, allows one to relate the ground-state properties of the non-interacting and unitary gases. For instance, in uniform space, we have the chemical potential $\mu = \xi E_F$, where the Fermi energy $E_F = k_F^2/2m$ and Fermi momentum $k_F$ corresponds to those for a non-interacting Fermi gas of the same density. 
Using the LDA then yields $\mathcal{E}=\sqrt{\xi} \mathcal{E}_0$ in the trapped system, where $\mathcal{E}_0 = \frac\omega4(3N)^{4/3}$ is the energy of the non-interacting Fermi gas within LDA. In principle, one can extract the Bertsch parameter directly from the energy of the trapped few-body system using $\xi=(\mathcal{E}/\mathcal{E}_0)^2$. However, in order to minimize the effects due to the shell structure and zero-point energy, it is convenient to subtract and add the exact non-interacting energy $\mathcal{E}_{\mathrm{NI}}$ for the $N$-body system, yielding $\xi=\left[1+(\mathcal{E}-\mathcal{E}_{\mathrm{NI}})/{\mathcal{E}_0}\right]^2$~\cite{Levinsen2017}. Here, $\mathcal{E}_{\mathrm{NI}} \to \mathcal{E}_0$ in the limit $N \gg 1$.

\begin{figure}[t]
\centering
\includegraphics[width=\textwidth]{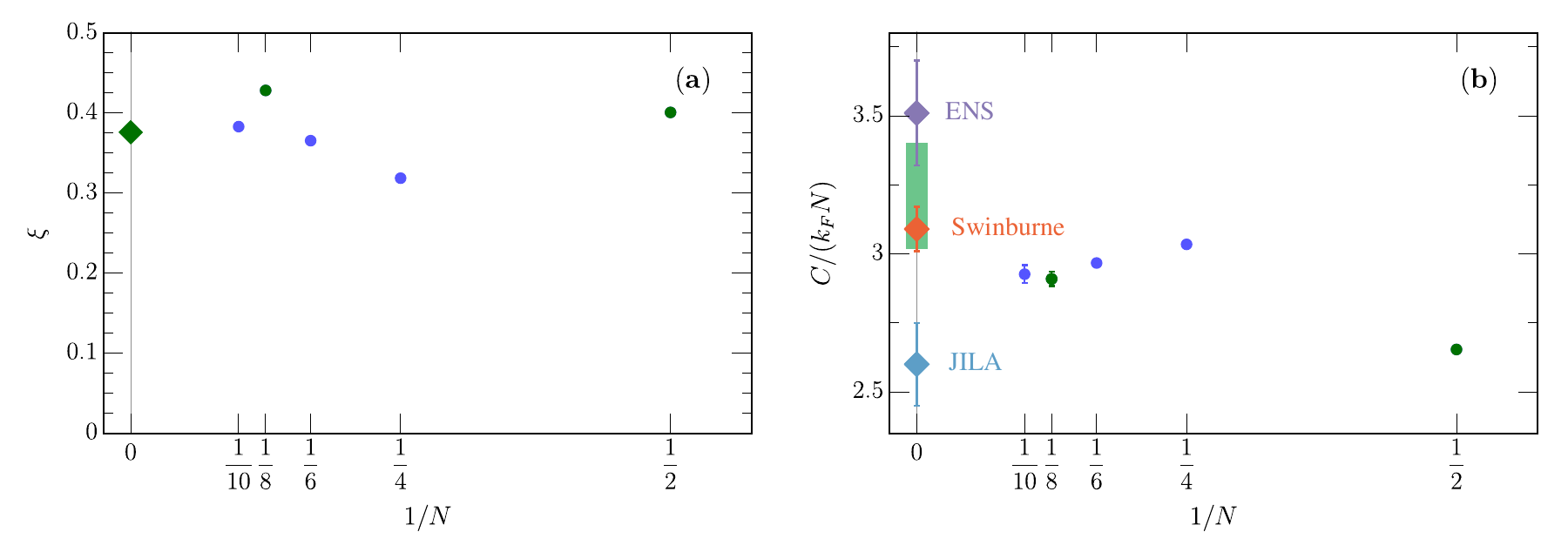}
\caption{\label{fig:bertsch} (a) Bertsch parameter and (b) contact of a unitary Fermi gas in a uniform system, calculated from the energy and contact of few-body systems in an isotropic harmonic trap. The green circles indicate closed shells of the corresponding non-interacting problem in the harmonic oscillator.
The diamonds show the experimental data points from~\cite{Ku2012} in (a), and from \cite{Navon2010} (purple), \cite{Sagi2012} (blue), \cite{Hoinka2013} (orange) in (b). The bar in (b) indicates the region of recent theoretical results. Figure adapted from Ref.~\cite{Levinsen2017}.}
\end{figure}

The results of this analysis are shown in Fig.~\ref{fig:bertsch}(a) up to $N=10$, where the few-body results are taken from Refs.~\cite{Busch1998,Stecher2009,yinTrappedUnitaryTwocomponent2015}. Remarkably, we see that the two-body system already gives a Bertsch parameter that is close to the experimentally determined many-body limit of $\xi=0.376(4)$~\cite{Ku2012} (see also QMC~\cite{Carlson2011}). As $N$ is increased, there are only small changes in $\xi$, with some oscillations and shell effects still present. 

Another key parameter of the unitary Fermi gas is the contact~\cite{Tan:2008a,Tan:2008b}, which gives a measure of the pair correlations at short distance.  In the trapped system, this takes the form~\cite{Blume2012}
\begin{align}
\mathcal{C}   = - 4\pi m \left(\frac{\partial \mathcal{E}}{\partial a^{-1}} \right)_{N},
\end{align}
which, within LDA, is equivalent to trap averaging the contact $C$ for a uniform system~\cite{Werner2012a}. This leads to the following relation for the contact at unitarity~\cite{Werner2009,Hoinka2013}
\begin{align}
    \frac{\mathcal{C}}{N \kappa_F} = \frac{256}{105\pi \xi^{1/4}} \left(\frac{C}{Nk_F}\right),
    \label{eq:Crelation}
\end{align}
where $C$ is the contact for a uniform gas with Fermi momentum $k_F$, while the trapped Fermi momentum $\kappa_F = (24 N)^{1/6} \sqrt{m\omega}$. Note that, since the contact involves a derivative with respect to the interaction strength, we expect it to be less sensitive to shell effects than $\xi$. Figure~\ref{fig:bertsch}(b) shows the result for the uniform-gas contact, extracted from trapped few-body results for the contact~\cite{Busch1998,yinTrappedUnitaryTwocomponent2015}. Once again, we see that the results appear to converge rapidly, and that they are in reasonably good agreement with both experiment and recent many-body calculations.

Recent measurements map out the temperature dependence of the contact at unitarity \cite{mukherjee2019} as well as the interaction dependence at low temperature \cite{jager2024}. At unitarity, this agrees well with theoretical predictions from bold diagrammatic Monte Carlo \cite{rossi2018} and, near $T_c$, from Luttinger-Ward computations \cite{Enss:2011}.  Out of equilibrium, the time evolution of the contact can be inferred from the spectrum \cite{Bardon:2014,Luciuk:2017,Xie:2025} or from the momentum distribution \cite{dyke2021,yi2025quantum}.

The close relationship between few- and many-body physics discussed above clearly warrants further study, and there is the prospect of applying similar approaches beyond the ground-state unitary Fermi gas. It has already been shown~\cite{Levinsen2017} for 
the so-called Fermi polaron~\cite{Massignan2014pdm,Scazza2022}, where $N_\downarrow=1$ and $N_\uparrow>1$, that one can extract not only the thermodynamic properties from few-body calculations but also the polaron effective mass, a dynamic property. 

\subsection{Experimental overview}\label{sec:exp_overview}

\subsubsection{Observation of elliptic flow in a mesoscopic Fermi gas}\label{sec:flow}

\begin{figure}[ht]
\includegraphics[width=\linewidth]{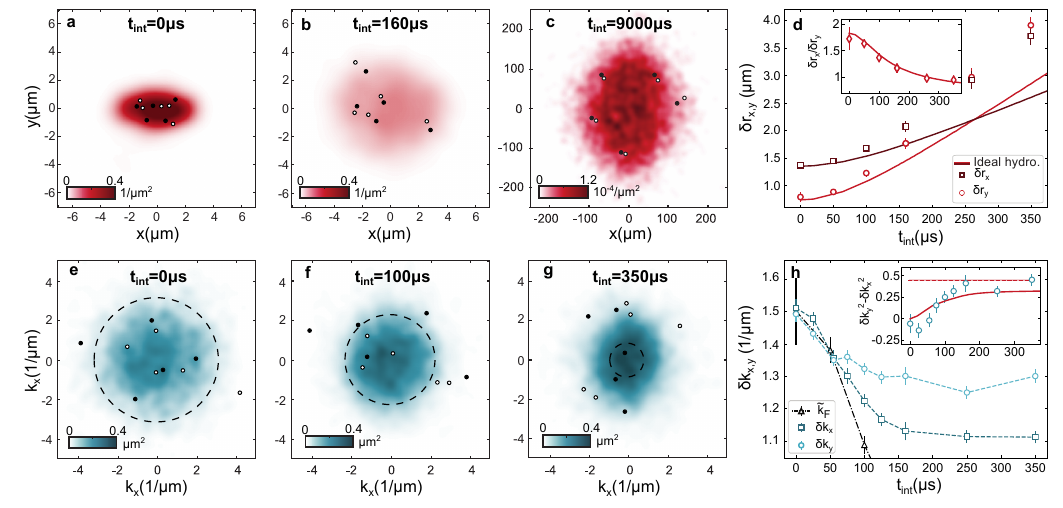}
\caption{\textbf{Elliptic flow of ten fermions.} We prepare 5+5 strongly interacting spin up and down atoms (black/white dots) in the ground state of an elliptically shaped trap. We measure their positions (\textbf{a-c}) or momenta (\textbf{e-g}). The two dimensional histograms show the density distribution, obtained from averaging over many experimental realizations of the same quantum state. The initial system has an elliptic density distribution in real space and a round Fermi surface in momentum space (see \textbf{a} and \textbf{e}). We study the expansion after switching off the trap (\textbf{b-c}, \textbf{f-g}) and observe the inversion of the initial aspect ratio in real space and the build-up of momentum anisotropy. The dashed black circle in \textbf{e-g} shows the Fermi momentum calculated from the real space peak density. 
\textbf{d} Ratio of Root mean square width of the atom positions $\delta r_\text{x}/\delta r_\text{y}$ as a function of $t_\text{int}$. 
\textbf{h} Root mean square value $\delta k_\text{x},\delta k_\text{y}$ of the momenta of the atoms as a function of $t_\text{int}$. The triangles show the Fermi momentum $\Tilde{k}_\text{F}$, rescaled to the geometric mean of $\delta k_\text{x},\delta k_\text{y}$ at initial time $t_\text{int} = \SI{0}{\micro s}$.  The connecting lines serve as a guide to the eye.
All error bars show the \SI{95}{\percent} confidence interval of the mean, determined using a bootstrapping technique. Figure adapted from Ref.~\cite{Brandstetter:2023jsy}\label{fig:expansion}}
\end{figure}

Ultracold atom experiments are well suited for the exploration of complex quantum physical questions by distilling them to their basic microscopic constituents while retaining their essential characteristics. The precise control over microscopic parameters available in these experiments allows for the incremental addition of complexity, enabling us to understand the fundamental principles that govern complex phenomena. Inspired by high-energy nuclear collisions \cite{Floerchinger:2021ygn}, where collective behavior (i.e. elliptic flow) is observed even though standard criteria for hydrodynamics do not hold, we studied the emergence of elliptic flow in mesoscopic
systems. This section provides a short summary of the experimental apparatus and the observed elliptic flow in a few-fermion system. A more detailed account can be found in~\cite{Brandstetter:2023jsy}.

In the cold atom experiment, we prepare a discrete quantum many-body state, composed of N spin up and N spin down atoms (denoted N+N) in the ground state of a two-dimensional optical harmonic trap, utilizing a technique developed in previous works~\cite{Serwane_2011, Bayha_2020}. A broad Feshbach resonance~\cite{Zuern_2013} allows us to tune the strength of the contact interactions between particles of different spin. We study the quantum dynamics of few contact-interacting fermionic $^6$Li atoms after release from an elliptically shaped trap. We take snapshots of the many body wave function at different time steps during the expansion in both real \cite{Brandstetter:2023jsy} and momentum space \cite{Holten2022} with single particle and spin resolution~\cite{Bergschneider_2018}. This allows us to obtain the real and momentum space density by averaging over $\approx 1000$ snapshots. Beyond that, as we have access to the position or momentum of every particle, we can also extract the correlations of arbitrary order.

We measure the density evolution of a system of 5+5 atoms, initially confined in an elliptically shaped trap.
\Cref{fig:expansion} shows the measured density profiles in real (a-c) and momentum space (e-g). Panel d) shows the evolution of the aspect ratio of the rms width of the system in x- and y- direction, h) shows the evolution of the momentum space rms widths. Initially, the system exhibits an elliptic density profile in real space, while it is isotropic in momentum space, as expected from the many-body limit of a degenerate Fermi gas. In real space, the system expands anisotropically, with faster acceleration in the initially tightly confined direction, leading to an inversion of the initial aspect ratio after long interacting expansion times. In momentum space, the system is initially isotropic. After release from the initial trap, the width in momentum space decreases in both x- and y- direction. For times smaller than $t_\text{int}<\SI{70}{\mu s}$, the system follows the decrease in the Fermi momentum, calculated from the peak density at the corresponding time. At longer times, we observe an anisotropy---elliptic flow---that subsides at $t_\text{int} \approx \SI{150}{\mu s}$.

The deterministic control over the atom number available in our experiment allows us to study the emergence of hydrodynamic behavior from the single particle limit. 
For a single particle in an elliptic trap, we also observe an inversion of the initial aspect ratio, however, solely due to the initially elliptic momentum distribution in contrast to the 5+5 particle case. 
As we increase the number of atoms, an initially isotropic Fermi surface builds up in momentum space \cite{Floerchinger:2021ygn}, while we observe a build-up of momentum space anisotropy during the expansion starting from systems of 3+3 atoms.

In conclusion, we observed elliptic flow in a system of ten strongly interacting atoms, which is considered a smoking gun for hydrodynamic behavior, even though the system size, interparticle spacing and mean-free path are not separable.

\subsubsection{Measuring diffusivity and viscosity on the atomic scale }\label{sec:diffusivity}

Understanding the hydrodynamics of strongly interacting quantum fluids has brought together disparate subfields of physics, including black holes and string theory, quark-gluon plasmas, quantum fluids, and cold atoms \cite{Schafer:2009dj,Taylor:2010,Adams:2012th}. The shared physical characteristics of these systems are that they are strongly interacting Fermi gases, in a regime without sharp quasiparticle excitations. The transport characteristics of these various systems appear to be bounded. Most famously, the KSS conjecture $\eta/s \geq \hbar/(4 \pi k_B)$ \cite{Kovtun:2005}; see Ref.~\cite{Son:2007b} for a review. Since this conjecture, a number of counterexamples have been found---models in which the bound does not hold---but there are no known experimental violations of the KSS bound, nor a quantitative understanding. One can frame many hydrodynamic bounds in terms of a diffusivity (of momentum, energy, or spin): the lower bound on  transport speed in metallic systems is then $D\gtrsim v^2 \hbar/k_B T$, where $D$ is a diffusion coefficient and $v$ is a typical velocity scale of the system \cite{Hartnoll:2015,Blake:2016wvh}. Finite propagation of information also poses an upper bound: $D\lesssim \tau_{\text{eq}}v_{\text{LR}}^2$ \cite{Hartnoll:2015,Lucas:2017ibu} where $v_{\text{LR}}$ is the Lieb-Robinson velocity (the upper bound for propagation speed of local perturbations) and $\tau_{\text{eq}}$ is the thermalization time to reach local equilibrium. In both limits, understanding hydrodynamic bounds requires understanding the local relaxation of a strongly interacting system. However, as we outline in this Section, direct local measurements on a sufficiently fast time scale are so far absent. 

Cold-atom experiments on hydrodynamics are typically based on global (system-wide) dynamics, which are then interpreted with a hydrodynamic model. Teams report best-fit transport coefficients (such as $D$ or $\eta$), which can further be interpreted via a local relaxation time, $\tau_\mathrm{eq}$. The pioneering measurements of shear viscosity by Thomas et al.\ studied time-of-flight expansion dynamics~\cite{Cao:2011,Joseph:2015}; see also reviews of Refs.~\cite{Schafer:2009dj,Adams:2012th}, which emphasize the connection to quark-gluon plasmas. 
In these systems hydrodynamic behavior was observed in the regime 
of sufficiently strong interaction. For the system sizes studied in the
original experiments, $N\sim 10^5$, this corresponds to the condition
$|k_F a|^{-1} \lesssim 1$, where $k_F$ is the Fermi momentum and $a$ 
is the s-wave scattering length. It was also observed that even at 
unitarity hydrodynamics breaks down in the dilute corona of the atom cloud.
This implies that a quantitative description of the experiments requires 
theories that interpolate between hydrodynamics at the center of the
cloud and ballistic expansion in the corona \cite{Bluhm:2016}.

A second generation of experiments has used the in-situ dynamics of 
box traps in which the density of the gas is approximately constant.
This implies that no unfolding of the density dependence of transport
coefficients is required, and that there are no complications due to 
non-hydrodynamic modes in the dilute corona. A typical experiment
uses external potentials to excite modes of a fixed wave-vector $k$. 
Experiments have studied first and second sound, as well as thermal 
relaxation. Ref.~\cite{Patel:2019udb} used the observed damping 
rate $\Gamma$ of ordinary (first) sound to extract the sound 
diffusivity $D = \Gamma/k^2$. For a scale-invariant fluid ($\zeta = 0$),
$D=D_\eta + D_\kappa$, with  $D_\eta = 4 \eta / (3\rho)$ and 
$D_\kappa = \alpha^2c^2\kappa_T T/(\rho c_P^2)$. Here, $\rho$ is the
mass density, $\alpha$ is the thermal expansion coefficient, $\kappa_T$
is the thermal conductivity, and $c_P$ is the specific heat at constant
pressure. Using either the measured viscosity $\eta$, or constraints
on $\eta$ from theory, one can extract the thermal conductivity. Wang 
et al.~\cite{Wang:2022} performed similar measurements after a quench, 
where both an underdamped sound mode and a purely diffusive thermal
mode are excited. This enabled an independent quantification of $\kappa$ 
and $\eta$. In the high temperature regime, the results are consistent
with the kinetic theory prediction $\kappa_T=(15/4)(k_B/m) \eta$.
The hydrodynamics of uniform Fermi 
gases is discussed further in \cref{sec:uniformFG}.

Another line of work has investigated the spin diffusivity, $D_s$, in
strongly interacting Fermi gases. Longitudinal diffusivity was probed 
using spin-dipole dynamics: first, separating up- and down-spin clouds, 
and then letting them collide \cite{Sommer:2011, Valtolina:2017}. 
Eventually, the spin-polarized clouds merge diffusively. 
Transverse spin diffusivity has been probed through the relaxation dynamics 
of a spin texture, such as a helical spin spiral. Once again, the locally
polarized cloud relaxed diffusively towards an unpolarized equilibrium 
\cite{Koschorreck2013,Bardon:2014,Trotzky:2015}. 
Among the measurements to date, spin diffusivity was observed to respect the 
conjectured bound $D_s \gtrsim \hbar/m$ in 
Refs.~\cite{Sommer:2011,Bardon:2014,Trotzky:2015}, but not in 
Ref.~\cite{Koschorreck2013}; see Ref.~\cite{Enss:2019b} for further 
discussion. 

Transport coefficients have not only a DC value, but also a frequency dependence: diffusivity $D(\omega)$, shear viscosity $\eta(\omega)$, bulk viscosity $\zeta(\omega)$, charge conductivity $\sigma(\omega)$, and spin conductivity $\sigma_s(\omega)$. To date, most cold-atom measurements of hydrodynamics have probed the $\omega\to 0$ limit. In contrast, theoretical work over the last decade has indicated a wealth of interesting spectral characteristics.  Ref.~\cite{Enss:2012} showed that the high-frequency tail of $\sigma_s(\omega)$ follows the contact. The high-frequency tail of $\zeta(\omega)$ is also governed by the contact parameter \cite{Hofmann:2011,Dusling:2013,Enss:2019}. Sum rules for the frequency integral of $\zeta(\omega)$ and $\eta(\omega)$ are discussed by Refs.~\cite{Taylor:2010,Enss:2011,Hofmann:2011,Goldberger:2012}, finding closed expressions for both. 
None of these universal spectral characteristics have been probed experimentally.  

Predictions of spectral functions require a microscopic theory. Kinetic theory for viscosity spectra, for instance, has been pursued in Refs.~\cite{Braby:2011,Dusling:2013,Chafin:2013,Nishida:2019,Hofmann:2020,Fujii:2020,Fujii:2023}  primarily in the high-temperature limit, and by \cite{Escobedo:2009,Stringari:2013} for the superfluid state. An interesting complication arises: non-analytic corrections have caused discrepancies between the low-frequency values of $\zeta(\omega)$ and its known DC values  \cite{Nishida:2019,Enss:2019,Martinez:2017}. Subtle issues such as this indicate that a measurement of $\eta(\omega)$, $\zeta(\omega)$, or $D(\omega)$ would test a challenging aspect of many-body theory. 

The experimental challenge in measuring spectral properties is that most measurements to date are limited to the time scale of global dynamics, which probe only the low-frequency limit. As with any over-damped system, a faster local $\tau_\mathrm{eq}$ leads to slower global relaxation---one can picture bubbles rising slowly in shampoo or molasses. Alternatively, recall that the two over-damped modes of a simple harmonic oscillator relax with time scales $\tau_\mathrm{eq}$ and $1/(\omega^2 \tau_\mathrm{eq})$, becoming increasingly disparate. Similarly, studying global dynamics reveals progressively less about the spectral character of dissipation in the limit of strong interactions. 

An alternative is to pursue local measurement, which can be rapid. For instance, one can measure the temporal evolution of the contact parameter $C(t)$, which provides a near-instantaneous measure of local correlations. In Refs.~\cite{Bardon:2014,Luciuk:2017}, contact dynamics provided an alternate witness for relaxation of a spin-polarized gas towards equilibrium, without a direct measurement of magnetization. In that case, the relaxation was governed by spin diffusivity: as the locally spin-polarized Fermi gas relaxed towards an equilibrium mixture, $C$ increased due to stronger local particle-particle correlations. 

Even more promising is the measurement of dynamical viscosity through contact dynamics. Refs.~\cite{Fujii:2018,Enss:2019} have shown that $\zeta(\omega)$ can be accessed through  modulation of the scattering length, followed by rapid measurement of $C(t)$. An experimental technique to measure the contact within $5\mu$s, faster than the Fermi time, has recently been demonstrated using dimer projection \cite{Xie:2025}. With a similar motivation, \cref{sec:attractor} discusses particular time-series of the scattering length, followed by time-resolved measurements of the contact. In both cases, fast local measurements will probe the spectral character of many-body relaxation dynamics. 

\subsubsection{Quantum dynamics and hydrodynamics in uniform Fermi gases} \label{sec:uniformFG}

Landau’s Fermi liquid (FL) theory has for decades been a cornerstone in our understanding of fermionic quantum many-body systems, providing a robust framework that requires only a few phenomenological parameters (the so-called Landau parameters). Direct experimental tests of Landau’s argument from first principles have been largely out of reach due to the strong interactions in existing Fermi system candidates (such as $^3$He), or their complexity (such as typical electronic systems). Ultracold atoms provide a textbook platform for studying Landau’s FL physics. First, in the dilute limit they accurately implement the Hamiltonian of fermions interacting with contact interactions---characterized by the s-wave scattering length $a_\mathrm{s}$. This model Hamiltonian has played a central role in quantum many-body physics and is amenable to a first-principle FL description in the limit of weak interactions (including the determination of the Landau parameters and the scattering amplitude). Secondly, the tunability of $a_\mathrm{s}$ provides a physical realization of the theoretical idea of the switching-on of interactions.

We presented a recent experiment where we prepared a spatially uniform spin-balanced two-component Fermi gas of $^6$Li atoms in the normal phase ($T>T_\mathrm{c}$), and studied how sound emerges as interactions are gradually turned on. After preparing an equilibrium Fermi gas at the desired interaction strength, we turned on a spatially uniform shaking force along the $y$ axis and measured the density response of the gas. This shaking potential predominantly coupled to the longest wavelength excitation in the momentum space ($|\mathbf{q}_\mathrm{ext}|=\pi/L$, where $L$ is the box length along the shaking direction). At the longest wavelength, the sound mode is related to the collisional parameter $\omega_0\tau_\mathrm{coll}$, where $\omega_0=v_\mathrm{F}q_\mathrm{ext}$, and $\tau_\mathrm{coll}$ is the typical collision time of the system; for a weakly interacting Fermi gas, $\tau_\mathrm{coll}=\hbar^3/(2ma_\mathrm{s}^2 k_\mathrm{B}^2 T^2)$. In our experiment, we observed a narrow peak emerging in the density response function as the $|k_\mathrm{F}a_\mathrm{s}|$ increases, signaling the onset of sound and the system entering the hydrodynamic regime. For more details see Ref.~\cite{Huang:2024qdj}.

We explored different regimes: (I) For $|k_\mathrm{F}a_\mathrm{s}|\ll 1$, the gas is weakly interacting and the response is in good agreement with the prediction derived from a first-principle transport equation. In particular, for $k_\mathrm{F}a_\mathrm{s}=0$, we measured for the first time the Lindhard function---the density response for a noninteracting Fermi gas. (II) Around $|k_\mathrm{F}a_\mathrm{s}|\gtrsim 0.7$, the first-principle theory breaks down, which coincides with the system entering the regime of strong interactions. In this regime, no existing kinetic theory has been developed at low $T$, and our data provides a testing ground for more advanced theories. (III) By measuring the sound damping rate $\Gamma$ as a function of $T/T_\mathrm{F}$ at an intermediate interaction strength ($|k_\mathrm{F}a_\mathrm{s}|\approx 0.7$), we extrapolated the attenuation coefficient $\Gamma T^2$ down to $T\rightarrow 0$---the hydrodynamic Fermi liquid regime.

Finally, we measured the excited quasiparticle distribution $n(k)$ in the momentum space for the collisionless regime ($\omega_0\tau_\mathrm{coll}\gg 1$) and the near-hydrodynamic regime ($\omega_0\tau_\mathrm{coll}\approx 1$). In the collisionless regime, we observed anisotropic distributions due to the existing poles in $\delta n(k)=n(k)-n^\mathrm{eq}(k)$ of single particle-hole excitations. In contrast, the distribution in the hydrodynamic regime is isotropic, showing that collisions tend to restore the equilibrium distribution faster over a drive period.

In the future, it would be interesting to attempt the observation of zero sound in the normal phase of the polarized Fermi gas. Theoretical refinements, particularly incorporating higher-order corrections into the Landau parameters and in the collision integrals, would not only allow one to take into account renormalization effects ab initio, but also provide a controllable path toward understanding the transport properties (such as sound velocity, viscosity, and thermal conductivity) of the normal degenerate strongly interacting Fermi gas.

\subsubsection{Experimental prospects with multi-orbital Fermi gases }\label{sec:mixedFG}

Ultracold atoms offer a rich playground for the investigation of strong correlation phenomena in fermionic quantum systems. Their long coherence times and an ever-increasing degree of controllability, along with unique probing tools, have enabled over the last decade a variety of exciting experiments in which the state of each and every atom in fermionic few- and many-body systems could be measured with remarkable precision (see for instance~\cite{Holten2022,Gross2021}). Experiments until now have largely focused on systems of fermions belonging to a single dispersion band, such as those realized with alkali atoms in a harmonic trap or the lowest band of an optical lattice. Furthermore, explorations have been typically limited to spin-1/2 fermions that realize SU($2$)-symmetric Hamiltonians, while single-atom-resolved studies of systems with SU($N>2$) symmetries have yet to be performed. 

A strong motivation to advance ultracold atom experiments beyond the realization of single-band and SU(2) symmetric models comes from both condensed matter and high-energy physics. For example, many instances of collective behavior in quantum materials, such as magnetism or unconventional superconductivity, arise from interactions between electrons occupying distinct atomic orbitals, i.e.~featuring different dispersion relations. The electrons' orbital degrees of freedom often intertwine with their spin, leading to strong correlations that govern low-temperature transport properties, as observed in heavy fermion materials and transition metal oxides. 

To effectively mimic ``orbitals'' with distinct dispersion relations that interact within the same quantum system, it is advantageous to employ atomic species with multiple long-lived electronic states (e.g. alkaline-earth atoms) or mixtures of different atomic species, trapped in state- or species-dependent optical potentials. A particularly intriguing case is the possibility to fully localize one orbital while another remains delocalized, a configuration known to give rise to the physics of Anderson's impurity model and the Kondo effect. Particles occupying the localized orbital act as impurities, simultaneously lending themselves as local probes of the many-body dynamics. Adding fermionic atoms around an impurity one by one \cite{Wenz2013} provides unique opportunities to study the build-up of strong fermionic correlations in impurity problems, especially out of equilibrium, with a level of control that is clearly unattainable in the solid state. 

In a new experiment in Trieste, we utilize ytterbium atoms (${}^{171}$Yb and ${}^{173}$Yb) to build mesoscopic fermionic systems with single-particle resolved read-out of the wavefunction~\cite{Abdel_Karim_2025}. Ytterbium atoms in the ground and metastable clock states, ${}^1$S$_0$ and ${}^3$P$_0$, can be selectively trapped and manipulated to explore two-orbital problems \cite{Gorshkov2010}, using the optical control toolbox that has been developed in the last two decades to build the most precise optical-lattice clocks. Pioneering experiments have already carefully characterized inter-orbital interactions at ultracold temperatures, and have demonstrated the key building blocks for new exciting explorations of two-orbital many-body dynamics  \cite{Scazza2014,Cappellini2014,Hoefer2015,Pagano2015,Riegger2018,Ono2019,Abeln2021,Bettermann2020}. With the capability of controlling and detecting the number of particles in both orbitals, while gaining coherent control on their nuclear spin, a variety of interesting experiments will be within reach, from impurity-mediated quantum transport \cite{You2013} and correlated dissipative dynamics \cite{Sponselee2019} to the non-equilibrium orthogonality catastrophe \cite{Knap2012} and the Kondo impurity problem~\cite{Amaricci:2025qwl}.

\subsubsection{Experimental prospects with composite degrees of freedom }\label{sec:composite}

Binary mixtures of ultracold fermionic atoms with resonant interactions represent versatile platforms with which to investigate a wealth of fundamental transport phenomena with unparalleled cleanliness, both in the continuum and lattice configurations, both in the normal and superfluid phases, spanning from many-particle environments down to mesoscopic and few-particle systems, see Refs. \cite{Bloch2008mbp,Varenna2007, Zwerger2012tbb, Gross2017qsw} and references therein. Yet, most experiments so far focused on homonuclear spin mixtures, with the few exceptions of $^6$Li-$^{40}$K, $^{40}$K-$^{161}$Dy and $^6$Li-$^{53}$Cr systems – where first studies of collisional hydrodynamics have been explored, both in expanding clouds \cite{Trenkwalder2011, Ravensbergen2020} and through the study of collective modes in trapped samples \cite{Finelli2024ulc}.  

Indeed, heteronuclear Fermi mixtures offer appealing new opportunities to study strongly-interacting fermionic matter and the associated many-body dynamics, owing to the different optical properties and unequal masses featured by the two atomic constituents. On the one hand, the different response of each component to laser fields allows one to engineer species-selective optical potentials to control the confinement, motion, and dimensionality of each species almost independently from the other one. More fundamentally, a sizable mass asymmetry $M/m$ between the two elements of the mixture can dramatically and qualitatively modify the dynamical properties of the system, both at the few- and many-body level. 
Celebrated examples include the predicted emergence of exotic superfluid states promoted by the two different dispersions \cite{Baarsma2013isp} and the disappearance of well-defined polaron quasiparticles in the heavy-impurity limit \cite{Massignan2014pdm}. Another example is the transition from quantum-limited transport—characteristic of equal-mass systems \cite{Sommer:2011, Valtolina:2017}—to the complete absence of diffusion when light particles evolve in a bath of infinitely massive ones. In this case, the massive particles act as an (almost static) landscape of point-like random scatterers \cite{Massignan2006}, which can potentially lead to localization phenomena \cite{Anderson1958}.

Additionally, for intermediate mass ratios, 8.17$<M/m<$13.6, non-Efimovian cluster states made by one light particle bound to two, three, or four heavy identical fermions are expected to appear \cite{Kartavtsev2007let, Endo2012, Naidon2017epa, Bazak2017}, paralleled by the emergence, at the many-body level, of new types of quasi-particles \cite{Mathy2011tma,Liu2022}, and superfluid states beyond the Cooper pairing paradigm \cite{Liu2023}. 

In this context, the $^6$Li-$^{53}$Cr Fermi mixtures realized in our lab \cite{Neri2020roa, Ciamei2022}, and the availability of suitable Feshbach resonances for the control of Li-Cr interactions \cite{Ciamei2022B}, open appealing prospects for the investigation of many-body dynamics of strongly-correlated fermionic matter in the presence of non-perturbative few-body effects. Indeed, the chromium-to-lithium mass ratio, $M/m\!\sim$8.8, may uniquely enable access, within the resonantly-interacting regime, to exotic long-lived three- and four-body cluster states \cite{ Bazak2017, Liu2024}. 
Our recent activities focused on the production of weakly-bound LiCr Feshbach dimers and the thorough characterization of their properties \cite{Finelli2024ulc}. On the one hand, the outcome of our studies unveiled a new promising pathway to experimentally realize quantum gases of paramagnetic polar molecules---as LiCr ground-state molecules are predicted to feature a sizable electric dipole moment $d_e\!\sim$3.3 Debye on top of a large electronic spin $S=5/2$. On the other hand, and more importantly for the RRTF topics, this survey allowed us to realize large samples of up to $5 \times 10^4$ ultracold $^6$Li$^{53}$Cr bosonic Feshbach dimers, at phase-space densities exceeding 0.1, which represent a promising starting point for the exploration of novel many-body dynamics within mass-asymmetric Fermi mixtures. 

As an illustrative step forward in this direction, we are currently exploring the diffusion dynamics of light Li impurities, initially confined in a species-selective, vertically-oriented optical trap, and then released within a large sample of fermionic $^{53}$Cr atoms, acting as a medium of heavy point-like scatterers. After turning off the species-selective dimple, lithium atoms are let free to expand along the axial direction of the main optical dipole trap---confining both our mixture components---in the presence of a weak harmonic axial confinement, set by our magnetic-field curvature, and with a tunable interaction with the surrounding chromium gas. The latter is arbitrarily adjusted by setting the bias magnetic field around a narrow, 0.5 G-wide inter-species Feshbach resonance centered around $B_0$=1414 G \cite{Ciamei2022B, Finelli2024ulc}, occurring between the lowest Zeeman sublevels of the two species, hereafter denoted Li$|1\rangle$ and Cr$|1\rangle$, respectively. 
The mixture is initially prepared at a target field $B_T$ in the non-resonant combination Li$|1\rangle$-Cr$|2\rangle$, with Cr$|2\rangle$ denoting the second-to-lowest Zeeman level. A 0.9 ms-long radio-frequency pulse, resonant with the Cr$|1\rangle \leftrightarrow$Cr$|2\rangle$ transition, is employed to rapidly switch the internal state of chromium atoms with almost 100$\%$ transfer efficiency, thereby making the resulting Li$|1\rangle$-Cr$|1\rangle$ mixture experience an arbitrarily large (or weak) interaction. Right after the end of the pulse, the optical dimple is removed, and the subsequent Li dynamics is monitored as a function of time, through in-situ absorption images, and the analysis of their second moment along the axial direction, $\langle x^2(t) \rangle$. In particular, we focus on the 30 ms time window of the early-time dynamics, corresponding to half a period of the breathing mode of free Li atoms in the final trap. This allows us to: (i) mitigate the Cr inhomogeneity experienced by the Li atoms during their expansion; (ii) reduce detrimental effects of inelastic collisions affecting the atomic mixture in the strongly-interacting regime (roughly corresponding to $|B_T-B_0|\leq$ 10 mG); (iii) ensure a peak-to-peak stability of the magnetic-field bias of about 2 mG over the entire time window. 

These measurements, and the associated theoretical analysis, still ongoing, will be subject of a future publication and will not be shown nor discussed here in detail~\cite{Zaccanti:2023txb}. Yet, we anticipate that our data provide a thorough characterization of the early-time  expansion of lithium atoms within the Cr host gas, clearly revealing the crossover from ballistic to diffusive dynamics as the interspecies interactions are tuned from their (small) background value towards the resonant regime. There, we measure extremely small diffusion constants $D$ down to only a few $\hbar/m$ quanta. This is consistent with what is reported for unitary Fermi gases of lithium atoms \cite{Sommer:2011}, despite the large mass asymmetry and the narrow nature of the Feshbach resonance featured by the Li-Cr mixture \cite{Ciamei2022B, Finelli2024ulc}. Most surprisingly, we reveal an unforeseen, qualitative change in the expansion dynamics in the resonant regime, as the system temperature is further reduced: From a normal (albeit slow) diffusive expansion---where the mean squared displacement grows linearly in time, $\langle x^2(t) \rangle =2 D t$---to a \textit{subdiffusive} one at our lowest temperatures, where we find $\langle x^2(t) \rangle \propto t^\alpha$ with $\alpha<$1. The origin of such an anomalous behavior is not explained yet, and it is currently subject of theoretical analysis. Nonetheless, this latter experimental study, only briefly summarized here, highlights how unequal-mass Fermi mixtures, thanks to the additional degrees of freedom they naturally offer, provide exciting new possibilities to explore many-body dynamics of ultracold fermionic matter, possibly leading to qualitatively new phenomena beyond their homonuclear counterparts.

\subsection{Theoretical overview}\label{sec:theory_overview}

\subsubsection{Microscopic approaches to few-body systems}\label{sec:microscopic}

For bosons and fermions alike, the question of how many (or rather, how few) atoms must conspire to form a precursor of a quantum phase transition upon changing a system’s order parameter (or, to be more precise, its finite-size correspondence) is an intriguing open problem, that only more recently came into the reach of experimental exploration. In fermionic systems, the bottom-up formation of a Fermi sea and pairing, as well as the signaling of the quantum phase transition by Higgs amplitude mode precursors are vivid examples~\cite{Wenz2013,Zuern_2013,Bjerlin:2016,Bayha_2020}. The motivation for these more recent experimental achievements conceptually leans on the ancient Sorites’ paradox, as referred to at the pages of S. Jochim’s group in Heidelberg \cite{sorites}: ``1,2,3, many - how few particles turn into a heap?"
How does a BCS state evolve from single pairs of fermions? Or how many atoms do we need to form a “condensate”? And how do finite-size effects change the system properties? Can a few-body system show superfluid behavior? 
Large and weakly interacting systems can usually be treated rather accurately by mean-field approaches, as discussed in the first chapter of this report. To answer the above questions, however, finite-size effects often make it necessary to turn to more exact methods, that also give access to excitations in the system. A direct way to obtain such exact solutions is to diagonalize the many-body Hamiltonian numerically. This approach is briefly discussed in what follows in this subsection. 

To summarize it in most simple terms, exact diagonalization means what the name implies---one builds the many-body Hamiltonian matrix (for an appropriately truncated Hilbert space) and diagonalizes it exactly. To set up this matrix, one needs to consider all the many possible configurations of single-particle state occupancies in the Fock states forming the many-body basis of Slater determinants (for fermions) or permanents (for bosons) in the physically relevant parts of Hilbert space. The procedure is also called the method of “configuration interaction” (CI) (see~\cite{Sherill1999} for a review), a terminology that also serves as an umbrella for more sophisticated ways to construct the appropriate Fock states and linear combinations thereof. The CI method has its roots in quantum chemistry and nuclear physics, where many of the leading concepts were developed.   
The CI scheme as such is straight-forward to write down and implement.
Computationally, however, it is often beyond reach---the degree of complexity of the problem grows factorially with the number of constituents, rendering the task often impossible to solve with reasonable accuracy.  Despite the nowadays available computational power, the method can provide physically reliable answers often only for a rather small range of system parameters. This poses a dilemma---while for moderate interactions and  large particle numbers the systems can be treated rather well by the above-mentioned mean-field approaches (such as Gross-Pitaevskii methods for bosons, and fermionic density functional theory), exact solutions including access to excitations can so far only be obtained for the very smallest particle numbers at relatively moderate interaction strength. Very little is known so far on how to treat the intermediate regime between “few” and “many”. 

Let us now look closer into the procedure of exact diagonalization of a many-body Hamiltonian (as explicitly defined in Eq.~\ref{eq:many-body} below). Having constructed the Hilbert space, where the many-body basis states are built from a single-particle basis  often selected to be the eigenbase of the single-particle part of the Hamiltonian, { i.e.}, the first two terms in Eq.~\ref{eq:many-body}, and introducing an appropriate truncation, one constructs the matrix representation of the full Hamiltonian in this many-body basis and diagonalizes it. Algorithms such as the Lanczos~\cite{Lanczos1950} and Arnoldi~\cite{Arnoldi1951}  methods are useful tools to find the most important eigenstates within a certain range of the spectrum. 
This sounds simple and appealing since the exact eigenstates (ground states and excitations) allow an easy access to the physical observables. However, as mentioned above, there is a significant drawback: the numerical complexity of the problem at hand puts a severe restriction on the computational feasibility, concerning both the number of particles and the strength of interactions (correlations) that one can consider. The approach very often necessitates various approximations---of the simple, and of the more complicated kind. 
A natural simplification is to make use of all symmetries such that the Hamiltonian matrix falls into a block-diagonal structure. For the subsequent appropriate truncation of the Hilbert space, there are several options---a straightforward one is to simply set an energy cutoff. It is here important to respect the relevant energy scales of the problem when considering which particle-hole and two-particle two-hole excitations {\it etc.} to allow. A simple brute-force choice of cut-off often fails. A cleverer way out is the so-called {\it importance-truncated} configuration interaction method (ITCI) (as described in Ref.~\cite{Roth2009} and in the context of bosonic few-body systems in \cite{Chergui2023}), where one makes use of the fact that it is often only a small fraction of Hilbert space that is relevant for a specific eigenstate. The degree of importance is hereby determined by a given threshold of the relevant observable, such as the energy, where the contribution of subsets of states in relation to the exact or converged result is appropriately evaluated. For a pedagogical introduction to the ITCI method, see for example the recent work by Chergui~\cite{Chergui:phd}.

The problems of convergence that one may face for strong interactions are also widely known from nuclear structure calculations, where a remedy was the Lee-Suzuki transformation~\cite{Suzuki1980} and renormalization group methods~\cite{Bogner2007}  that may greatly improve the convergence, see also the review by Bogner {\it et al.}~\cite{Bogner2010}. A discussion of effective interactions, renormalization schemes, and related computational approaches in the context of ultra-cold atoms is provided in the introduction of the recent work by Brauneis {\it et al.}~\cite{Brauneis2025}, or for example in chapter 9 of the review by Mistakidis {\it et al.}~\cite{Mistakidis2023}.

To give some examples in this brief survey of exact diagonalization calculations, for which we above only discussed the elementary concept, we now turn to bosonic trapped systems that are well suited to exemplify the few- to many-body transition in the spirit of the above-mentioned Sorites’ paradox.

{\bf From mean field to exact solutions - rotating bosons in the lowest Landau level}. We first consider a simple toy system of weakly interacting bosons confined in a rotating trap, which allows us to directly map out the differences between the limit of large yet finite-size systems and the thermodynamic limit. We here mainly refer to the early works by Mottelson~\cite{Mottelson1999}, Bertsch and Papenbrock~\cite{Bertsch1999}, and Kavoulakis {\it et al.}~\cite{Kavoulakis2000}, and the later work by Cremon {\it et al.}~\cite{Cremon2013,Cremon2015} who addressed this example in greater detail. (For a review on rotating bosonic condensates and analogies to the quantum Hall effect, see for example the reviews by Fetter~\cite{Fetter2009} and Saarikoski {\it et al.}~\cite{Saarikoski2010}). 

In the following we summarize a few of the key results of Cremon et al.~\cite{Cremon2013, Cremon2015}. We consider the Hamiltonian of $N$ interacting bosons, as defined in Eq.~\ref{eq:many-body} in \cref{sec:ECG} below, with a quasi-2D isotropic harmonic trapping potential with trapping frequency $\omega $ and two-body interactions of the usual short range type, i.e., $V_{\mathrm{2b}}({\bf r}_i - {\bf r }_j)=g \delta ({\bf r}_i - {\bf r }_j)$. 
For sufficiently weak interactions, at given total angular momentum $L>0$ 
only single-particle oscillator states with no radial nodes and single-particle angular momenta $m\ge 0$ play a role~\cite{Mottelson1999,Bertsch1999}, and the many-body basis in this so-called ``Lowest Landau Level'' (LLL) is spanned by the $F$-dimensional Fock space $\left\{ \mid 0^{N_0}, 1^{N_1}, 2^{N_2},\dots , m^{N_m} \rangle \right\} _{j=1}^F$  with the constraints that 
$\sum _m N_m=N$ and $ \sum _m m N_m =L $ with good total angular momentum $L$ as a consequence of the circularly symmetric trap. 
The many combinatorial ways to distribute $L$ units of angular momentum to $N$ bosons in the Fock states constitute a large degeneracy~\cite{Mottelson1999} which is increasing with $L$ and $N$.   
The Hamiltonian of the rotating system is then diagonalized in this subspace of degenerate states, in the spirit of degenerate perturbation theory, since interactions are weak. 
Since
\begin{equation}
    \hat H_{LLL} = \hbar \omega N + \hbar (\omega -\Omega )L + \frac g2 
    \sum _{ij}\delta ({\bf r}_i - {\bf r }_j)~,
\end{equation}
as here written in the rotating frame for a trap rotation frequency $\Omega \le \omega $, the diagonalization is performed only on the interaction part.  (Note that the well-defined subspace of the LLL also circumvents all regularization issues that are usually connected with the exact solutions for contact-interacting systems in dimensions larger than one).  
For moderate rotation it is often sufficient to work with a single-particle basis cutoff, $m\le m_{\mathrm{max}}$, which significantly reduces the many-body space. 
\begin{figure}[t]
\begin{center}
\includegraphics[width=\linewidth ]{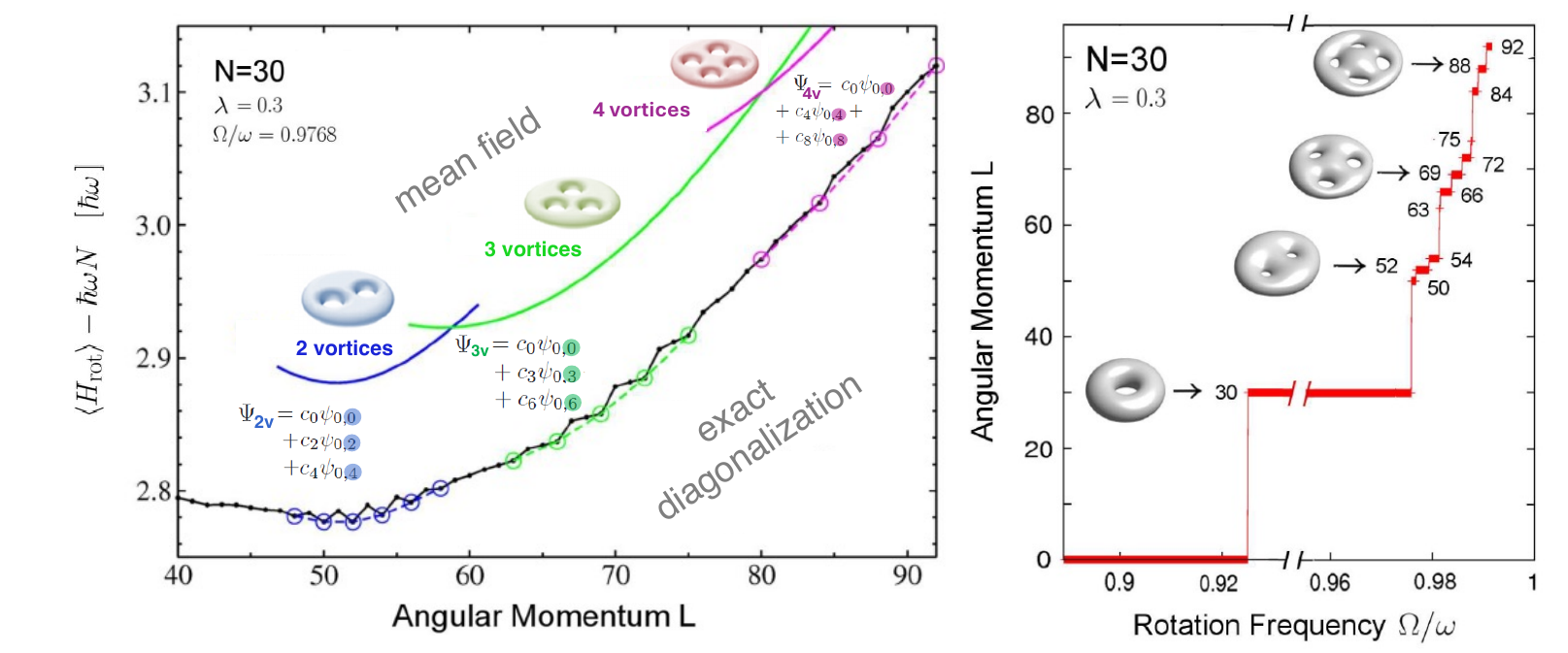}
\end{center}
\caption{
(Left) ``Yrast" states of maximal angular momentum $L$ (in units of ${\hbar }$) at a given energy $\langle H_{\mathrm{rot}}\rangle - \hbar \omega N$
in the rotating frame, resulting from exact diagonalization (lower line) and a Gross-Pitaevskii analysis (upper intersecting parabolae). The colors blue, green and magenta indicate the numbers of vortices exemplified by the given order parameters $\Psi _{\mathrm{2V}}, \Psi _{\mathrm{3V}}$ and $\Psi _{\mathrm{4V}}$. The rotational frequency is $\Omega = 0.98 \omega $
and $g=0.02$ (setting $\hbar ^2 / (2M) =1$). 
(Right)  Angular momentum (in units of ${\hbar }$) resulting from the minimization of the Yrast energy obtained by exact diagonalization (lower line shown in the left panel), as a function of rotation frequency $\Omega $ (in units of the trap frequency $\omega $). The insets show the iso-surfaces of the pair-correlated densities at half-maximum value, for a reference point of $(x,y)=(1,0)$a.u. These figures are adapted
from Cremon {\it et al.}~\cite{Cremon2013}
Copyright \copyright\, 2013 by the American Physical Society.
}
 \label{fig:Reimann_Figure1}
\end{figure}

The left panel of Fig.~\ref{fig:Reimann_Figure1} shows the so-called ``yrast line", i.e. the ground state energy as a function of angular momentum, for the example of $N=30$ bosons determined by exact diagonalization in the LLL (here calculated with $m_{\mathrm max}=12$) (lower line) and compares it with the results of Gross-Pitaevskii mean field theory (upper intersecting parabolae)~\cite{Kavoulakis2000}.  This comparison clearly shows the difference between the mean-field approach and exact solutions, where finite-size corrections appear as oscillatory patterns as a consequence of the broken symmetry in the solutions with higher vortex numbers.  In the thermodynamic limit, these patterns disappear. 
As shown in the right panel of Fig.~\ref{fig:Reimann_Figure1}, upon rotation the system at first does not acquire angular momentum as long as the rotation frequency is below a certain value. This is characteristic of a superfluid and is observed here also in the case of a finite-sized system with $N=30$ bosons in a trap. Upon increasing rotation, a first vortex can be accommodated for $L/(\hbar N)=1$. It has been shown that this single-vortex state corresponds to a condensate in the $m=1$ single-particle state~\cite{Mottelson1999}. To leading order in $N$ the state is of the form $\mid \Psi _{1V}\rangle= \sum _k (-1)^k C_k \mid 0^k,1^{N-2k},2^k\rangle $ with $C_k=1/(\sqrt2 ^{k+1})$, see Ref.~\cite{Wilkin1998,Kavoulakis2000}. Only the $m=1$ single-particle state is macroscopically occupied.  Finite-size effects yield occupancies of the $m=0$ or $m=2$ neighboring orbitals and energy corrections that are of lower order in $N$.
This systematics continues for higher rotation frequencies, where additional vortices penetrate the cloud, forming a vortex cluster one by one (see the insets showing the vortices as holes in the equi-density surfaces).

\subsubsection{Fluid dynamic description of the few-fermion system}\label{sec:CAfluid}

The experiments of Ref.~\cite{Brandstetter:2023jsy} discussed in \cref{sec:flow} provide an explicit demonstration of the emergence of hydrodynamic-like behavior in a mesoscopic Fermi gas. Having shown that the observed elliptic flow is a genuine interaction-driven effect, the data offers the unique opportunity of testing the many-body description of these systems against data from a mesoscopic sample. 

In this section, we review the rationale behind the theoretical calculations performed in Ref.~\cite{Brandstetter:2023jsy}. The question we want to answer is the following: does the many-body description of a $^6$Li gas at zero temperature, which is based on superfluid hydrodynamic equations, enable us to capture the features observed in the experiment? To answer this, we build a hydrodynamic model tailored to match the physical parameters of the experimental setup, and whose predictions are then compared to the observations.

\paragraph{\textbf{Real space analysis.}} Our starting point is the construction of the initial condition of the trapped system. This amounts to introducing a mass density, $\rho$, at $t=0$. 
Although we lack theoretical guidance for the one-body density of a strongly-interacting mesoscopic system, we empirically observe that an excellent fit of the trapped gas for 5+5 fermions ($N=5$, see \cref{fig:expansion}) can be achieved through a generalized Gaussian distribution in two dimensions
\begin{equation}
    \rho({\bf x},t=0) \propto \exp \left ( - \left(\frac{|x|}{a_x}\right)^{b_x} \right ) \exp \left ( - \left(\frac{|y|}{a_y}\right)^{b_y} \right ), \hspace{30pt}  \int d^2{\bf x}~ \rho({\bf x},t) = 2mN ,
\end{equation}
with parameters $a_x=\SI{2.21}{\micro m}$ , $b_x=3$, $a_y=\SI{1.04}{\micro m}$, $b_y=2$, and where $m$ is the mass of $^{6}$ Li. 

Subsequently, we evolve this initial density in time by solving ideal (superfluid) hydrodynamic equations. This means that we have to solve mass and momentum conservation equations, that is   \cite{Landau_1987}
\begin{equation}
\begin{split}
\label{eq:ih}
    \partial_t \rho + \nabla \cdot (\rho {\bf v}) &= 0 \, , \\
  \rho   (\partial_t + {\bf v} \cdot \nabla) {\bf v} &= -\nabla P \, ,
\end{split}
\end{equation}
where the fluid velocity is denoted by ${\bf v}=(v_x,v_y)$. To close the set of equations, we need an equation of state. To consistently match to the many-body description of the zero-temperature two-dimensional Fermi gas, we utilize a parametrization of the pressure, $P$, as a function of the mass density, $\rho$, that is motivated by experimental results obtained with macroscopic samples \cite{2014PhRvL.112d5301M,2016PhRvL.116d5303B}. 
More specifically, the pressure is typically plotted against an interaction strength parameter, $\eta = \ln (k_F a_{2D})$, where $k_F=\sqrt{2\pi \, n}$ is the Fermi momentum corresponding to the total number density $n=\rho/m$, and $a_{2D}$ is the 2D $s$-wave scattering length. The non-interacting limit with infinite interaction parameter corresponds to the scenario of an ideal Fermi gas, where pressure is only a consequence of Pauli repulsion and the system is not in a local equilibrium state \cite{2015arca.book....1L}
\begin{equation}
   P_{\rm ideal} \equiv P (\eta\rightarrow\infty) = 
   \frac{\pi \hbar^2}{2 m^3} \rho^2 .
\end{equation}
The variation of $P$ towards the BEC limit, $\eta \rightarrow -\infty$ has then been mapped out experimentally. In the experiment of Ref.~\cite{Brandstetter:2023jsy}, the value of $\eta$ goes from a maximum of about 1.2 at the center of the gas at $t=0$, and then decreases steeply with the expansion time. We find that an exponential form for the decrease of the pressure with decreasing $\eta$ captures well the experimental results \cite{2014PhRvL.112d5301M}. 
In particular, we obtain that
\begin{equation}
\label{eq:Pfit}
    P (\eta) = \alpha e^{\beta \eta} P_{\rm ideal},
\end{equation}
with $\alpha=\num{0.216\pm0.008}$ and $\beta=\num{0.67\pm0.05}$ provides a very satisfactory fit of the measured equation of state. For the relation to the mass density, we have thus $P = g \, \rho^{\kappa}$, which corresponds to a polytropic equation of state. Here $\kappa=2+\beta/2$ while $g$ is proportional to $\alpha$ and its expression can be found in Ref.~\cite{Heyen:2024xxb}.

To solve the hydrodynamic equations in time, we use the \texttt{pyro} code \cite{pyro}. The temporal dependence of the spatial dispersions of the system
\begin{equation}
    \langle \delta r_x^2 \rangle (t) = \frac{\int d^2{\bf x}~ \rho({\bf x},t) \, x^2}{\int d^2{\bf x} ~ \rho({\bf x},t)}, \hspace{30pt}     \langle \delta r_y^2 \rangle (t) = \frac{\int d^2{\bf x}~ \rho({\bf x},t) \, y^2}{\int d^2{\bf x} ~ \rho({\bf x},t)},
\end{equation}
are then compared to the experimental observations. For the absolute magnitude, the hydrodynamic simulation leads to an expansion that is significantly slower than the observed one. On the other hand, the aspect ratio of the system, $\delta r_x /\delta r_y$, as a function of time turns out to be in excellent agreement with the data shown in Fig.~\ref{fig:expansion}.

\paragraph*{\textbf{Second-order hydrostatic approach to the initial condition.}}

When dealing with small system sizes or small particle numbers, quantum corrections to ideal hydrodynamics are expected to appear, even in a superfluid scenario. Here we discuss quantum corrections to the trapped density that should naturally play a role in the mesoscopic system investigated in the experiments.

\begin{figure}[t]
    \centering
    \includegraphics[width=\linewidth]{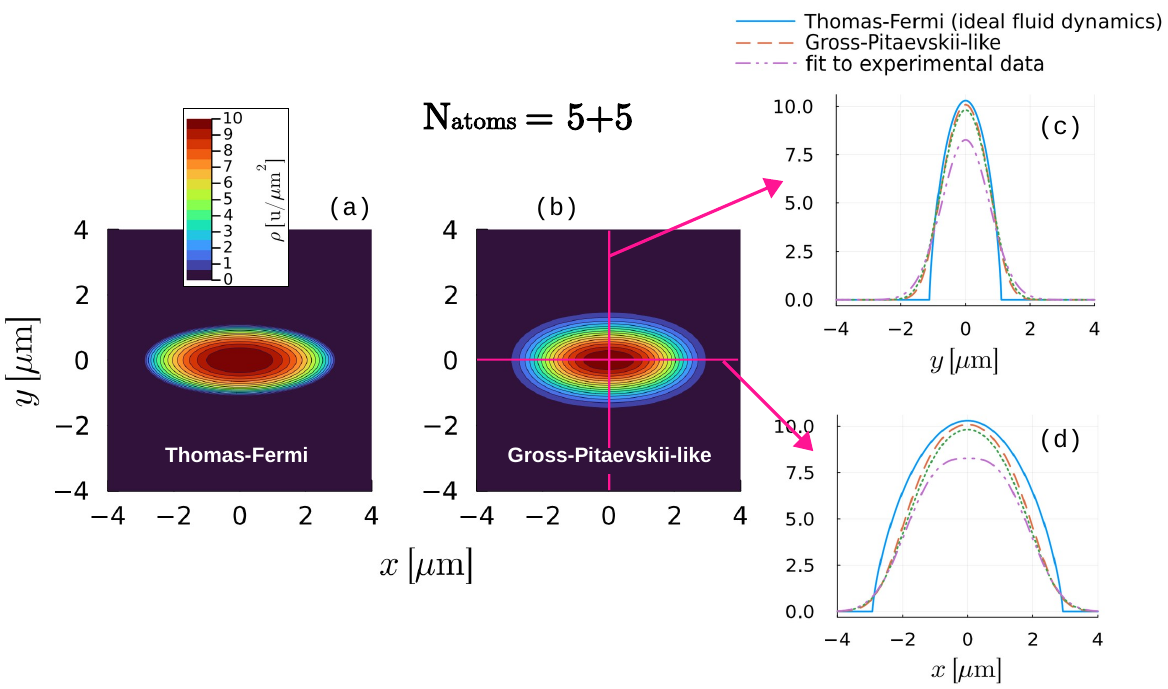}
    \caption{Initial condition for the hydrodynamic expansion of the mesoscopic Fermi gas. Predictions from many-body theory for a system of 5+5 $^{6}$Li atoms are shown both for the ideal hydrodynamic, or Thomas-Fermi case [panel (a)] and with the inclusion of second-order corrections, or quantum pressure [panel (b)] resulting in a Gross-Pitaevskii-like stationary equation. Slices of the initial density for $x=0$ [panel (c)] and $y=0$ [panel (d)] are compared to a fit of the experimental measurement of the initially trapped density shown in Fig.~\ref{fig:expansion} (shown as dot-dashed curves). The green dotted lines in panels (c) and (d) represent the results obtained with $\lambda=1.5$ in Eq.~(\ref{eq:initialDensityEquationWithYTerm}).}
    \label{fig:}
    \label{fig:5+5dens}
\end{figure}

We recall that, for a given equation of state, it is possible to obtain a hydrostatic solution for the density of the trapped system. This amounts to solving
\begin{equation}
   \frac{1}{m} \rho   \boldsymbol{\nabla} V = -   \boldsymbol{\nabla} P ,
\end{equation}
where $V$ is in our case the external harmonic oscillator potential. Using, then, $\nabla P = \kappa g \rho^{\kappa-1} \nabla \rho$, where $g$ and $\kappa$ come from the equation of state of the Fermi gas in Eq.~(\ref{eq:Pfit}) as discussed above, one arrives at the so-called Thomas-Fermi solution
\begin{equation}
    n(\mathbf{x}) = \left( \frac{(\kappa-1)(\mu_0 - \frac{m}{2} \omega^2_{jk} x_j x_k)}{g \, \kappa} \right)^{\frac{1}{\kappa - 1}} \, ,
    \label{eq:TFDensityProfile}
\end{equation}
which describes the system within the area defined by $\mu_0 > \frac{m}{2} \omega^2_{jk} x_j x_k$, where $\omega_{jk}$ are the frequencies of the confining potential. Here the value of $\mu_0$ determines the total mass, or particle number. The example of a two-dimensional Thomas-Fermi profile for a system that matches the experimental parameters is shown in Fig.~\ref{fig:5+5dens}. The inverse-parabola shape turns out to present tails that are too sharp compared to those experimentally observed. This suggests the potential impact of finite size corrections.

We discuss, thus, quantum corrections to the hydrostatic problem, which in a fluid dynamic setup amount to second-order corrections in the derivative expansion. Following the derivations of Ref.~\cite{Heyen:2024xxb}, if one does not neglect the quantum pressure (or von Weizs\"acker term) in the derivation of the superfluid hydrodynamic equations, a correction that is second-order in the derivatives of the density appears
\begin{equation}
     \frac{1}{m}\rho \boldsymbol{\nabla} V = -\boldsymbol{\nabla} P + \lambda \frac{\hbar^2}{2m^2} \rho \boldsymbol{\nabla} \left(\frac{\boldsymbol{\nabla}^2 \sqrt{\rho}}{\sqrt{\rho}} \right).
\label{eq:initialDensityEquationWithYTerm}
\end{equation}
This equation can in general be derived from the variation of the effective action for a superfluid order parameter field. For $\lambda=0$, one recovers ideal hydrodynamics, and the solution of this equation is the Thomas-Fermi profile discussed above. For $\lambda=1$, this is equivalent to a generalized Gross-Pitaevskii equation with an equation of state chosen suitably to match that of the $^{6}$Li gas.

We now want to test the correction induced by this term on the predicted density of the trapped system. This is computed numerically, following the method discussed in Ref.~\cite{Heyen:2024xxb}. The results are shown in panel (b) of \cref{fig:5+5dens}. We note a significant enhancement in the tails of the mass distribution. Interestingly, for a Gross-Pitaevskii-like scenario, the predicted tails are found to closely match those of the experimental system. We stress that this comes naturally and without adding any extra model parameters.

We conclude by noting that the Heidelberg group has recently performed the measurement of the initial density of a strongly-interacting system of 6+6 fermions with a separation of scales between the trap size and the fermion-fermion pair size \cite{Brandstetter:2024gur}.  As this is closer to the dilute type of system that the Gross-Pitaevskii equation is meant to describe, it would be of great interest to study whether the measured trapped density can be captured by the solution of the second-order hydrostatic problem. This is left for a future analysis.

\subsubsection{The correlated Gaussian approach to few-body gases }\label{sec:ECG}

The explicitly correlated Gaussian (ECG) technique is a numerical method based on the Ritz variational principle, which seeks to minimize the energy of a quantum system using trial wave functions. This approach has found broad applications in solving atomic, molecular, and nuclear few-body systems quantum-mechanically. In the context of cold atoms, the ECG approach has been used to study Efimov physics in Boson systems, BCS-BEC crossover in unitary Fermi gases, and properties of Fermion clusters, among others~\cite{mitroyTheoryApplicationExplicitly2013}.

The typical Hamiltonian that can be solved by ECG is the following:
\begin{align}
H = \sum_{i=1}^N\left[\frac{\mathbf{p}_i^2}{2m}+V_{\text{2D}}(\mathbf{r}_i)\right]+\sum_{i<j}V_{\text{2b}}(\mathbf{r}_i-\mathbf{r}_j),
\label{eq:many-body}
\end{align}
where $N$ is the total number of particles, $m$ the single-particle mass, $\mathbf{r}_i$ and $\mathbf{p}_i$ the 2D position and momentum vectors of the $i$-th particle, $V_{\text{2D}}(\mathbf{r}_i)$ the 2D single-particle harmonic trapping potential, and $V_{\text{2b}}(\mathbf{r}_i-\mathbf{r}_j)$ the two-body interactions. After separating the center-of-mass degrees of freedom from the relative degrees of freedom, the system Hamiltonian can be written in terms of $N-1$ Jacobi coordinates $\mathbf{x}_i$ $(i=1,2,\cdots, N-1)$~\cite{mitroyTheoryApplicationExplicitly2013}. Jacobi coordinates are linear combinations of particle coordinates that separate the center-of-mass motion from the relative motion, simplifying the mathematical treatment of the system.

In 3D, the explicitly correlated Gaussian approach uses a Gaussian model interaction potential:
\begin{eqnarray}
V_{\text{2b}}(\mathbf{r}_i-\mathbf{r}_j) = V_0 \exp\left(-\frac{|\mathbf{r}_i-\mathbf{r}_j|^2}{r_0^2}\right).
\end{eqnarray}
The Gaussian potential depth $V_0$ and width $r_0$ are adjusted to reproduce the desired 2D $s$-wave scattering length used in the experiment. However, in 2D, one needs to include the effective range in the low-energy expansion,
$$\cot[\delta(k)] = \frac{2}{\pi} \left[ \gamma + \ln \left( \frac{ka_{2D}}{2} \right) \right] + \frac{1}{2} k^2 R_s + ..., $$
where $\gamma$ is Euler's constant~\cite{yinFewBodyPerspectiveQuantum2020} and $k$ is the relative wave vector. The scattering length and effective range can then be calculated from the asymptotic behavior of the zero-energy scattering solution $u(r)$ by matching the phase shift in the asymptotic limit:
$$
a_{2D} = \lim_{r \to \infty} r \exp\left(-\frac{u(r)}{r u'(r)}\right), \quad R_s = 2 \int_0^\infty \left[ \ln^2 \left( \frac{r}{a_{2D}} \right) - u^2(r) \right] r dr.$$

This choice of potential not only allows us to model the short-range interactions effectively but also enables analytical evaluation of matrix elements when combined with Gaussian trial basis functions.

In experiments, however, the effective range is typically negative, which cannot be reproduced by a single attractive Gaussian potential. This could be solved by superposing two Gaussians together to form an attractive potential with a repulsive barrier~\cite{yinFewBodyPerspectiveQuantum2020}

We use correlated Gaussians, $\phi(\mathbf{x}, A)=\exp\left(-\frac{1}{2}\mathbf{x}A\mathbf{x}\right)$, as the basis functions to solve for the $N$-fermion systems. Here $\mathbf{x}$ collectively denotes the $N-1$ Jacobi coordinates, and $A$ is a symmetric and positive definite variational matrix that characterizes the correlation between particles. These basis functions are chosen for their flexibility in describing particle correlations and their analytical tractability in evaluating kinetic, potential, and interaction terms.

For fermionic systems, we must anti-symmetrize the full wave function using a Slater determinant. The basis functions are randomly generated from a set of trial wave functions designed to cover all relevant length scales of the system, from the range of the interaction potential to the characteristic length of the harmonic trap.

While the ECG method has been successfully applied to systems of up to about 10 fermions in spherically symmetric traps~\cite{yinTrappedUnitaryTwocomponent2015}, the anisotropy of our system increases the computational complexity. This is because we need two sets of parameters to describe the $x$ and $y$ directions instead of a single radial direction. Consequently, we expect to be limited to 8 or fewer fermions in such an anisotropic trap.

\subsubsection{Dissipation in small systems}\label{sec:dissipation}

Transport processes can lead to dissipation. For example in fluid flow
(cf. \cref{sec:CAfluid}), shear and bulk friction as well as thermal conductivity
give rise to local dissipation, and the entropy increases as the fluid
approaches equilibrium. Experiments can measure the relaxation rates
of collective modes of harmonically trapped gases \cite{Riedl:2008} or the
sound attenuation in uniform systems \cite{Patel:2019udb}. Alternatively, in
continuously driven systems transport processes give rise to
dissipative heating, and a heating rate or phase shift of the response
can be measured to quantify transport coefficients (\cref{sec:diffusivity}). The
resulting \emph{transport time} (different for each transport channel) is
well-defined also in small and strongly correlated systems, even when it
may not be possible to define a mean free path. Within the memory
function formalism, the transport coefficients result as the product
of the transport time, which is a dynamical property, and a
thermodynamic factor related to the equation of state \cite{Frank:2020}.

Computationally, the transport time is obtained by a number of popular
approaches. Boltzmann kinetic theory follows the evolution of the
single-particle distribution function in phase space, which is
affected both by collisions and by mean-field interactions. It is
justified when scattering is nearly elastic (no significant broadening
of spectral functions) and often assumes molecular chaos (the joint
distribution function of both scattering partners factorizes into a
product of single-particle distributions). The collective modes of a
trapped system are then obtained by solving the equation of motion in
phase space, for instance using the method of moments \cite{Riedl:2008, Chiacchiera:2009}.

In the strongly correlated fermion system, local pair correlations
play a crucial role (\cref{sec:fewatoms,sec:ECG,sec:attractor}). These violate the
assumption of molecular chaos: with short-range attraction it is highly likely to have an up-spin
particle near a down-spin particle, and their distributions are no
longer independent. Recent Boltzmann calculations \cite{Dusling:2013}
take this into account, and it has been proposed to include the pair
distribution as a separate degree of freedom in order to account for
these correlations \cite{Fujii:2023}. A full quantum transport computation
becomes necessary when scattering occurs so rapidly that successive
collisions can no longer be treated as independent, and there are
substantial inelastic effects that lead to broadening of the spectral
functions. In this case, linear response theory expresses the
transport properties via the Kubo formula as a response to an external
perturbation. These response functions of the stress tensor (for
viscosity) or the energy current (for thermal conductivity) are
computed by various many-body approaches, ranging from Quantum Monte
Carlo \cite{Wlazlowski:2012} via diagrammatic approaches \cite{Enss:2011} to holography \cite{Schafer:2014viscosity}. An important technical aspect is that transport requires
the response in real time or frequency, which either requires the
analytical continuation of numerical imaginary time data or the
solution of Dyson-Schwinger equations directly in real frequency \cite{Johansen:2024, Enss:2024, Dizer:2024}.  Theoretical results for the
transport coefficients in uniform systems can be applied to trapped
gases via the local density approximation (LDA).

Small systems provide new ways to probe transport properties. One can
observe pairing directly in real space, see how correlations build up in
time, and study their effect on transport \cite{Brandstetter:2023jsy}. An aim
is to disentangle the effects of collisions, mean-field interactions,
and pairing correlations to reveal the origin of collectivity. Furthermore, small systems exhibit precursors to
critical fluctuations near phase transitions and link them to
few-particle excitations \cite{Bjerlin:2016}. In particular, local dissipation
that originates from local correlations (\cref{sec:diffusivity}) does not require
large systems or long-wavelength excitations, in particular for bulk
viscosity where the scattering rate is found to scale predominantly
with temperature but not with density, as a consequence of pair
excitations \cite{Enss:2019, Enss:2025}. Fully spatially resolved dynamics of Fermi gases, including superfluids, has recently been achieved using time-dependent density functional theory \cite{Barresi:2023}.

\subsubsection{Non-hydrodynamic modes in ultra-cold atoms}\label{sec:nonhydro}

Hydrodynamics describes the universal evolution of 
near-equilibrium systems on long timescales, when their dynamics are governed only by conservation laws
like conservation of energy and momentum. 
Out-of-equilibrium systems have transient, non-universal features of their dynamics, characterized by non-hydrodynamic modes, that vanish in the hydrodynamic regime
but can play a major role far from equilibrium.

Relativistic hydrodynamics has enjoyed remarkable success in
the phenomenology of heavy-ion collisions (see \cref{sec:HIattractor}). 
The geometry in the initial state of nuclear collisions, along with the pressure-driven expansion
characteristic of hydrodynamics, generate long-range correlations in the momenta of produced particles 
that have been used in tandem with hydrodynamic simulations to constrain transport coefficients like viscosity of the quark-gluon plasma~\cite{Nijs:2020ors,JETSCAPE:2020shq}. 
Much more surprisingly, hydrodynamic-like correlations have also been observed in much smaller systems like proton-nucleus collisions, where
the assumptions of hydrodynamics seem to break down due to large gradients. 

Groundbreaking theoretical work~\cite{Heller:2015dha} discovered attractor solutions for relativistic hydrodynamics. Due to the decay of non-hydrodynamic modes, different initial conditions
can reach a common solution long before the naive regime of applicability of hydrodynamics. Significant recent work in the field of heavy-ion collisions has focused on
understanding whether attractors may be part of the surprising applicability of hydrodynamics in small collision systems, but it is extremely challenging to access the non-equilibrium
structure in these collisions. Cold atomic gases provide an avenue to study the onset of
hydrodynamics and non-hydrodynamic modes in a strongly interacting system with an experimental setup that is under much better control.

Hydrodynamic transport in cold atomic gases has been widely studied through collective oscillations, where one deforms a trapped gas and studies the frequency and damping rate of its subsequent oscillations. 
There are two independent collective oscillations, one associated with radial deformations (called the breathing mode) and one for shear deformations (quadrupole mode). For a two-dimensional gas
with radii $b_x$ and $b_y$, the sum $\delta B = (\delta b_x + \delta b_y)/2$ and difference $\delta Q = (\delta b_x - \delta b_y)/2$ of their deformations obey
\begin{equation}
	\delta Q(\bar{t}) = \gamma
\exp(-\Gamma_{Q,0} \bar{t}) \cos(\omega_Q \bar{t} + \text{const}),
\end{equation}
\begin{equation}
\delta B(\bar{t}) = \beta \exp(-\Gamma_B \bar{t}) \cos(\omega_B \bar{t} +
\text{const}).
\end{equation}
In the hydrodynamic regime, $\Gamma_B = 0$, $\omega_B = 2$ and $\omega_Q =\sqrt{2}$. 
On the other hand, if interactions between particles are weak, $\omega_Q = 2$ while other quantities stay the same.
Fits to these coefficients have been used to measure the shear viscosity of Fermi gases near unitarity~\cite{Kinast:2005zz, Vogt_2012}.

However, the arguments in the previous paragraph are based purely on hydrodynamic transport, and neglect transient, non-hydrodynamic effects that may be dominant out of equilibrium.
Ref.~\cite{Brewer:2015hua} studied collective oscillations in simulations of cold
atomic gases beyond the regime of hydrodynamics using a lattice truncation of the Boltzmann equation, and
found that the quadrupole mode also has a contribution from a purely-damped, non-hydrodynamic component, 
\begin{equation}
\label{eq:fullQ} \delta Q(\bar{t}) = \gamma \exp(-\Gamma_{Q,0} \bar{t}) \cos(\omega_Q \bar{t} + \text{const}) + A \exp(-\Gamma_{Q,1}
\bar{t}).\\ 
\end{equation} 
This provides a much better fit of the form of the early-time behavior of cold atomic gases released from an anisotropic trap within non-linear kinetic theory. The breathing mode is not found to have any discernible non-hydrodynamic component. \Cref{eq:fullQ} can be found analytically
in kinetic theory near (but outside) the hydrodynamic limit, and is confirmed in numerical simulations. However, none of these approaches is fully controlled for quantum gases
in the far out-of-equilibrium regime which motivates the study of non-hydrodynamic transport in experiments on cold atomic gases.

To take a first step in this direction, Ref.~\cite{Brewer:2015ipa} reconsidered the experimental measurements of hydrodynamic
collective modes in 2D Fermi gases \cite{Vogt_2012} to search for signatures of non-hydrodynamic excitations. Though Ref.~\cite{Vogt_2012}
did use a purely damped term in their fits as in \cref{eq:fullQ}, they did not report the values for the damping rate of the non-hydrodynamic
component, $\Gamma_{Q,1}$. In Ref.~\cite{Brewer:2015ipa} the raw data was re-analyzed to extract the non-hydrodynamic mode damping rate. Uncertainties in the 
re-analysis are large, primarily because the time resolution of the experiments was not fine enough to capture the fast decay of the transient modes. However, 
it finds a hint of a non-zero non-hydrodynamic component of the quadrupole oscillation, and motivates further experiments to study the detailed structure of non-hydrodynamic modes in
cold atomic gases.

\subsubsection{Hydrodynamic attractor in ultra-cold atoms } \label{sec:attractor}

\begin{figure}[t]
\begin{minipage}{0.5\hsize}
\begin{center}
\includegraphics[width=\linewidth]{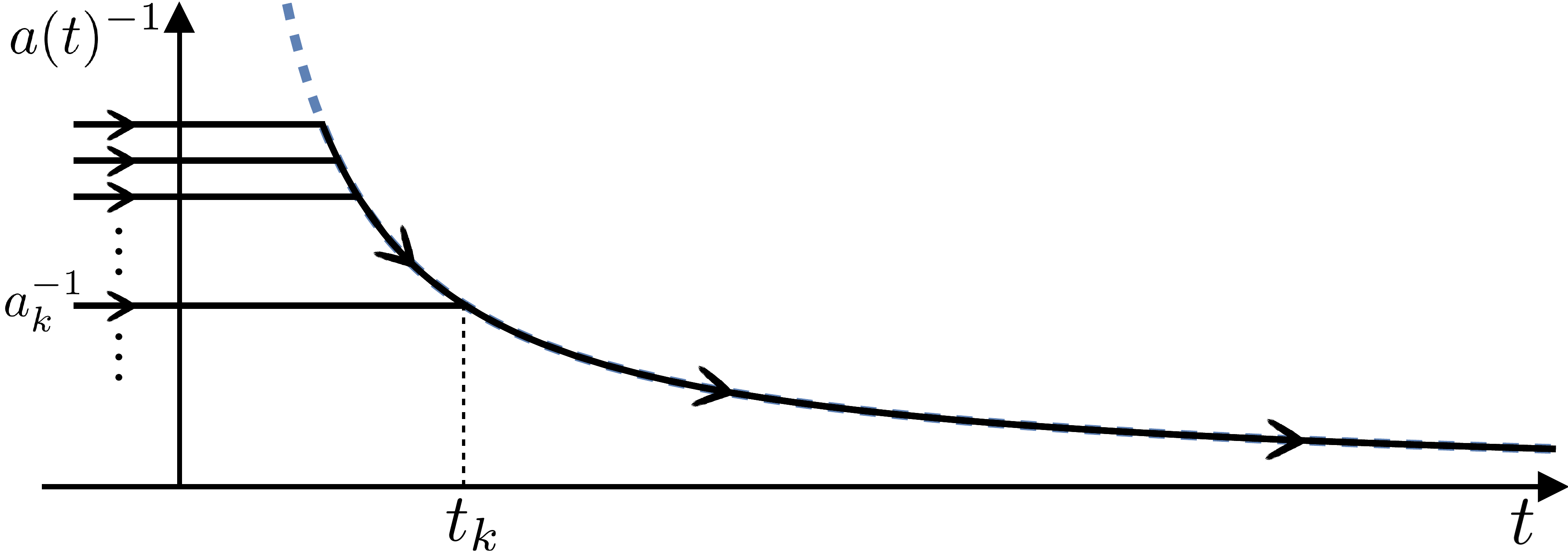}
\end{center}
\end{minipage}
\hspace{0.6em}
\begin{minipage}{0.45\hsize}
\begin{center}
 \includegraphics[width=0.9\linewidth]{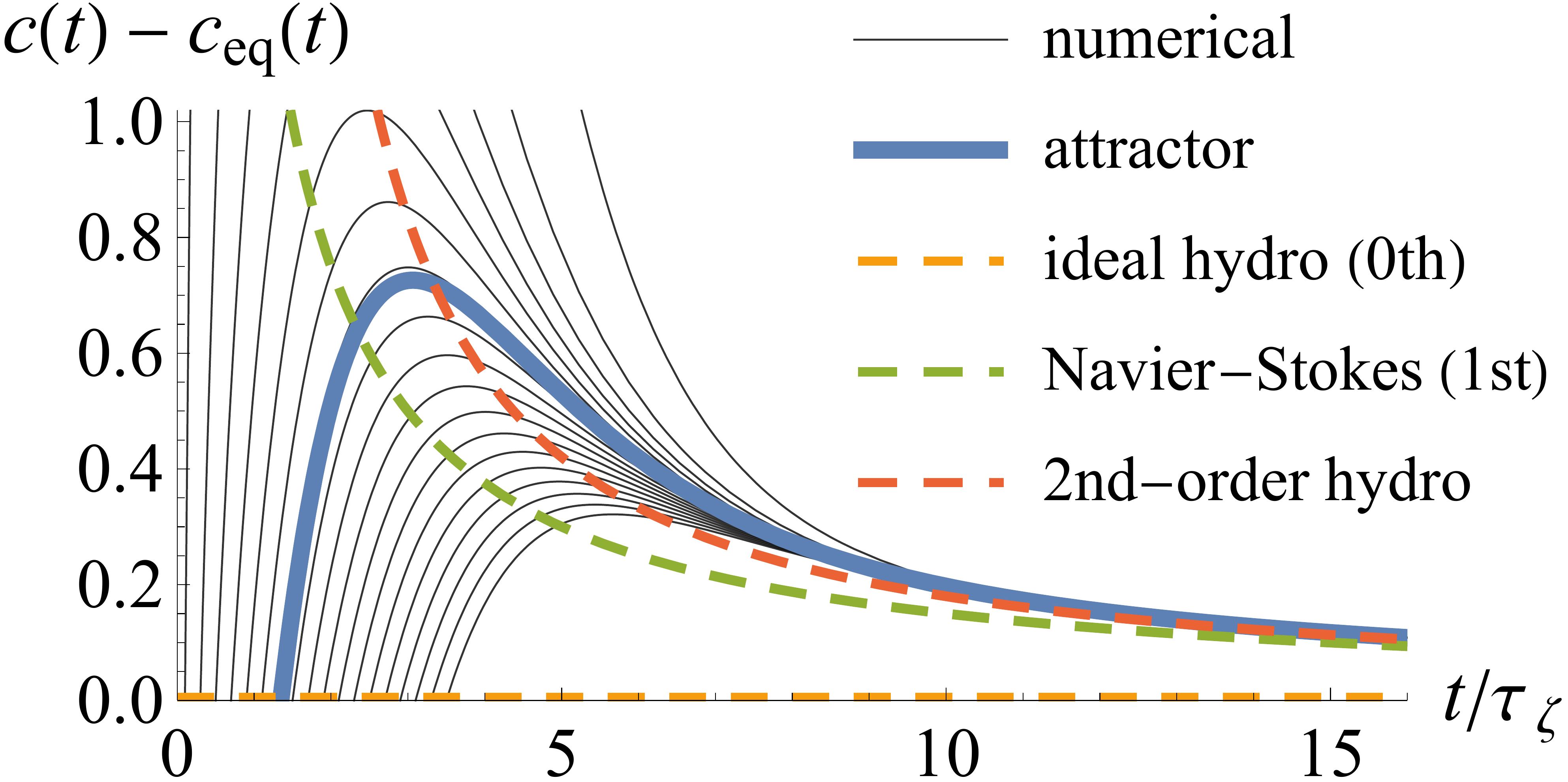}
 \end{center}
\end{minipage}
\caption{
(Left) The driving protocol for the scattering length to realize the hydrodynamic attractor in a uniform system without fluid velocities.
We keep the scattering length at a constant value $a_k$ until time $t=t_k$ and then approach it to the unitary limit $a^{-1}(t)\to0$ asymptotically with a power law of time, \cref{eq:protocol}. 
To realize various initial conditions, the initial scattering length $a_k$ and time $t_k$ are varied while fixing $\tilde{a}=a_{k}(\tau_{\zeta}/t_{k})^{\alpha}$ for $k=1,2,3,\ldots$.
Due to this fixing, the power-law drive of all drives is on a single curve.
(Right) The plot of the deviation of the dimensionless contact density from its equilibrium value, $c(t)-c_{\textrm{eq}}[a(t)]\equiv(\mathcal{C}(t)-\mathcal{C}_{\textrm{eq}}[a(t)])/(12\pi m a(t))\times (\tau_{\zeta}/\zeta[a(t)])$, as a function of $t/\tau_{\zeta}$ under power-law drive with $\alpha=1/2$. 
While black thin lines give numerical solutions for $t_k/\tau_{\zeta}=0.1,0.3,0.5,\ldots,3.5$, the blue thick line represents the hydrodynamic attractor solution.
We also plot hydrodynamic results of zeroth-order (orange), first-order (green), and second-order (red dashed line) from the expansion of $\pi_{\textrm{att}}(t)$ with respect to $\tau_{\zeta}/t$.
These figures are adapted
from K.~Fujii and T.~Enss, ``Hydrodynamic attractor in Ultracold Atoms'' Phys.~Rev.~Lett.~\textbf{133(17)} 173402 (2024)~\cite{Fujii:2024},
Copyright \copyright\, 2024 by the American Physical Society.
}
 \label{fig:attractor_coldatom}
\end{figure}

In this section, we propose a setup to observe hydrodynamic attractors in ultra-cold atomic gases~\cite{Fujii:2024}.
We consider a two-component Fermi gas at any temperature in three dimensions whose interaction is characterized only by the two-body $s$-wave scattering length.
In this system, the time variation of the scattering length at fixed gas volume allows the realization of phenomena equivalent to isotropic fluid expansions of the gas, as there are no other intrinsic reference scales~\cite{Fujii:2018}.
Taking advantage of this equivalence, we show that hydrodynamic attractors can be investigated in the Fermi gas by driving the scattering length to the strongly interacting limit, i.e., the unitary limit, over time, as schematically depicted in the left panel of \cref{fig:attractor_coldatom}.
It is worth noting that our method applies to a wide range of ultracold atomic systems whose interaction is fully characterized by the $s$-wave scattering length, such as fermions in the whole BCS-BEC crossover and repulsive Bose gases.

To realize the universal hydrodynamic attractor, we drive the scattering length so that the system is initially brought out of equilibrium and then gradually approaches thermal equilibrium.
Specifically, we drive the scattering length as
\begin{align}
a_{k}(t)^{-1}=
\left\{
\begin{array}{ll}
a^{-1}_{k} & t<t_k, \\[0.8ex]
a^{-1}_{k}(t/t_k)^{-\alpha} & t>t_k,
\end{array}
\right.\qquad \textrm{for}\quad k=1,2,\ldots
\label{eq:protocol}
\end{align}
with $\alpha>0$.
Namely, we fix the scattering length at $a_k$ until time $t_k$, after which we drive it according to a power law with exponent $\alpha$.
The subscript $k$ is used to distinguish the drives corresponding to different initial conditions.
To make drives with different initial conditions the same over a long time, we fix $\tilde{a}:=a_{k}(\tau_{\zeta}/t_{k})^{\alpha}$, which represents the scattering length at the relaxation time in the power-law drive.
Consequently, at long times, all the driving protocols lie on a single curve, as depicted in the left panel of Fig.~\ref{fig:attractor_coldatom}.

According to the linear response theory, the modulation of the scattering length causes a change in its conjugate quantity, the contact density.
We express the deviation of the contact density from its equilibrium value as $\pi(t)\equiv \left(\mathcal{C}(t)-\mathcal{C}_{\textrm{eq}}[a(t)]\right)/(12 \pi m a(t))$.
Here, $\mathcal{C}(t)$ is the contact density at time $t$, while $\mathcal{C}_{\textrm{eq}}[a]$ is the instantaneous contact density determined in thermal equilibrium with the scattering length $a$.
The correlation function for $\pi(t)$ corresponds to the complex bulk viscosity~\cite{Fujii:2020} and is well described by a Drude form at low frequencies~\cite{Nishida:2019,Enss:2019,Hofmann:2020,Fujii:2023}, so that $\pi(t)$ exhibits relaxation dynamics characterized by the relaxation time $\tau_{\zeta}$ at long times.

The right panel of \cref{fig:attractor_coldatom} plots the dimensionless version of the deviation, $\pi(t)\times (\tau_{\zeta}/\zeta[a(t)])$, under the drive of the scattering length with a power $\alpha=1/2$.
Since the bulk strain rate tensor with time-dependent scattering length is given by $-3\dot{a}(t)/a(t)$~\cite{Fujii:2018} and becomes smaller as time increases, the system approaches thermal equilibrium, and thus, the deviation tends toward zero.
Importantly, the numerical solutions for $t_k/{\tau_{\zeta}}=0.1,0.3,\ldots,3.5$ (black thin lines) converge rapidly to the universal attractor solution $\pi_{\textrm{att}}(t)$ (blue thick line) before being reduced to the hydrodynamic solutions (dashed lines).
This is because the long-time expansion of $\pi_{\textrm{att}}(t)$, i.e., the expansion with respect to $\tau_{\zeta}/t$, is asymptotic with convergence radius zero, whereas its expansion provides hydrodynamic solutions (dashed lines).
Due to this asymptotic nature of $\pi_{\textrm{att}}(t)$, the universal attractor behavior emerges before the time scale $t\gg \tau_{\zeta}$ at which hydrodynamics becomes accurate; this is called the hydrodynamic attractor.

The attractor behavior that we have proposed can be probed in current cold atom experiments, and thus, our results establish cold atom systems as a new platform for exploring hydrodynamic attractors. While experimental studies of rapid hydrodynamization are currently actively pursued in 1D integrable systems \cite{Le:2022ntg}, in higher dimensions new explorations could be envisaged based on existing platforms on thermal gases (e.g. \cite{2018PhRvA..98a1601F,Li:2024ivj}).
Recently, the attractor behavior has been also studied in periodically driven systems~\cite{Mazeliauskas:2025jyi}. Such systems never approach the Navier-Stokes behavior and instead exhibit a cyclic attractor even at late times. Further considerations on the emergence of an attractor for a Fermi gas near unitarity can be found in \cite{Heller:2025yxm}.

\subsubsection{Entanglement and thermalization in small closed systems }\label{sec:entanglement}

A closed quantum system initialized in a pure state $|\psi(0)\rangle$ and subject to a Hamiltonian $H$ evolves unitarily in time $t$ as $|\psi(t)\rangle = e^{-iHt}|\psi(0)\rangle$. Since the global state $\rho(t) = |\psi(t)\rangle \langle \psi(t)|$ remains pure, it can strictly not evolve into a mixed thermal state $\rho(\beta) \propto e^{-\beta H}$. Nevertheless, isolated quantum many-body systems---including heavy-ion collision and ultracold atom experiments---generally relax to thermal equilibrium~\cite{Berges:2020fwq}. This apparent dichotomy can be resolved by considering a (small) subsystem $A$ and its (larger) complement $B$. All physical observables $\mathcal{O}_A$ accessible through measurements in $A$ are completely determined by the reduced density matrix $\rho_A(t) = \text{Tr}_B \left[|\psi (t) \rangle \langle  \psi (t)|\right]$, where $\text{Tr}_B [\dots]$ denotes a partial trace. A local observable $\langle \mathcal{O}_A \rangle (t) = \text{Tr}_A \left[ \mathcal{O}_A \rho_A(t) \right]$ can thus approach a thermal value $\langle \mathcal{O}_A \rangle_\beta = \text{Tr} \left[ \mathcal{O}_A \rho_\beta \right]$ as $t \rightarrow \infty$ while the global state remains pure. Intuitively, the complement $B$ serves as an environment which makes $A$ appear ``thermal''.
	
	This intuition can be made more precise through the notion of \emph{quantum entanglement}. For a bi-partition $A:B$ and an associated factorization of the Hilbert space $\mathcal{H}$ as $\mathcal{H} = \mathcal{H}_A \otimes \mathcal{H}_B$, a pure quantum state $|\psi\rangle \in \mathcal{H}$ is called (bi-partite) \emph{entangled} if it cannot be written as a product state, i.e., there do not exist pure states $|\psi_{A/B}\rangle \in  \mathcal{H}_{A/B}$ such that $|\psi\rangle = |\psi_A\rangle \otimes |\psi_B \rangle$. One can show that $|\psi \rangle$ is entangled if and only if the von Neumann \emph{entanglement entropy}
	\begin{align}
		S_A = - \text{Tr}_A \left[\rho_A \log \left(\rho_A\right) \right]
	\end{align}
	 is non-zero~\cite{horodecki2009quantum}. In other words, the mixedness of the reduced state $\rho_A$ measures the entanglement among $A$ and $B$. Related measures include the $n$-th order Rényi entropies $S_A^{(n)} = \frac{1}{1-n} \text{Tr}_A \log \left(\rho_A^n\right) $, which are often simpler to access in quantum simulation experiments~\cite{daley2012measuring}.
	
	For a closed, but (locally) thermalized system, it is the entanglement entropy $S_A$ that mimics the role of a thermodynamic entropy. As typical thermal systems exhibit an extensive scaling of thermodynamic entropy, one expects the entanglement entropy to grow with the volume $V_A$ of $A$, i.e., $S_A \propto V_A$. Such volume-law scaling is indicative of highly entangled quantum many-body states, in contrast to ground states of gapped Hamiltonians that typically exhibit area-law scaling of the entanglement entropy~\cite{eisert2010colloquium}.  When initializing a system with small or even vanishing $S_A$, the build-up of entanglement over time thus constitutes a characteristic fingerprint of thermalization that indicates an effective loss of information in the subsystem. It is generally believed that such information scrambling is driven by Hamiltonians that exhibit quantum chaos, a property that is often analyzed by studying eigenenergy level statistics in the context of the eigenstate thermalization hypothesis~\cite{d2016quantum}.
	
	In the last decade, progress in the control and manipulation of synthetic quantum systems has made these theoretical ideas experimentally accessible. For example, local thermalization and the simultaneous build-up of entanglement has been observed in a remarkably small system of six atoms trapped in an optical lattice~\cite{kaufman2016quantum}. More recently, programmable quantum simulators and digital quantum computers are being used to implement real-time evolution for studying thermalization~\cite{andersen2024thermalization}, where entanglement can be quantified through protocols involving randomized measurements~\cite{brydges2019probing}. However, since the unbiased measurement of entropies typically requires a measurement budget that scales exponentially with the (sub)system size, alternative protocols exploiting the structure of the underlying quantum state are being developed to study large-scale entanglement in quantum simulation experiments~\cite{joshi2023exploring}.
	
	In summary, studying entanglement properties can help to shed light on whether and how quantum systems thermalize. In synthetic quantum systems, this view-point led to the discovery of so-called quantum many-body scars, exceptional states of low entanglement that exhibit anomalously slow relaxation~\cite{bernien2017probing,serbyn2021quantum}. In high-energy collider physics, quantum entanglement may explain the observed thermal excitation spectra in $e^+ e^-$ collisions~\cite{berges2018thermal}. Finally, we are just beginning to explore the role of entanglement for the dynamics of general (lattice) gauge theories, e.g.~\cite{mueller2022thermalization,halimeh2023robust,ebner2024eigenstate}.

\section{Status and prospects with high-energy collisions}\label{sec:high-energy}

\subsection{What is all the fuss about small systems in high-energy collisions?}\label{sec:smallpuzzle}

For completeness, here we recall and elaborate some of the key points discussed in \cref{sec:hydro_2}.

\paragraph{QGP in heavy-ion collisions.}

The formation of QGP has been established in experiments at RHIC on Au+Au collisions at $\sqrt{s_{\rm NN}} =$ 200 GeV and on Pb+Pb collisions at a center-of-mass energy of $\sqrt{s_{\rm NN}} =$ 2.76 and 5.02 TeV at the LHC at CERN. A relativistic hydrodynamic framework proves extremely successful in the description of the experimental observables, mainly in two respects.

First, state-of-the-art hydrodynamic simulations naturally capture the steepness and normalization of the spectrum of soft hadrons emitted in the collisions (radial flow) when taking as input the equation of state of high-temperature QCD \cite{HotQCD:2014kol}, that is, including de-confined quark and gluon degrees of freedom in the medium. The average momentum of the emitted hadrons in combination with the measured particle multiplicities indicate, in particular, that the system reaches temperatures of order trillion kelvins in its early stages, fully consistent with the formation of a QGP \cite{Gardim:2019xjs}. The inclusion of initial-state fluctuations in the initial energy density profiles via fluctuating nucleon-nucleon collision positions enables the model to capture, in addition, the fluctuations of the radial flow \cite{Bozek:2012fw,Gale:2012rq,Bozek:2017elk,Schenke:2020uqq,Giacalone:2020dln,Samanta:2023amp,Samanta:2023kfk,Parida:2024ckk}, which have been extensively analyzed experimentally \cite{STAR:2019dow,ATLAS:2024jvf,ALICE:2024apz,ATLAS:2025ztg,ALICE:2025rtg}. 

Not only the radial (isotropic) flow of the system is captured by a hydrodynamic picture. As discussed, the particle distribution in momentum space, as a function of the azimuthal angle $\varphi$, can be parameterized by a Fourier cosine series, 
\begin{equation}
\frac{dN}{d\varphi} \sim 1 + \sum_{n = 1}^{\infty} 2 \, v_{n}\cos[n(\varphi-\Psi_{n})],
\label{eq:singleparticlevn}
\end{equation}
where $\Psi_n$ are the $n^{\rm th}$-order planes of symmetry.  The coefficients $v_n$ describe the magnitude of the system response to the anisotropies of the initial energy density distribution, and the lower-order $v_n$ coefficients ($n = 2, 3$) are in fact directly proportional to the initial-state spatial eccentricities, $v_n \propto \varepsilon_n$ (for $n > 3$, the $v_n$ coefficients contain contributions from the $n^{\rm th}$-order eccentricity as well as lower orders)~\cite{Heinz:2013th}.  Extensive measurements have been made of the $v_n$ coefficients as a function of the collision system, size (centrality), and energy, as well as the particle species, momentum ($p_{T}$), etc.~\cite{ALICE:2022wpn}, in order to characterize properties of the QGP medium. This vast ensemble of observables can all be quantitatively captured by the hydrodynamic framework of high-energy collisions \cite{Gale:2012rq}, and are today utilized in comprehensive Bayesian analyses \cite{Paquet:2023rfd} to pin down values of parameters encoding microscopic QCD dynamics (such as the shear and bulk viscosities). Measurements of the $v_n$ fluctuations as well as correlations between $v_n$ coefficients have led to further constraints on our understanding of QGP properties. 

\paragraph{QGP in small systems?}

The first observations of collectivity in small systems came from two-particle correlations in $(\Delta\varphi,\Delta\eta)$, in particular those measured in collisions which produce large multiplicities.  The multiplicity is related to the ``activity'' of the collision, and monotonically increases with the size of the interaction region (in heavy-ion collisions) or the number of parton-parton interactions (in small collision systems). Correlations in $\Delta\varphi$, the relative azimuthal angle between two particles, over a large range in $\Delta\eta$, the relative pseudorapidity, are a natural consequence of collective flow in a hydrodynamic description.  If particles are correlated with respect to a global symmetry plane as \cref{eq:singleparticlevn} (e.g., a plane dictated by the initial spatial anisotropies, $\varepsilon_n$), then the two-particle distribution in $\Delta\varphi$ has the form
\begin{equation}
\frac{dN}{d\Delta\varphi_{ab}} \sim 1 + \sum_{n = 1}^{\infty} 2 \, v_{n,a}v_{n,b}\cos[n\Delta\varphi_{ab}],
\label{eq:pairvn}
\end{equation}
where the two particles in the pair, $a$ and $b$, may be drawn from different classes defined by their properties (e.g.\ momentum, species) and thus have different azimuthal flow coefficients, $v_{n,a}$ and $v_{n,b}$ (cross-terms proportional to $v_{n,a}v_{m,b}$, for $n\neq m$, may also arise as the symmetry planes of different orders are correlated).  Furthermore, these correlations are expected to persist over wide ranges of $\Delta\eta$ since the particles respond to the flow field defined by the global symmetry of the system.  While such characteristic correlations in $(\Delta\varphi,\Delta\eta$) were observed in heavy-ion collisions, it was surprising to observe them in collisions where the system was expected to be too small to produce a QGP or to build up a flow field defined by a global plane of symmetry.  

Since the first observation of the ridge in p+p and p+Pb collisions, experimental and theoretical efforts have been undertaken to investigate and explain this phenomenon.  Despite significant experimental challenges related to subtracting short-range (local) non-flow-like correlations \cite{Feng:2024eos}, and although alternative theoretical advances have demonstrated that a hydrodynamic approach is not strictly necessary to produce azimuthal correlations over wide ranges in pseudorapidity \cite{Grosse-Oetringhaus:2024bwr}, a rather convincing picture has emerged whereby the observed flow and momentum anisotropies originate from a pressure-gradient-type response of the created matter to the shape of its initial geometry. This picture is expected to become even more established thanks to upcoming data from collisions of highly-deformed $^{20}$Ne nuclei, collected by the LHC in 2025, offering a means to fine-tune the initial-state geometry of a small system.

\paragraph{Prominent open questions.}

The observation of flow-like signals in small collision systems  has prompted a reassessment of the conventional interpretation of both heavy-ion and $p$+$p$ collisions leading to several fundamental open questions, including:
\begin{enumerate}
\item How does initial state energy deposition (geometry) transfer into a final-state anisotropic particle distribution? Is hydrodynamics the explanation in large systems? What is the microscopic description, on the level of particle interactions, for this hydrodynamic behavior?
\item What is the origin of anisotropic flow in small systems? Is it hydrodynamics?  How many scatterings or interactions does it take to generate flow, to equilibrate, and to thermalize?
\item Is there a QGP medium produced in small systems? What are its properties?
\item What is the interplay between momentum scales (hard and soft probes)? Is there quenching (suppression or modification) of hard probes in small systems?
\end{enumerate}

The recent experimental and theoretical work, and remaining open questions, clearly demonstrate the complementarity of studies of large and small collision systems. Therefore, making progress in understanding both small and large collision systems can inform our understanding of high-energy nuclear and hadron collisions as a whole, thus enabling us to attack the following questions:  
\begin{itemize}
\item Does the presence of long-range collective effects in $p$+$p$ collisions indicate the presence of a hydrodynamic medium?  Is a QGP formed in small collision systems?  {\em What can heavy-ion collisions teach us about more fundamental collisions at high energies?}  
\item Are there alternative explanations that can fully describe the observed flow-like signals in small collision systems?  If so, is our understanding of anisotropic flow in heavy-ion collisions incomplete?  {\em What can small collision systems teach us about collisions of heavy nuclei?}  
\end{itemize}

As we review and discuss in the next section, thanks to the great breadth of colliding systems that have been and will be analyzed, combined with the great quality of the data that a machine such as the LHC can deliver, we expect by the end of this decade and the beginning of the High Luminosity phase of the LHC that much progress will be made in finding an answer to these overarching problems.

\subsection{Experimental overview}

\subsubsection{Multi-particle correlations as a probe of collectivity in finite systems }\label{sec:multi}

\begin{figure}[t]
\centering
\includegraphics[width=0.35\textwidth]{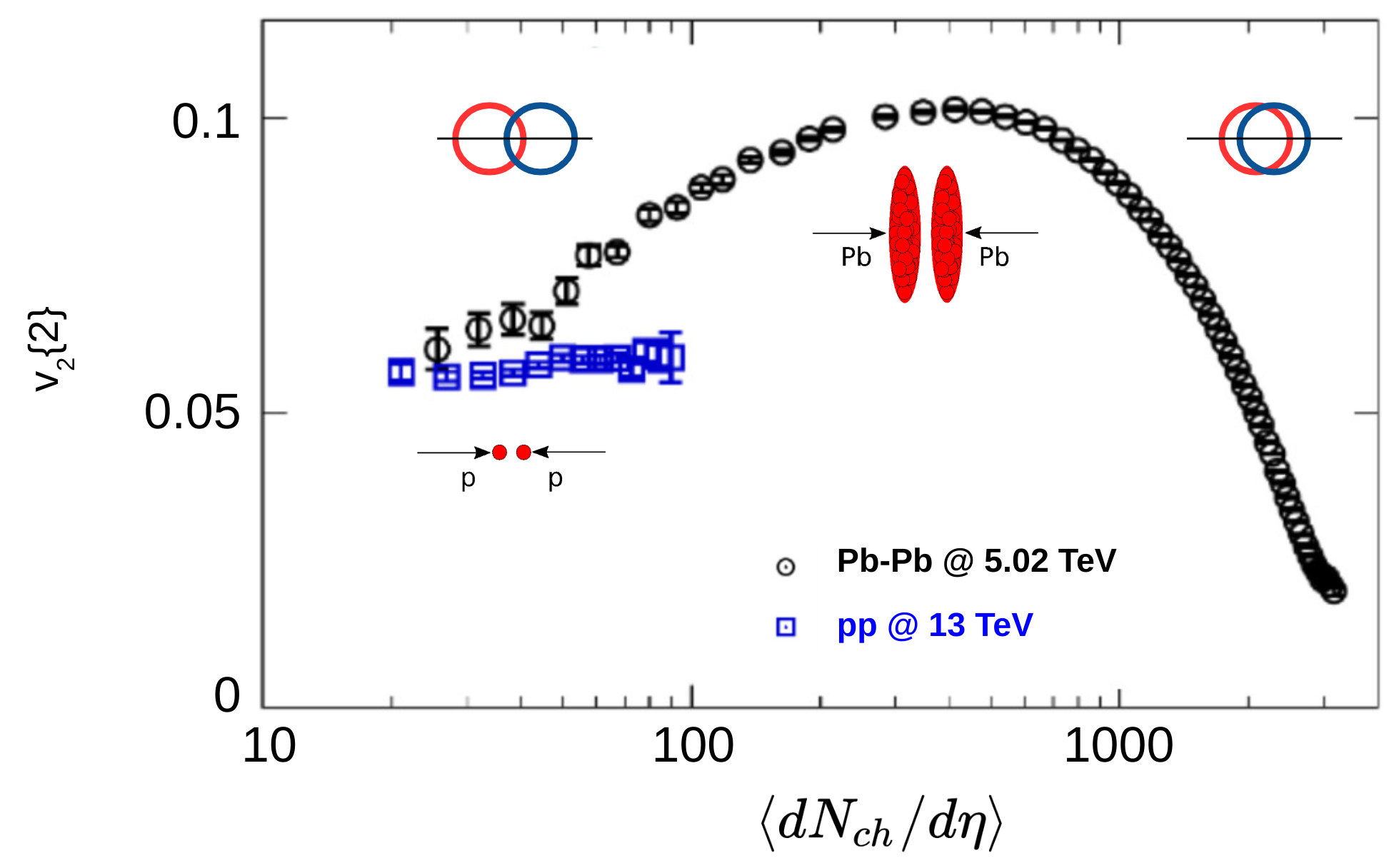}~~~%
\includegraphics[width=0.25\textwidth]{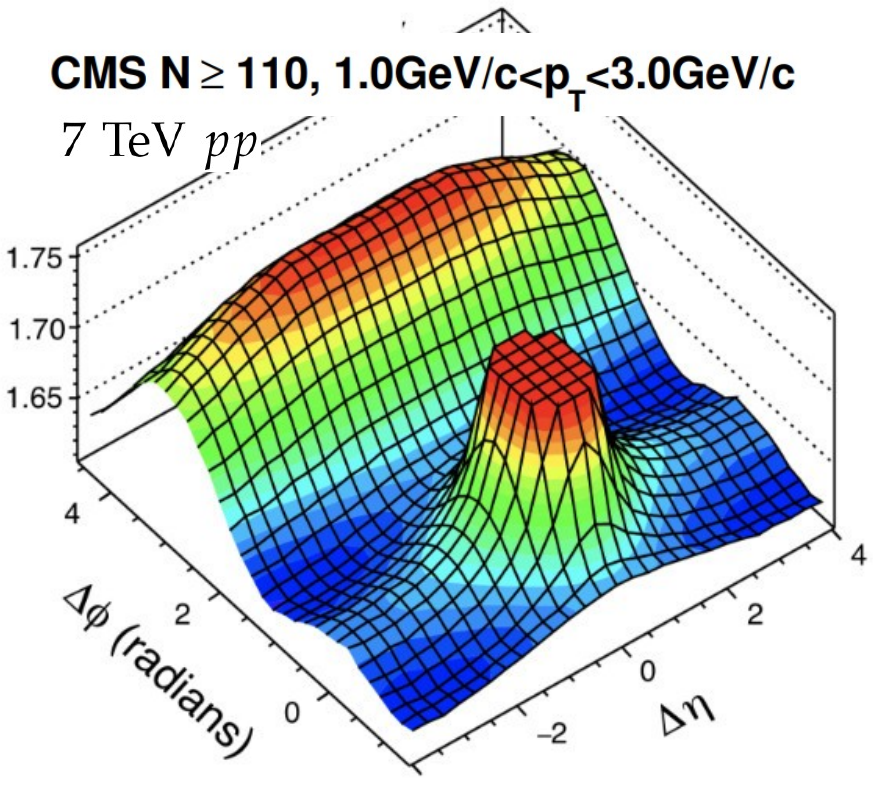}~~~%
\includegraphics[width=0.3\textwidth]{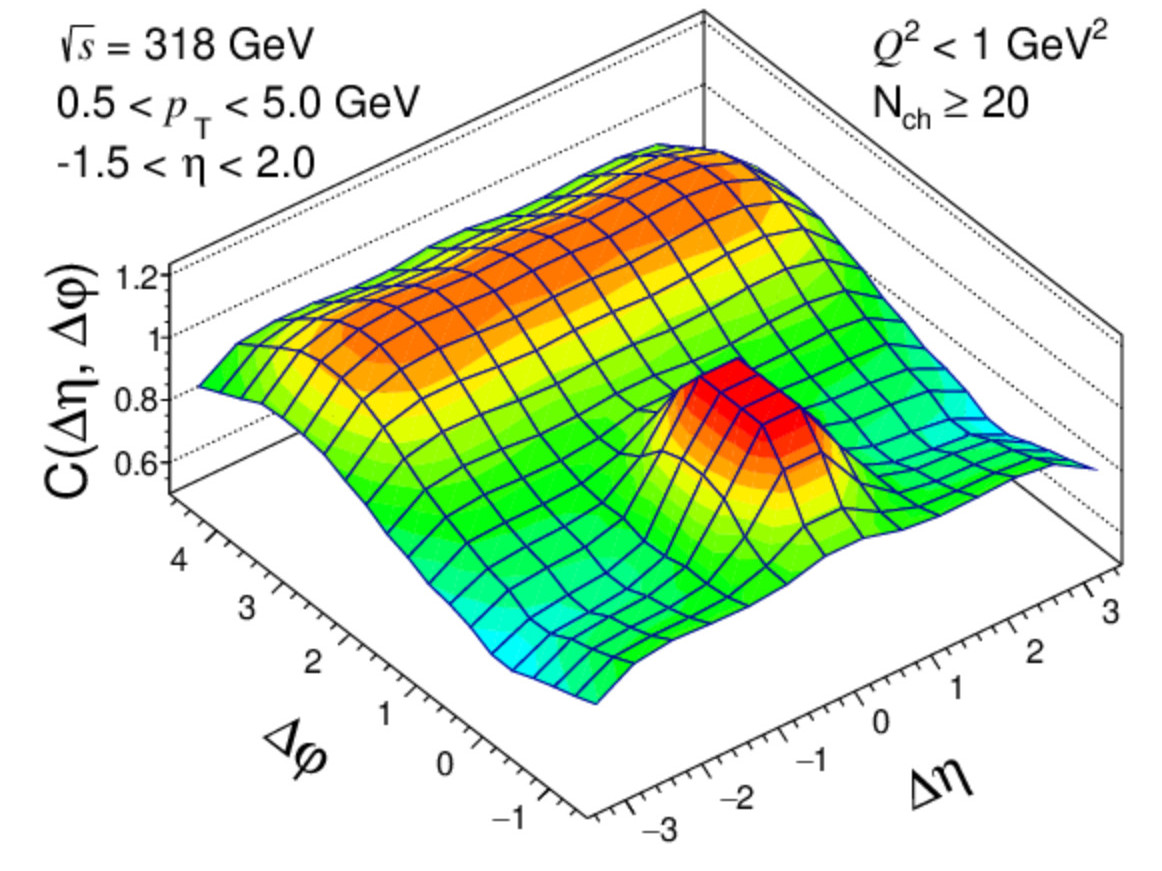}%
\caption
{
(left panel) Two particle correlation $v_2\{2\}$ vs.\ charged particle density in Pb--Pb and $pp$ collisions (figure adapted from~\cite{ALICE:2019zfl}).
Two-particle correlations $C(\Delta\eta,\Delta\varphi)$ in (middle panel) high-multiplicity proton-proton collisions at $\sqrt{s}=7$~TeV, adapted from Ref.~\cite{CMS:2010ifv} and  (right panel) $ep$ photoproduction reactions at $\sqrt{s}=318$~GeV~\cite{ZEUS:2019jya,ZEUS:2021qzg}.
}
\label{fig:2-part-correlation}
\end{figure}

The properties of the quark-gluon plasma created in heavy-ion collisions and its collective behavior are reflected in numerous observables.
The RHIC and the LHC experiments ~\cite{CMS:2010ifv,CMS:2013jlh,ALICE:2019zfl} reveal striking similarities among many observables across different systems, from heavy-ion to proton-proton collisions. Multi-particle azimuthal correlations prove a powerful tool to study the evolution of these collisions~\cite{Grosse-Oetringhaus:2024bwr}.
The few- and many-body correlations among the produced particles originate at different stages of the collision evolution~\cite{ALICE:2013vrb} and are indicative of different sources of hadron production.
Different production mechanisms result in short- or long-range structures in the distribution of the angular separation between particles~\cite{ALICE:2011svq,ALICE:2014dwt}.
These structures can be studied multi-differentially using multi-particle cumulants. The second- and fourth-order cumulants of the distribution of the elliptic flow magnitude ($v_2$) are measured as~\cite{Borghini:2000sa,Borghini:2001vi,Bilandzic:2010jr}: 
\begin{align}
    \label{eq:v22}
    c_n\{2\} & =\langle\langle{\cos n(\varphi_1-\varphi_2)}\rangle\rangle,\\
    \label{eq:v24}
    c_n\{4\} & = \langle\langle\cos n(\varphi_1-\varphi_2+\varphi_3-\varphi_4)\rangle\rangle-2\langle\langle\cos n(\varphi_1-\varphi_2)\rangle\rangle^2.
\end{align}
Here $\varphi_i$ ($i=1\dots4$) are the azimuthal angles of the produced particles, $\langle\ldots\rangle$ is the average over all pairs and quadruplets of particles in a collision, while  $\langle\langle\ldots\rangle\rangle$ is the average over all collisions in the selected centrality or multiplicity class.
An example of the charged-hadron multiplicity dependence of the two-particle cumulant ($v_2\{2\}=\sqrt{c_2\{2\}}$) measured for different colliding systems~\cite{ALICE:2011svq,ALICE:2012eyl} is shown in the left panel in Fig.~\ref{fig:2-part-correlation} for collision energies relevant to the LHC.

Two- and multi-particle correlation techniques have been successfully applied to search for similar patterns and collective behavior in even smaller systems, such as electron-proton collisions~\cite{ZEUS:2019jya,ZEUS:2021qzg}.
Among the observables used to analyze the electron-proton data is the double-differential two-particle correlation:
\begin{equation}
C(\Delta\eta,\Delta\varphi) = 
\frac{ S(\Delta\eta,\Delta\varphi)}{BG(\Delta\eta,\Delta\varphi)}.
\label{Eq:CDetaphi}
\end{equation}
Here $\Delta\eta$ and $\Delta\varphi$ are particle pair separations in pseudo-rapidity and azimuthal angle, respectively, $S$ is the number of particle pairs in the same collision and $BG$ is the combinatorial background extracted with an event-mixing technique.

\Cref{fig:2-part-correlation} shows the structures in the double-differential two-particle correlations measured in (middle panel) high-multiplicity proton-proton collisions and (right panel)  electron-proton photoproduction reactions.
In both colliding systems, it is immediate to see a strong same-side peak at zero relative angles and the away-side ridge (the structure elongated in rapidity at opposite angles, $\Delta\varphi \sim \pi$), associated with the (di-)jet production.
The distinct feature in high-multiplicity proton-proton collisions, which is not present in the $ep$ interactions, is the same-side ridge at close azimuthal angles and large rapidity separation, which is typically associated with collective flow in heavy-ion collisions.

\subsubsection{Prospects for collectivity studies in relativistic collisions }\label{sec:HIprospects}

\begin{figure}[t]
\centering
\includegraphics[width=0.48\textwidth]{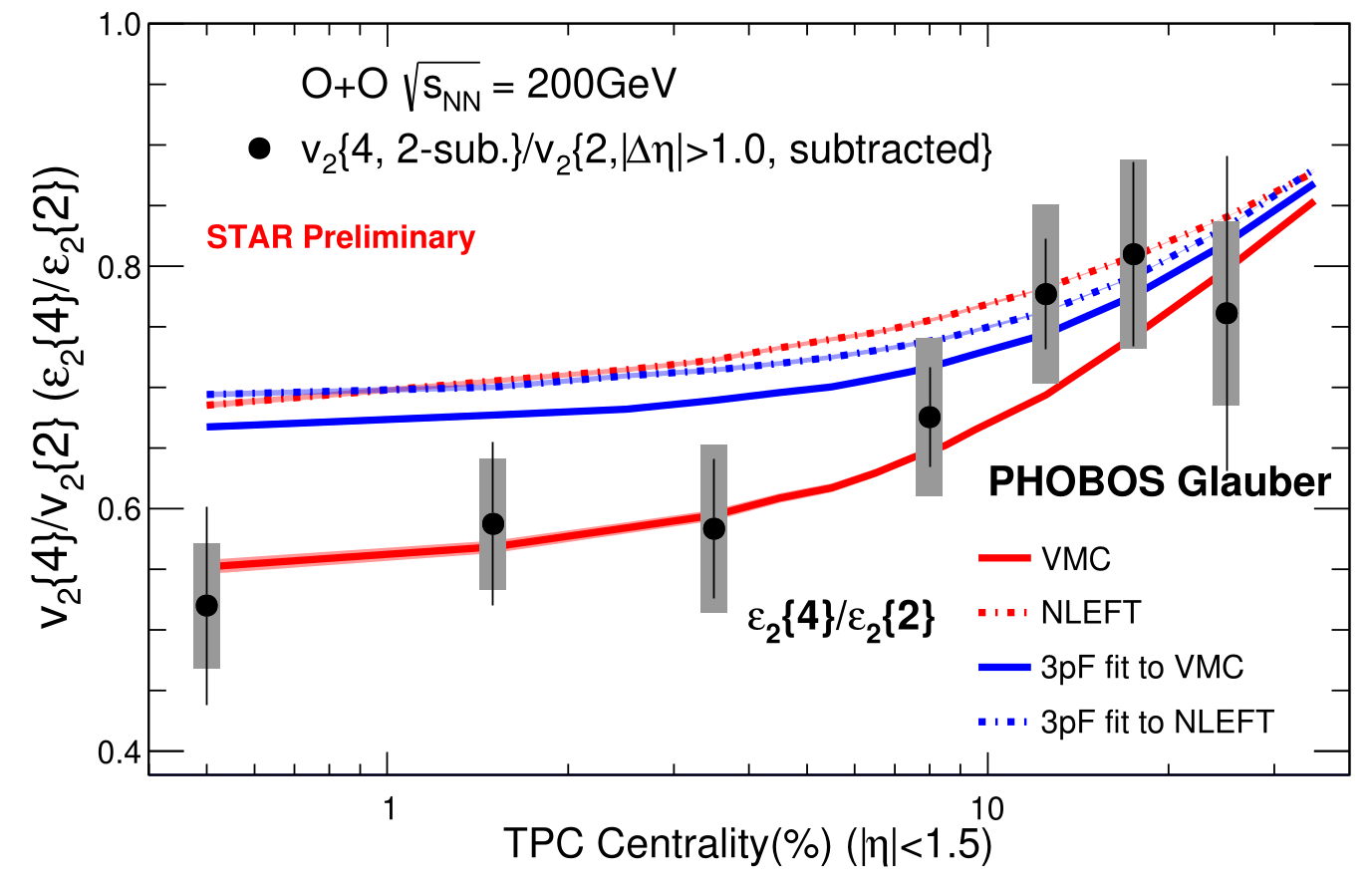}~~~~%
\includegraphics[width=0.48\textwidth]{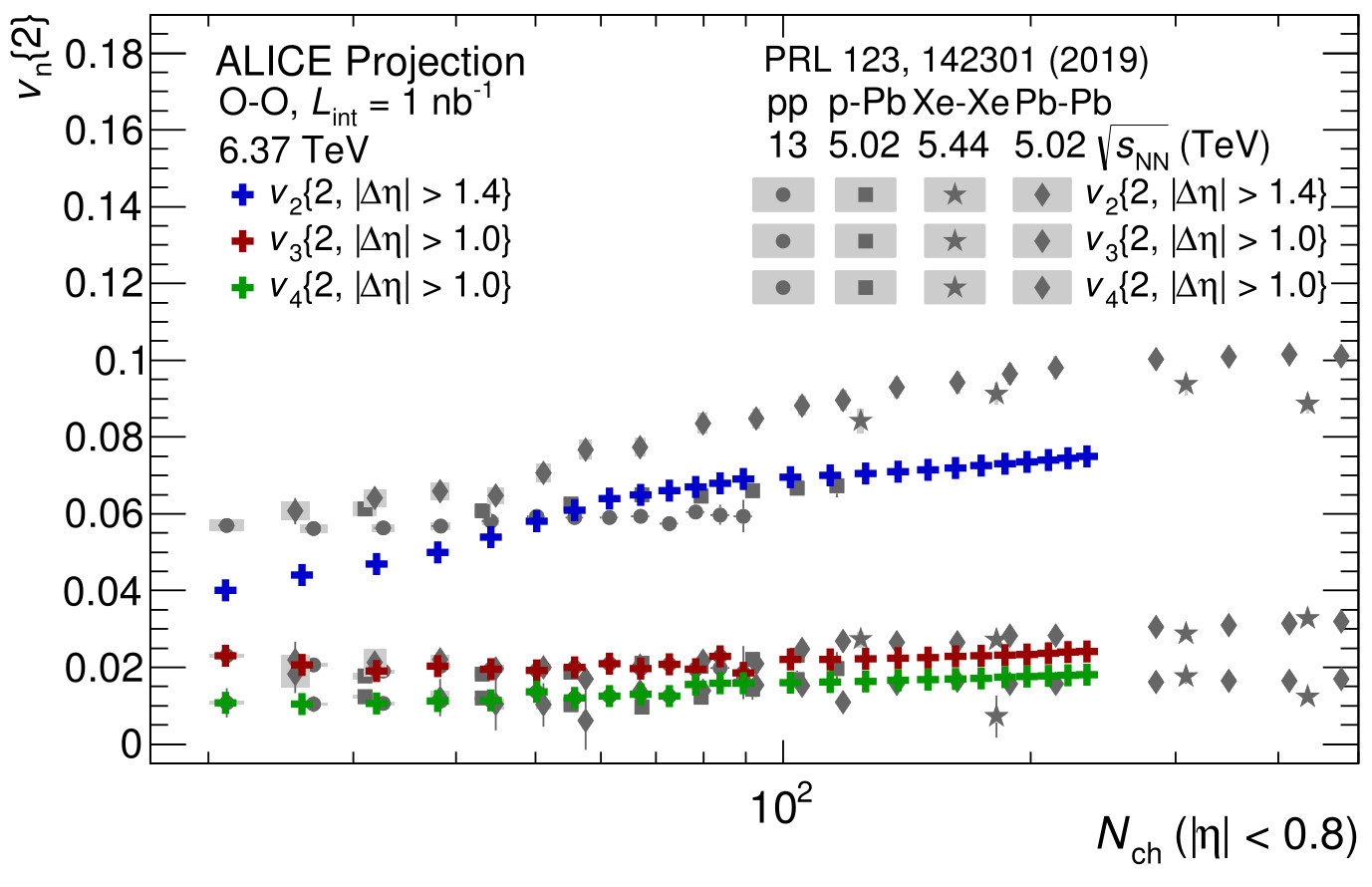}
\caption
{
(left panel)
Comparison of the ratio of the final state two- and four-particle anisotropies ($v_2\{4\}/v_2\{2\}$) in O+O collisions at $\sqrt{s_{NN}}=200$~GeV from the STAR preliminary data~\cite{Huang:2023viw} with the corresponding ratios of the initial state anisotropies ($\epsilon_2\{4\}/\epsilon_2\{2\}$) expected from different models of the oxygen nuclear-structure configurations.
(right panel)
Projections~\cite{ALICE-PUBLIC-2021-004} of the ALICE experiment at CERN for two particle correlations $v_n\{2\}$ for $n=2,3$, and $4$ vs. charged particle density in O-O collisions compared to existing data in pp, p-Pb, Xe-Xe, and Pb-Pb collisions.
}
\label{fig:Oxygen-STAR-ALICE}
\end{figure}

The studies of collectivity have bright perspectives in view of the near future programs with proton-nucleus and light-ion collisions which are gaining momentum at the LHC and SPS at CERN.
There are published results for the two-particle correlations in central Be+Be collisions from the beam energy scan program of the NA61/SHINE experiment at CERN SPS~\cite{NA61SHINE:2020tbf}, data from collisions of proton beams with light nuclei at the LHCb experiment~\cite{Bursche:2649878}, and recent preliminary data~\cite{Huang:2023viw} from the STAR experiment at RHIC for azimuthal correlations in O+O collisions (see left panel in \cref{fig:Oxygen-STAR-ALICE}).
A run with collisions of oxygen and neon nuclei took place at the LHC in summer 2025 and will allow to explore the effects seen in high-multiplicity $pp$ and $p$-Pb collisions, with a system that has a similarly small number of participating nucleons and final-state multiplicity as in $p$-Pb  but with a different sampling of the nuclei shapes in the initial state of the collision~\cite{Giacalone:2024luz}.
The projections of the ALICE experiment at CERN for two particle correlations $v_n\{2\}$ for different Fourier harmonics compared to existing data in $pp$, $p$-Pb, Xe-Xe, and Pb-Pb collisions are shown in the right panel of \cref{fig:Oxygen-STAR-ALICE}.
This data will be complemented by fixed target collisions with the upgraded LHCb's high-density gas target SMOG2 system~\cite{BoenteGarcia:2024kba,CERN-LHCC-2019-005}. Data from Ne--Ne, Pb--Ne, and Pb--Ar fixed-target collisions at $\sqrt{s}=69$ GeV (in the center-of-mass) is currently being analyzed, and theoretical calculations \cite{Giacalone:2024ixe} have highlighted the great opportunities offered by such data sets for collectivity studies.

In the long term future, the ALICE 3~\cite{ALICE:2022wwr} at the LHC and the ePIC experiment at the Electron Ion Collider~\cite{Willeke:2021ymc} will significantly extend these studies.
The large rapidity coverage of ALICE 3 will allow us to study the correlations between mid- and forward-rapidity regions of particle production, differentiating high-multiplicity events with multi-parton interactions (uniform particle density) from those with only a few high-multiplicity parton-parton interactions (strongly anisotropic).
The ePIC experiment at the EIC will allow to extend the studies of the multi-parton interactions and limits of collectivity started recently with re-analyses of the HERA~\cite{ZEUS:2019jya,ZEUS:2021qzg} and LEP~\cite{Badea:2019vey} data, and in photonuclear ultraperipheral $AA$ collisions at the LHC~\cite{ATLAS:2021jhn}.
By offering both high multiplicity $ep$ and $eA$ collisions, where the incoming virtual photon has a sufficiently long lifetime~\cite{Shi:2020djm}, the EIC has a unique position for high-precision study of collectivity in small systems and to explain its underlying mechanism.

\subsection{Theoretical overview}

\subsubsection{Applicability of hydrodynamic description in heavy-ion collisions}\label{sec:HIattractor}

A heavy-ion collision is a multi-phase process that requires a combination of different theoretical descriptions for various stages of the collision~\cite{Berges:2020fwq}. Essential parts of the standard model of heavy ion collisions are the formation of a far-from-equilibrium Glasma phase~\cite{Krasnitz:1998ns,Lappi:2006fp,Gelis:2006dv} which subsequently evolves towards viscous hydrodynamical expansion of Quark Gluon Plasma~\cite{Kolb:2003dz,Busza:2018rrf,Schenke:2021mxx}. The Glasma stage is described by solutions of 3+1-D Yang-Mills equations~\cite{Berges:2014yta}, where it is observed that the system flows to a non-thermal attractor whose dynamics is captured by an effective kinetic theory~\cite{Baier:2000sb}, leading to hydrodynamization and thermalization~\cite{Berges:2020fwq}. Remarkably, the non-thermal attractor observed is universal to that of expanding cold atomic gases, with identical real-time critical exponents describing its evolution~\cite{Berges:2014bba}. The dynamics of the attractor is described by an ultraviolet energy cascade and an infrared particle number cascade, with the latter possibly leading to the formation of a Bose-Einstein Condensate~\cite{Blaizot:2011xf,Berges:2014bba,Berges:2019oun}. We note that very similar behavior is seen in overoccupied relativistic scalar theories, where one can show explicitly that the endpoint of the infrared cascade is described by the Gross-Pitaevski equation~\cite{Deng:2018xsk}. The search for condensates in heavy-ion collisions, with a potential extension to small systems, is an important part of the scientific mission of the ALICE 3 detector~\cite{ALICE:2022wwr,Bailhache:2024mck,Florio:2025zqv}. 

The subsequent spatio-temporal evolution of QGP is obtained by numerically solving  relativistic viscous hydrodynamic equations. For causality and stability, relativistic Navier-Stokes equations are supplemented with relaxation-type equations for dissipative components of the energy-momentum tensor, i.e., Israel-Müller-Stewart equations~\cite{Romatschke:2017ejr,Kolb:2003dz,Schenke:2021mxx}. The initial energy-momentum tensor for hydrodynamic evolution is provided at some time $\tau_\text{hydro}$ after the collisions by a suitable initial state model. Important features of the initial conditions in heavy-ion collisions are spatial inhomogeneities and fluctuations of the initial energy density. These are mainly driven by the nucleus-nucleus overlap during the collision (impact parameter) and fluctuations in nucleon positions in the colliding nucleus wavefunction.
During the collision each colliding pair of nucleons produces a large number of particles (gluons and quarks), which convert the quantum fluctuations of nucleon positions to fluctuations of energy density, which are evolved classically according to hydrodynamics. Eventually the QCD matter cools down and the fluid degrees of freedom are converted into experimentally measurable hadrons.

The state-of-the-art hydrodynamic models can successfully and simultaneously describe extensive sets of experimental observables in heavy-ion collisions, such as low momentum particle spectra and harmonic flows~\cite{Bernhard:2019bmu,Nijs:2020ors}. These studies establish a strong connection between initial state geometry deformations and final state momentum anisotropies. One of the key properties of QGP extracted by these models is the viscosity over entropy ratio $\eta/s\approx 0.1-0.2$, which is close to the value computed in strongly-coupled supersymmetric quantum field theories~\cite{Policastro:2001yc} and is the smallest specific viscosity of all measured fluids~\cite{Bernhard:2019bmu}.

\begin{figure}[t]
\includegraphics[width=0.6\linewidth]{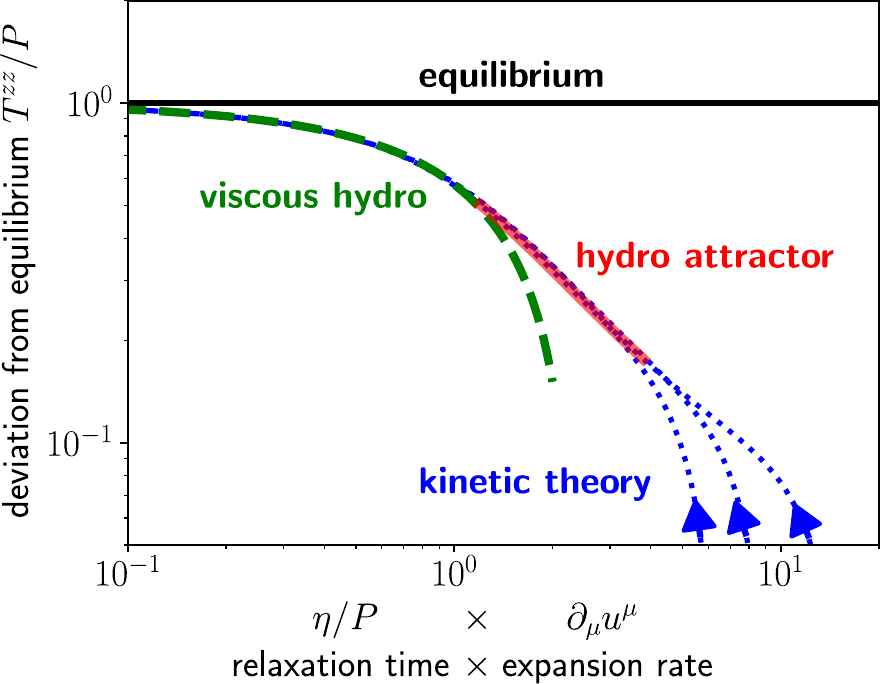}
\caption{Illustration of hydrodynamic attractor phenomena in QGP undergoing Bjorken expansion. Different QCD kinetic theory simulations (blue lines) start at an anisotropic initial state with large expansion rate. As the expansion slows down, the system approaches equilibrium by first merging to a hydrodynamic attractor (red line) and then joining Navier-Stokes hydrodynamics (green dashed line). Based on a figure from \cite{Giacalone:2019ldn}. \label{fig:hydroattractor}}
\end{figure}

The remarkable success of hydrodynamic models on sub-atomic scales
spurred the theoretical studies of new hydrodynamic theories~\cite{Schenke:2021mxx,Nagle:2018nvi}.
The applicability of hydrodynamics is helped by the specific shear viscosity being small, so that sizable velocity gradients $\partial_\mu u^\nu$ generate small corrections compared to the equilibrium pressure.
Secondly, Navier-Stokes equations provide a good description of the evolution outside the naive regime of applicability, i.e., even when corrections to equilibrium are of order 1. This is the consequence of the \emph{hydrodynamic attractor} phenomenon~\cite{Heller:2011ju,Florkowski:2017olj,Soloviev:2021lhs,Berges:2020fwq}. Figure \ref{fig:hydroattractor} depicts the pressure isotropization of QGP as described by microscopic QCD kinetic theory~\cite{Arnold:2002zm,Kurkela:2018oqw,Giacalone:2019ldn}. The expansion rate $\partial_\mu u^\mu=1/\tau$ is monotonically decreasing and the system approaches equilibrium, which is characterized by the deviation of the $zz$ component of the energy-momentum tensor from the equilibrium value. We observe that different kinetic evolutions collapse onto a common curve when the energy-momentum tensor component is 5 times smaller than the equilibrium value. Such rapid loss of initial state information and the universal far-from-equilibrium approach to equilibrium is called a hydrodynamic attractor (marked in red). Notably, the hydrodynamic attractor is different from the universal \emph{near-equilibrium} approach described by Navier-Stokes equations (marked in green). Nevertheless, the hydrodynamic attractor joins viscous hydrodynamic predictions when deviations from equilibrium are still a factor of 2 away from equilibrium. This is the main theoretical explanation for why viscous hydrodynamic simulations are successful in describing rapidly expanding QGP already $1\,\text{fm}/c$ after the collision~\cite{Berges:2020fwq}.

In recent years hydrodynamic attractors have been an active topic of theoretical research~\cite{Florkowski:2017olj,Romatschke:2017ejr,Soloviev:2021lhs,Jankowski:2023fdz}. 
They have been studied using different microscopic theories: generalization of Navier-Stokes, holographic models, and kinetic theories. Their mathematical structure (trans-series) reveal information about the non-convergence of hydrodynamics and gradient expansion~\cite{Heller:2018qvh,Aniceto:2018bis,Aniceto:2024pyc} (see also \cref{sec:nonhydro}). In heavy-ion collisions, the hydrodynamic attractors cannot be cleanly separated from other stages of the collision, and direct experimental evidence of this phenomenon is lacking. Possibly in smaller collision systems, such as light-ion collisions~\cite{Brewer:2021kiv}, where the pre-equilibrium stage is relatively longer, they could play a more significant role. High-momentum partons traveling through the QGP perturb the medium and therefore could be also used to study the non-equilibrium dynamics~\cite{CMS:2025dua}.
Recently there have been proposals~\cite{Fujii:2024,Mazeliauskas:2025jyi,Heller:2025yxm} to study hydrodynamic attractors in ultra-cold gases, see \cref{sec:attractor}. Highly tunable cold atom experiments are a promising avenue for discovering the hydrodynamic attractor phenomenon.

\subsubsection{Probing entanglement dynamics in ultra-peripheral collisions }\label{sec:UPCent}

Hanbury-Brown--Twiss (HBT) intensity interferometry, originally developed for astrophysical imaging, is a powerful technique to extract the spacetime structure of a chaotic source from measurements of correlations in  field intensities of identical particles emitted by the source in separate detectors. HBT can be understood as a wave interference phenomenon and has been demonstrated in classical and quantum optics. In addition to quantum optics,  it has become an essential tool across various fields of physics, from astronomy, nuclear physics, and quantum optics---a comprehensive recent review can be found in Alain Aspect's Les Houches lectures~\cite{aspect2020hanburry}. 

All these cases require that the intensity interferometry be that of indistinguishable particles.  Cotler and Wilczek, and subsequently, Cotler, Wilczek and Borish,  showed that intensity interference effects could be recovered in measurements of nonidentical particles by passing the emitted particles through a device that performs a unitary transformation entangling their wavefunctions, and subsequently filters the entangled state prior to their measurement in a detector. They dubbed this phenomenon “entanglement enabled intensity interferometry”, or $E^2 I^2$ for short, and with experimental colleagues demonstrated this effect  in optical intensity interferometry experiments utilizing distinguishable photons of two distinct wavelengths~\cite{qu_chromatic_2020}.

\begin{figure}[t]
\centering
\includegraphics[width=0.6\textwidth]{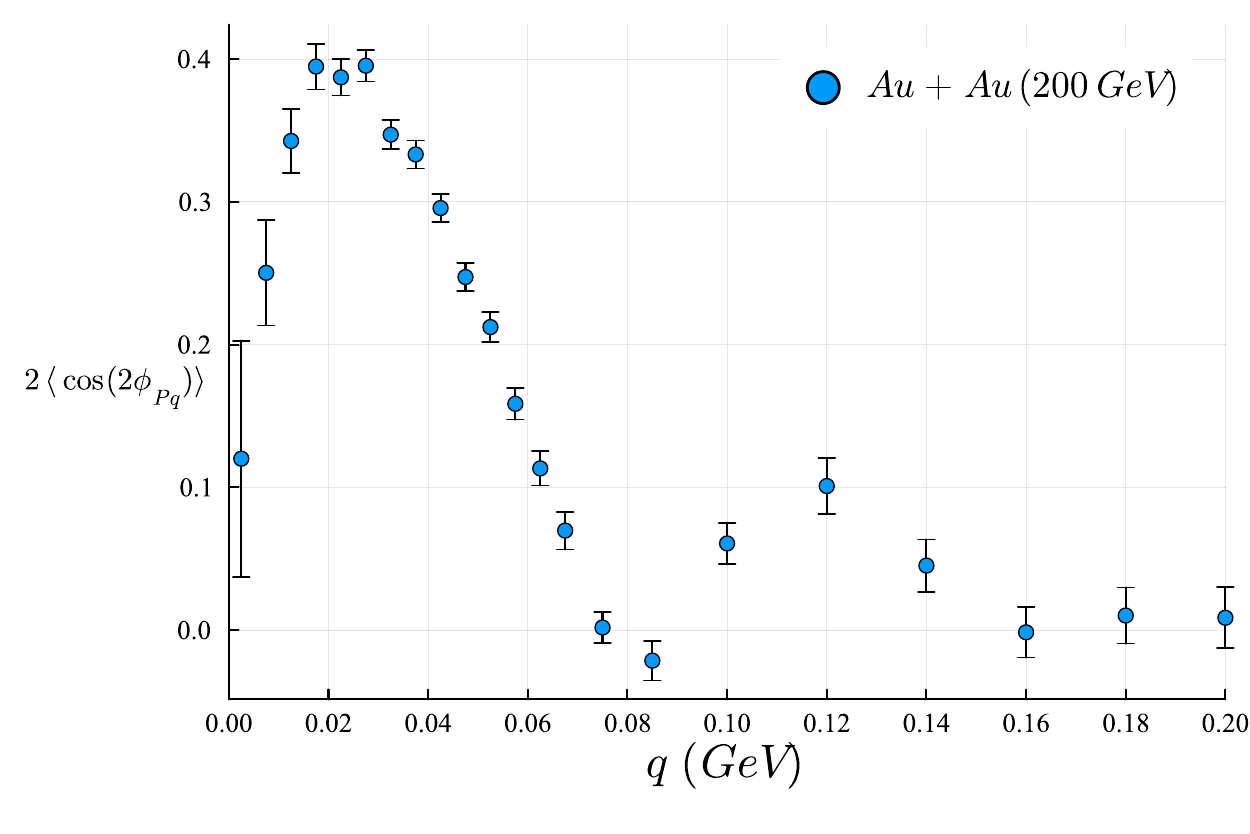}
\caption{\label{fig:UPC} 
Two-particle correlation data from the exclusive production of $\pi^\pm$-pairs arising from the decay of $\rho$-mesons in ultrarelativistic ultraperipheral heavy-ion collisions at RHIC~\cite{STAR:2022wfe}. The correlation data demonstrates the $\cos(2\Phi_{Pq})$ angular modulation in the azimuthal angle $\phi_{Pq}$ between vectors $q$ ($= (p_1 + p_2)$, corresponding to the $\rho$-meson momentum) and $P_\perp$ ($= (p_1 - p_2)$, the relative momentum), constructed from $p_1$ and $p_2$ corresponding to the momenta of the produced $\pi^\pm$-pairs in the laboratory frame, projected on the plane orthogonal to the beam axis.}
\end{figure}

In the context of high-energy physics experiments, probes of entanglement dynamics have only recently begun to attract significant attention \cite{Afik:2025ejh}. The STAR collaboration speculated in particular that $E^2 I^2$ explains  patterns observed in exclusive decays of vector mesons  measured in ultrarelativistic ultraperipheral nuclear collisions (UPCs) at the Relativistic Heavy Ion Collider (RHIC)~\cite{STAR:2022wfe}. In UPC experiments at RHIC, and at the Large Hadron Collider (LHC), beams of heavy ultrarelativistic nuclei generate extremely powerful electromagnetic fields. 
These allow for clean studies of strongly interacting matter because equivalent Weizs\"{a}cker-Williams (WW) photons from one of the nuclear beams can scatter (directly) off quark and (indirectly) off gluon fields in the other nucleus, producing strongly interacting subatomic particles in the final state. Of particular interest for $E^2 I^2$ are exclusive diffractive photoproduction processes where the decay products of a single final state (such as a vector meson) can be studied in isolation separated by a gap in rapidity from the struck nucleus. 
Specifically, STAR observed in exclusive $\rho^0$-meson decays into {\it distinguishable} $\pi^\pm$ pairs a strong $\cos2\phi$ and a modest $\cos4\phi$ interferometric modulation in the coherent cross-section (corresponding to intact nuclei). The two-particle correlation data are shown in Fig.~\ref{fig:UPC}. The angle $\phi$ is the azimuthal angle between $q = (p_1 + p_2)$ and $P_\perp = (p_1 - p_2)$, with $p_{1,2}$ being the momentum vectors of the daughter $\pi^\pm$ particles, projected along the plane orthogonal to the beam axes. 
The observed $\cos2\phi$ modulation is strongest at low transverse momentum, peaking with a value of $\approx 40\%$ at a $|q| \approx 20$ MeV/c and shows a wave interference structure exhibiting a minimum and second maximum around $|q| \approx 120$ MeV/c~\cite{STAR:2022wfe}.

The data show interesting similarities and differences to the two particle correlation patterns observed in HBT studies. The dip at $q\rightarrow 0$ is due to destructive interference of the identical amplitudes where the $\rho$-meson is produced off one nucleus or the other. (The interference is destructive because the polarization vectors in the two cases point in opposite directions, along the line of the impact parameter of the collision.)
The subsequent peak and minima carry however important information on the size and matter distribution inside the nuclei. In particular, once $q\sim 0.1$ GeV, the data are sensitive to few-body nuclear correlations inside the nuclei, and beyond this scale to strongly interacting dynamics inside the protons and neutrons themselves. 

The entire interference pattern (besides the ``trivial'' one at $q\rightarrow 0$) can be simply understood as a variant of the Cotler-Wilczek mechanism. The entire formation and decay of the $\rho$-meson is clearly a unitary (if poorly understood) process in QCD, with the $\rho$-meson acting as unitary transformation that entangles the wavefunctions of the $\pi^+$-meson with that of the $\pi^-$ before they are released in the decay. Indeed, if the $\pi^\pm$ pair were formed directly (and not via the decay of the $\rho$-meson), there would be no such interference pattern. (This is seen in experiment by requiring that the reconstructed invariant mass of the $\pi^\pm$-pair is close to the $\rho$-meson mass of $\sim 770$ MeV.)  More specifically, the $\rho$-meson is a spin-1 object, while the pions are spin-zero states. The $\pi^\pm$ bound state constituting the $\rho$ is then a P-wave angular momentum state. The cosine modulation then simply reflects the entanglement arising from the projection of the pions into their $L=1$ angular momentum eigenstates. 

The details of this computation are worked out in \cite{Brandenburg:2024ksp}. A quantitative comparison to data however requires more work, which is in progress. Of particular interest is the second peak and minimum in the range $q\sim 0.1-0.6$ GeV. As noted, this physics corresponds to gluon correlations within the nucleon. Further, in this kinematic range, the production of the $\rho$-meson causes the struck nucleus to break up into nucleons, a process called ``incoherent diffraction". Since the nucleons are still intact, the gluons that produce the $\rho$-meson must be in color singlet, so-called ``pomeron" configurations. The process in this kinematics is 
$\gamma+\mathcal{P}\rightarrow \rho\rightarrow \pi^++\pi^-$ where $\mathcal{P}$ represents the Pomeron. Thus $E^2I^2$ opens a novel window into understanding the structure and dynamics of the pomeron. Similar studies can be repeated for other vector meson two-body decays ($\phi\rightarrow K^+ K^-$, $J/\Psi\rightarrow e^+ e^-$), potentially providing further insight into the flavor structure of the pomeron. It is also conceivable that the $E^2 I^2$ framework could provide clear evidence for a conjectured ``odderon'' color singlet configuration of three gluons in the decay of $\chi_c$ mesons. With sufficient UPC data one can also explore i) $\rho-\omega$-meson mixing, with the latter, of comparable mass, decaying into a three-body $\pi^-\pi^+\pi^0$ final state, and ii) decays of higher mass resonances. All of these $E^2 I^2$ studies can also be carried out at the future Electron-Ion Collider~\cite{Aschenauer:2017jsk}, where polarized proton beams and polarized light nuclei will provide a further handle on nucleon and gluon correlations in appropriate kinematic ranges. 

From the perspective of a high energy physicist, the interesting question in the context of this report is whether a version of the $E^2 I^2$ process as described can be realized in few-body binary ultracold atom systems when the atoms are released from the trap. The phase information in transient bound states of the distinct atoms might similarly contain information on their dynamics in coincident measurements of the distinct atom species. We hope the above discussion will trigger ideas to be explored along these lines.

\section{Conclusions and outlook}\label{sec:conclusions}

In conclusion, the EMMI RRTF~\cite{EMMI_RRTF} gathered experts in cold-atom and high-energy physics to discuss newly emerging issues related to the puzzling observation of emergent many-body dynamics in mesoscopic quantum gases. Motivated by groundbreaking experiments in both fields, our goal is to establish a research program where Anderson's famous dictum \emph{More is different}~\cite{Anderson:1972pca} becomes the subject of quantitative studies. Our driving questions are: When does \emph{few} become \emph{more}? What if \textit{few} is also \textit{different}? Experiments with cold atoms or in high-energy particle colliders are ideally suited to probe and qualify this transition.

\begin{figure}[t]
    \centering
\includegraphics[width=0.65\linewidth]{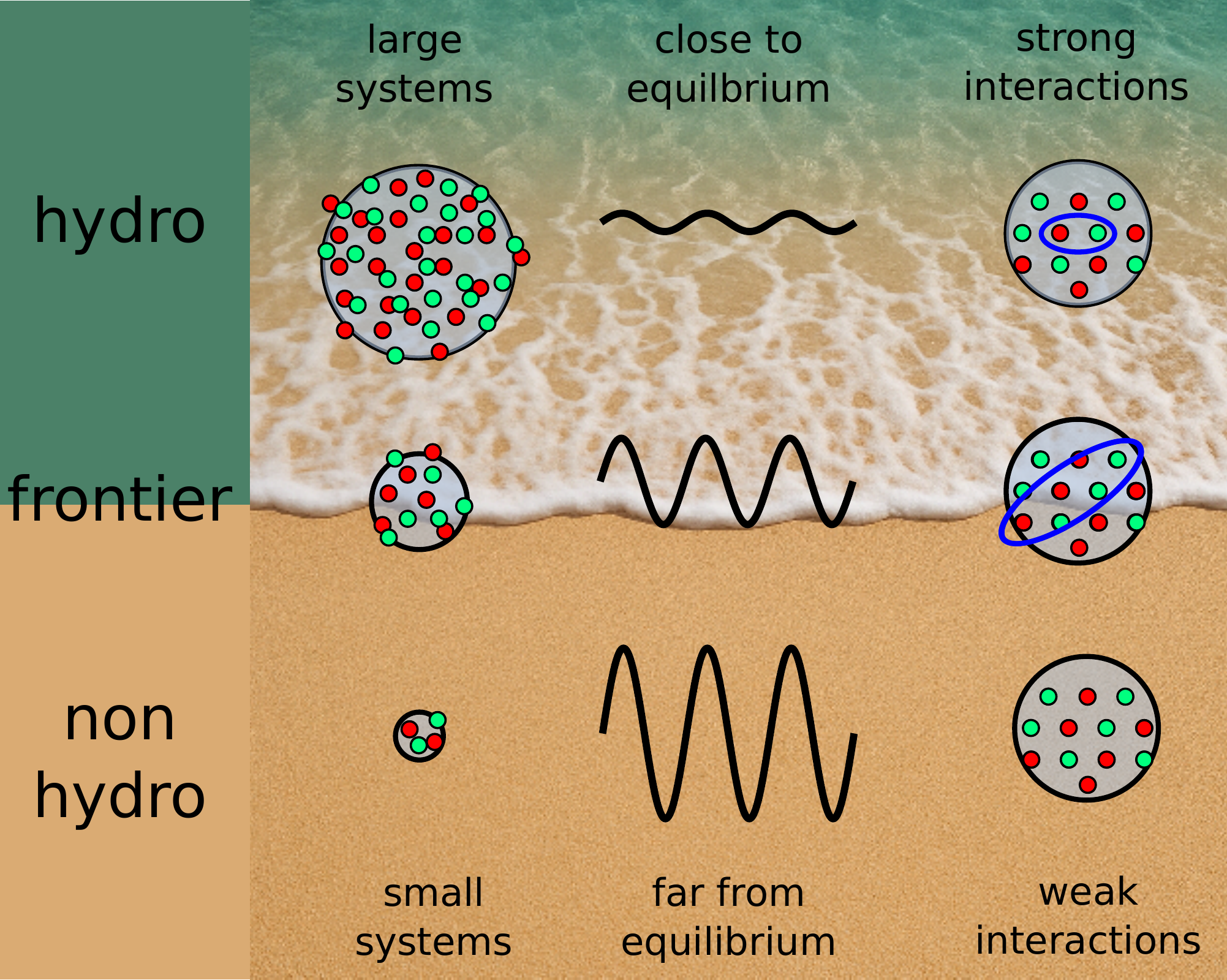}
    \caption{Frontiers of hydrodynamic (in)applicability as a function of the system size, closeness to equilibrium, and interaction strength. 
    The RRTF report summarizes the latest developments in experimental and theoretical research on the quantitative determination of the applicability region of hydrodynamics and other effective theories.}
    \label{fig:hydro_frontiers}
\end{figure}

More specifically, we have focused on the boundaries of applicability of effective many-body descriptions of matter, such as hydrodynamics, in regimes where longstanding textbook definitions should not be applicable. Traditionally, hydrodynamic approaches are meant to describe phenomena over long wavelengths compared to microscopic scales (large systems) and when their associated dynamics is slow enough (small frequencies) to permit microscopic interactions to keep the system close to local equilibrium. Yet, the recent theoretical developments and experimental results discussed in this report challenge these assumptions. We identify three principal frontiers that define the limits of (in)applicability of hydrodynamics and which are illustrated in \cref{fig:hydro_frontiers}: the size frontier, the equilibrium frontier, and the interaction frontier. In this section, we summarize how the theoretical and experimental developments discussed in this report inform us on these frontiers, and outline promising directions for future investigation.

\subsection{Size frontier: collective dynamics of few constituents}

Naively, one might expect that the ten-particle Schrödinger equation could be solved exactly, offering a definitive understanding of the transition from few- to many-body physics. In reality, few exact quantitative results are available from first principles for intermediate-size systems. For certain symmetric configurations, energy spectra can be computed using exact diagonalization, for instance, up to $N\sim 6$ fermions or $N\sim 30$ bosons in the lowest Landau level (see \cref{sec:microscopic}). Even approximate methods, such as the explicitly correlated Gaussian (ECG) technique, are currently limited to systems with around 10 fermions in symmetric traps (\cref{sec:ECG}). Moreover, real-time dynamics of just a handful of interacting atoms remains computationally intractable. Thus, a key challenge is to bridge the gap between few-body exact methods and effective continuum approaches like density functional theory (for fermions) or the Gross-Pitaevskii equation (for bosons), and to quantify the validity of these continuum theories in intermediate regimes (\cref{sec:dissipation}). Experimental observations provide essential benchmarks for tracking the crossover from few- to many-body behavior.

Despite these challenges, several theoretical and experimental results suggest that certain observables show remarkable insensitivity to the number of constituents. As discussed in \cref{sec:fewatoms}, quantities such as the Bertsch parameter, the contact, and the Fermi polaron energy exhibit only mild dependence on the number of atoms. On the experimental side, cold atom setups with up to ten interacting lithium atoms released from anisotropic 2D traps reveal geometry inversion patterns consistent with hydrodynamic expansion of the average particle number density (\cref{sec:flow,sec:CAfluid}). Similarly, in high-energy nuclear collisions, signatures of collectivity have been observed across all hadronic systems, even with relatively small final-state multiplicities (\cref{sec:smallpuzzle,sec:multi}). These findings highlight the relevance of identifying which observables can be reliably captured by many-body effective theories, even in systems with very few constituents.

Looking ahead, theoretical efforts should aim to systematically connect few-body models with hydrodynamic and macroscopic descriptions. They should help answer questions such as how many atoms are needed to form a condensate or whether one can test for superfluidity in a small system by detecting angular momentum upon rotation (\cref{sec:microscopic}). One potential way is to incorporate higher-order gradient corrections that are expected to play a sizable role in small systems (\cref{sec:dissipation,sec:CAfluid}). On the experimental side, continued exploration of few-atom gases, including mixed-species Fermi gases (\cref{sec:mixedFG}), will be crucial. Composite degrees of freedom provide an extension of the cold-atom toolbox that could help model aspects of the QGP such as specific mass ratios and cluster bound states and access new, subdiffusive transport regimes (\cref{sec:composite}). In high-energy physics, legacy data from $ep$ and $e+e$ collisions have been re-analyzed to place bounds on the requirements for the emergence of collective behavior (\cref{sec:multi}). Experiments using light-ion species in both collider and fixed-target modes at the LHC offer new platforms for testing models of collectivity with varying system sizes and geometries (\cref{sec:HIprospects}). In such cases, corrections associated with system size and lifetime become especially significant, and must be carefully accounted for in hydrodynamic modeling (\cref{sec:HIattractor}).

\subsection{Equilibrium frontier: short time scales and hydrodynamization}

Hydrodynamics is conventionally formulated to describe the long-time, low-frequency behavior of a system, capturing only modes near equilibrium. In contrast, nuclear collisions produce QCD matter that is both highly excited and extremely short-lived, existing for only about 100 yoctoseconds (or $10^{-22}\,\text{s}$ for the largest systems). It is therefore a nontrivial and striking feature of QCD that such systems appear to reach local thermal equilibrium quickly enough for a fluid description to become valid. This phenomenon of rapid hydrodynamization has been attributed to the existence of hydrodynamic attractors, which enable such a behavior to emerge as soon as non-hydrodynamic modes decay (\cref{sec:HIattractor}). 

Similar attractor dynamics and the role of non-hydrodynamic modes can also be explored in cold atom systems (\cref{sec:nonhydro,sec:attractor}). In these experiments, observing the early-time dynamics with high temporal resolution is critical to distinguishing genuine hydrodynamic behavior from transient non-equilibrium evolution.  In particular, fast, real-time probes of non-equilibrium dynamics, such as high-resolution measurements of the equation of state during evolution, are within the reach of modern platforms (\cref{sec:diffusivity,sec:dissipation}). These advances offer a novel laboratory for testing the onset of hydrodynamic behavior and exploring the frequency-dependent response of many-body systems.

Looking ahead for the theoretical counterpart, pushing into the high-frequency or short-time domain will require significant development. In particular, it will be necessary to go beyond conventional kinetic theory to understand the response of many-body systems to both rapid and large-amplitude perturbations (\cref{sec:attractor,sec:HIattractor}). These developments are especially timely in view of high-precision measurements of flow phenomena in collisions of $^{16}$O and $^{20}$Ne nuclei. This will enable a systematic mapping of collectivity from ultra-central light-ion collisions, which we expect to be dominated by hydrodynamic effects, to more peripheral collisions reaching into the small-system regime where out-of-equilibrium corrections become more and more significant.
These investigations will be complemented by precision measurements of high-momentum jets traversing the QGP in heavy-ion collisions, which perturb the medium throughout its entire evolution.

\subsection{Interaction frontier: strong coupling and many-body correlations} 

At both extremes of the temperature spectrum, having strong interactions among constituents is a crucial requirement for the onset of collective or hydrodynamic behavior. 

At finite temperature in the standard picture of a collisional fluid, a system can exhibit hydrodynamic behavior if its constituents interact frequently enough to establish local equilibrium. In the case of the QGP produced in heavy-ion collisions, strong interactions are in particular key to lead to low-viscosity behavior, which ultimately enables the emergence of collective dynamics despite the small size and short lifetime of such systems. Whether atomic gases below the critical temperature can surpass the QGP in achieving even lower specific viscosity remains an open question (\cref{sec:hydro_2}). An important direction to explore in the future involves quantifying transport coefficients and dissipative processes at strong coupling, where standard kinetic theory approaches may fail (\cref{sec:uniformFG,sec:composite,sec:diffusivity}).

At zero temperature, superfluid behavior, as dictated by the GPE, can emerge in a fermionic system if interatom interactions are strong enough to lead to the formation of fermionic pairs (the degrees of freedom that can form a condensate). The few-body experiments at Heidelberg University have, in particular, directly probed the onset of elliptic flow as a function of the interaction strength (\cref{sec:flow,sec:uniformFG}), showing that this emergent collective phenomenon is indeed driven by interactions. An important open question to clarify is whether the observed collective fluid behavior in microscopic systems can be ascribed to the effect of two-body interactions alone, or whether genuine higher-order correlations (e.g., three-body or beyond) are essential to explain the data. This is especially relevant in the regimes achieved in the few-body setup of the Heidelberg few-body lab, where the diluteness condition underlying the GPE is explicitly broken (pair size is on the same order as the system size). Therefore, a key experimental frontier is the direct measurement of two-body and higher-order correlations in the regime of strong interactions (\cref{sec:diffusivity,sec:mixedFG}). These measurements will be critical in elucidating how hydrodynamic behavior is related to underlying many-body quantum correlations. 

Finally, entanglement has been identified as a key mechanism underlying thermalization processes~(\cref{sec:entanglement}). Experiments with as few as six atoms have demonstrated that local thermalization coincides with the growth of entanglement, while atypical low-entanglement states exhibit remarkably slow relaxation. In high-energy collider physics, quantum entanglement has been suggested as a possible explanation for the surprisingly universal features of thermal hadron production, down to elementary processes such as $e^+ e^-$ collisions. Direct probes of quantum entanglement in particle collisions have only recently started to attract attention in the community. In particular, entanglement-enabled intensity interferometry has been proposed for future electron-ion collisions, and exploring analogous effects in ultracold atomic systems could open new avenues for testing the connection between emergent many-body dynamics and this genuinely quantum phenomenon~(\cref{sec:UPCent}).

\phantomsection
\addcontentsline{toc}{section}{Acknowledgments}
\section*{Acknowledgments}
We gratefully acknowledge the support of the ExtreMe Matter 
Institute EMMI at GSI, Darmstadt, which made the Rapid 
Reaction Task Force \emph{Deciphering many-body dynamics in
mesoscopic quantum gases} possible~\cite{EMMI_RRTF}. We also
thank the Institute for Theoretical Physics at Heidelberg 
University for hospitality and organizational support.

This work was supported in part by the DFG through the Emmy 
Noether Program (project number 496831614) (AM), through 
CRC 1225 ISOQUANT (project number 27381115) (AM, TE, RV), and 
through Germany’s Excellence Strategy EXC 2181/1-390900948 
(Heidelberg STRUCTURES Excellence Cluster) (TE).
JL and MMP acknowledge support from the Australian Research 
Council (ARC) through the ARC Discovery Projects DP240100569 
and DP250103746, and the ARC Future Fellowship FT200100619.
KF is supported by JSPS KAKENHI Grant No. JP24KJ0062.
TS is supported by the Office of Science, US Department of 
Energy, under contracts DE-FG02-03ER41260 and DE-SC0024622. RV is supported by the U.S. Department of Energy, Office of Science under contract DE-SC0012704 and within the framework of the SURGE Topical Theory Collaboration.
TVZ received funding under Horizon
Europe programme HORIZON-CL4-2022-QUANTUM-02-SGA via the Project No. 101113690 (PASQuanS2.1). FS acknowledges support from the European Research Council (ERC) under the European Union’s Horizon 2020 research and innovation programme (project OrbiDynaMIQs, GA No.~949438), and from the Italian MUR under the PRIN 2022 programme (project CoQuS, Prot.~2022ATM8FY). NN acknowledges support from the NSF (Grant No. PHY-1945324),
AFOSR (Grant No. FA9550-23-1-0605), and the David and Lucile Packard Foundation. 
YY acknowledges support from the Hong Kong RGC under Grant No. 14301425 and 24308323, and CUHK Direct Grant No. 4053731.
 S.M.R. was supported by the Swedish Research Council under 
grant No. 2022-03654VR and by the Knut and Alice Wallenberg 
Foundation under grant No. KAW2018.0217 and KAW2023.0322.
QG acknowledges support from the NSF under Grant No. PHY-2409600.

\newpage

\phantomsection
\addcontentsline{toc}{section}{References}
\printbibliography

\end{document}